\documentclass[final]{ws-ijmpa}
\usepackage[super,compress]{cite}
\usepackage{graphicx}

\newcommand{\pt}{\ensuremath{p_{\mathrm{T}}}\xspace}
\newcommand{\MeV}{\ensuremath{\,\text{Me\hspace{-.08em}V}}\xspace}
\newcommand{\GeV}{\ensuremath{\,\text{Ge\hspace{-.08em}V}}\xspace}
\newcommand{\TeV}{\ensuremath{\,\text{Te\hspace{-.08em}V}}\xspace}

\RequirePackage{xspace}
\RequirePackage{amsmath}

\begin{document}

\title{Review of bottomonium measurements from CMS}

\author{Z. Hu}
\address{Fermi National Accelerator Laboratory, Batavia, IL 60510, USA\\
zhenhu@fnal.gov}

\author{N. T. Leonardo}
\address{Laborat\'{o}rio de Instrumenta\c{c}\~{a}o e F\'{\i}sica Experimental de Part\'{\i}culas\\
Lisboa 1649-003, Portugal \\
nuno.leonardo@cern.ch}

\author{T. Liu}
\address{Fermi National Accelerator Laboratory, Batavia, IL 60510, USA}

\author{M. Haytmyradov}
\address{The University of Iowa, Iowa City, IA 52242, USA}

\maketitle


\begin{abstract}
We review the results on the bottomonium system from the CMS
experiment at the Large Hadron Collider.  Measurements
have been carried out at different center-of-mass energies in proton  
collisions and in collisions involving heavy ions. 
These include precision measurements of cross sections and polarizations,
shedding light on hadroproduction mechanisms, 
and the observation of quarkonium sequential suppression, 
a notable indication of quark-gluon plasma formation.   
The observation of the production of bottomonium pairs is also
reported along with searches for new states. We close with a brief
outlook of the future physics program.

\end{abstract}

\ccode{PACS numbers: 14.40.Pq, 25.75.Nq, 13.85.Ni}

\tableofcontents

\vspace*{8mm}

\begin{center}
{\footnotesize{[Published as {\it{Int. J. Mod. Phys. A}} {\bf{32}} (2017) 1730015]}} 
\end{center}

\thispagestyle{empty}

\clearpage

\section{Introduction}	

Bottomonia, bound states $b\bar{b}$ of a bottom quark and its antiparticle,
constitute the heaviest meson system. The only heavier quark, the top
quark, decays before it can hadronize. Light quark ($u$, $d$, $s$)
constituents in mesons typically move at relativistic speeds as their
masses are considerably smaller than the meson masses. Heavy quarkonia,
charmonia and especially bottomonia, are in contrast approximately
non-relativistic systems. This allows in turn the application of
effective theoretical approaches to describe nonperturbative effects in
Quantum Chromodynamics (QCD). As a result, heavy quarkonia provide a
valuable framework for probing the strong interaction as described by
QCD within the Standard Model (SM) and for searching for new phenomena.

The description of heavy quarkonium production is achieved through effective models of QCD. 
A most complete approach is provided by nonrelativistic QCD (NRQCD)~\cite{PLCho1,PLCho2}, 
which implements a factorization of the perturbative and nonperturbative terms. 
The latter are expressed in terms of long-distance matrix elements, 
assumed to be universal.
Several effective QCD 
models~\cite{PLCho1, PLCho2, GTBodwin, BGong, SBaranov, PFaccioli, pol1, pol2,
CSM-1, CSM-2, COM-1, COM-2, CEM} of quarkonium production 
predict different cross sections and polarizations. 
The thorough measurement of these observables 
plays a crucial role in advancing the theoretical understanding of 
the quarkonium production mechanisms. 

Under extreme temperature and density conditions, QCD calculations
predict a transition to a color-deconfined phase of matter, referred to as
the quark-gluon plasma (QGP). 
Heavy quarkonia are most promising probes of the QGP medium,
having become the focus of detailed scrutiny since the phenomenon
of color screening was proposed~\cite{Satz}. 
While charmonia has been extensively explored as a QGP probe~\cite{PHENIX, jpsiQGP}, 
until recently this had not been the case for bottomonia, even though  
the bottomonium family of states provides experimentally more robust and
theoretically cleaner probes.  

Heavy flavor states in general, and quarkonia in particular, are
``standard candles'', which are explored for detector calibration and used
in precision measurements of the SM as well as searches for phenomena
beyond the SM. Precision studies of observable decay rates may indicate
discrepancies that would be accommodated by new mediating heavy bosons.
They can also be used directly to search for new physics. Decays to
pairs of new light particles, or possible dark matter candidates, are
foreseen in certain new physics scenarios. Quarkonia may be further
explored to search for exotic states and rare physics
processes that are sensitive to new physics.

\begin{figure}[t]
\centerline{\includegraphics[width=0.7\textwidth]{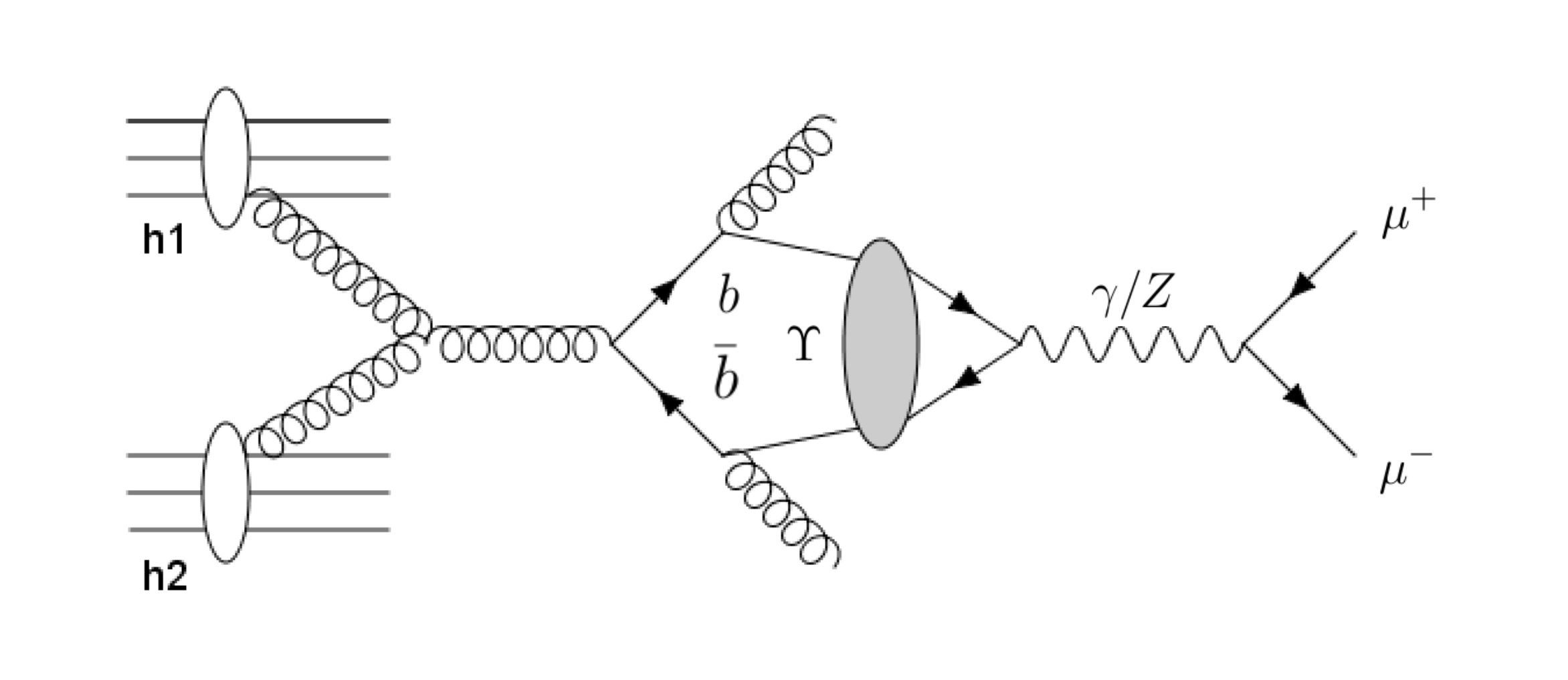}}
\caption{Schematic illustration of the production of a $b\bar{b}$ pair in a hadronic collision, followed by its evolution into a color-singlet bottomonium state,  involving gluon emission, and eventual electroweak decay into a muon pair.}
\label{Upsilonll}
\end{figure}

\begin{figure}[h]
\centerline{\includegraphics[width=0.9\textwidth]{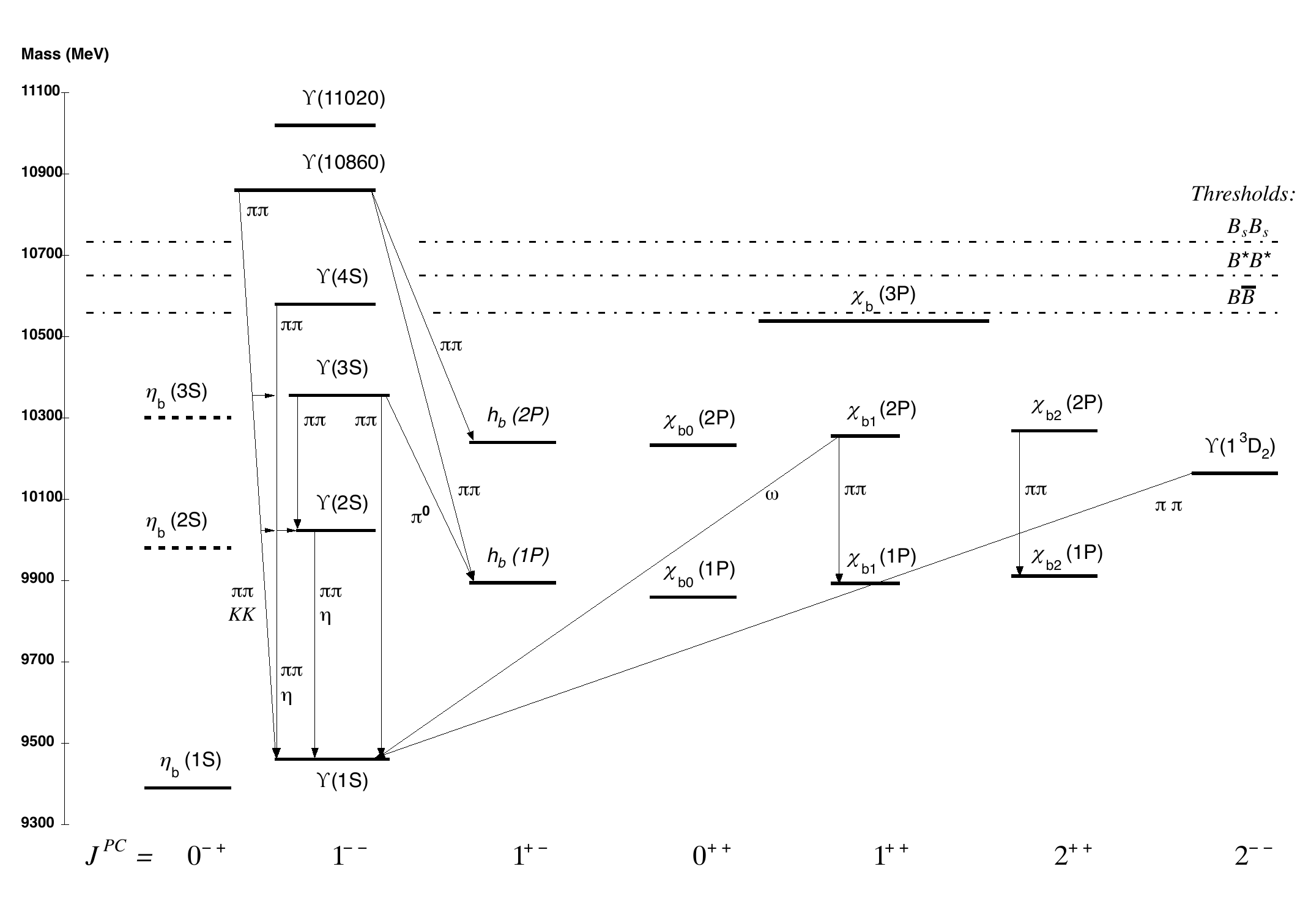}}
\caption{Representation of the bottomonium system, taken from the Particle Data Group 2012 Review~\cite{Bottomonium}. 
The dashed lines indicate $b\bar{b}$ states that have not yet been experimentally established. 
The single photon transitions $\Upsilon(nS)\rightarrow\gamma\eta_b(mS)$, $\Upsilon(nS)\rightarrow\gamma\chi_{bJ}(mP)$, 
and $\chi_{bJ}(nP)\rightarrow\gamma\Upsilon(mS)$ are omitted for clarity. 
Theoretically, the quarkonium system has been thoroughly studied with 
potential models~\cite{potential1, potential2, potential3, potential4, potential5, potential6, potential7}, 
effective field theory approaches~\cite{effective1,effective2}, and lattice gauge theory calculations~\cite{lattice1,lattice2}. \label{Bottomonium}}
\end{figure}

Bottomonia originate from $b\bar{b}$ pairs that are produced 
in the partonic interactions occurring in the hadronic collision, 
which evolve into color-neutral states, and are experimentally detected through their leptonic
decay channels, as illustrated in Fig.~\ref{Upsilonll}. The first
bottomonium state to be discovered, the $\Upsilon$, was observed in
1977 in the $\mu^{+}\mu^{-}$ spectrum produced in 400 GeV proton-nucleus
collisions by the E288 experiment at Fermilab~\cite{1977}. 
Later many other bottomonium states have been seen and the $b\bar{b}$ 
system has been experimentally established, as summarized in
Fig.~\ref{Bottomonium}~\cite{Bottomonium}. 
Since 2010, a new era of detailed bottomonium studies has been underway
at the Large Hadron Collider~(LHC)~\cite{ReviewOfLHCQuarkonium} with
higher energy and intensity. The center-of-mass energy of the
proton-proton ($pp$) collisions reached 13\,\TeV in 2015, which is about 7
times higher than that attained at the Tevatron ($p\bar{p}$). The
center-of-mass energy reached at the LHC with the Lead-Lead (PbPb) collisions
is more than 20 times that attained at the Relativistic Heavy Ion
Collider (RHIC). 

\begin{figure}[tbh!]
\centering
\includegraphics[width=0.7\textwidth]{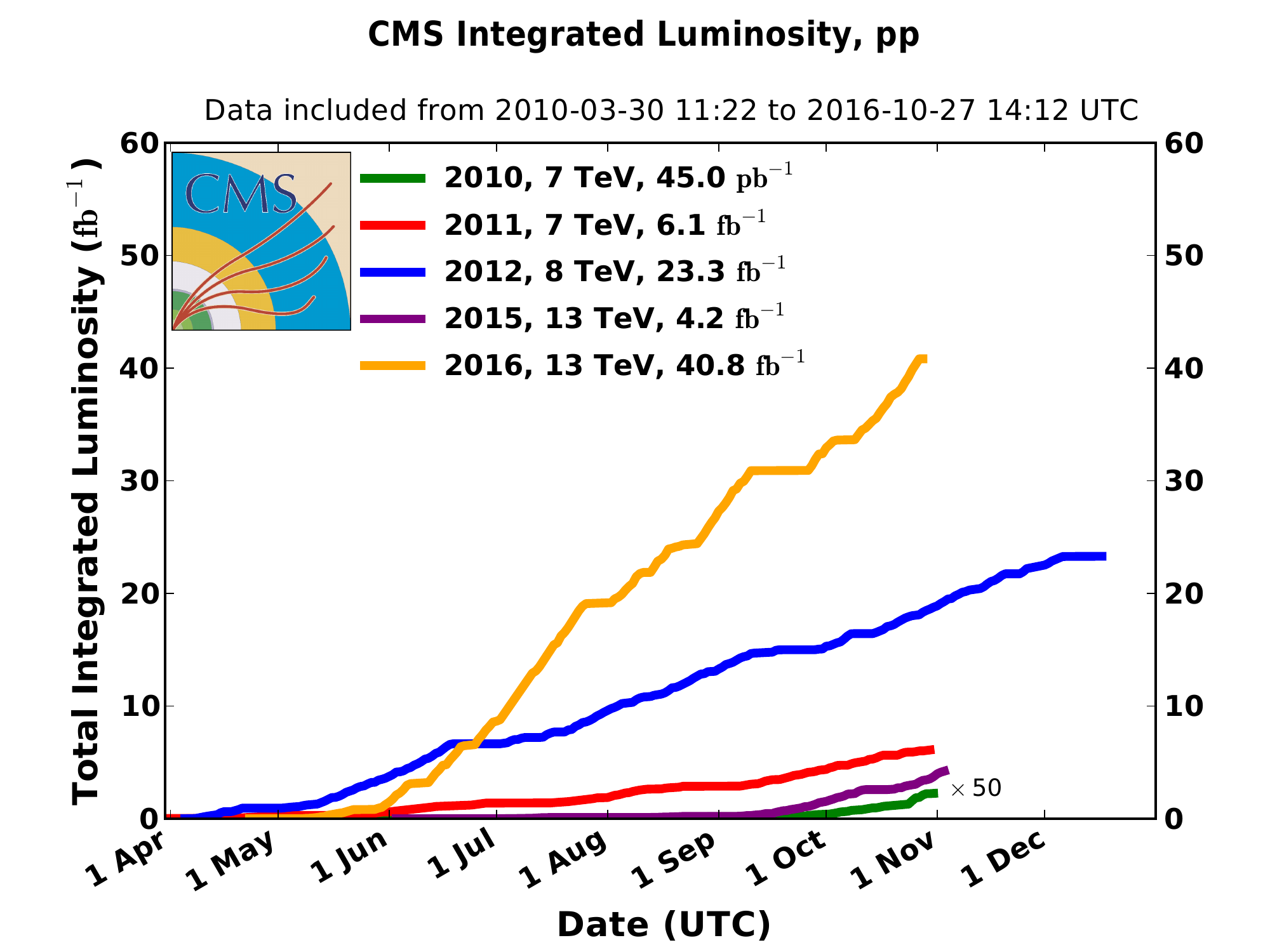}\\
\includegraphics[width=0.45\textwidth]{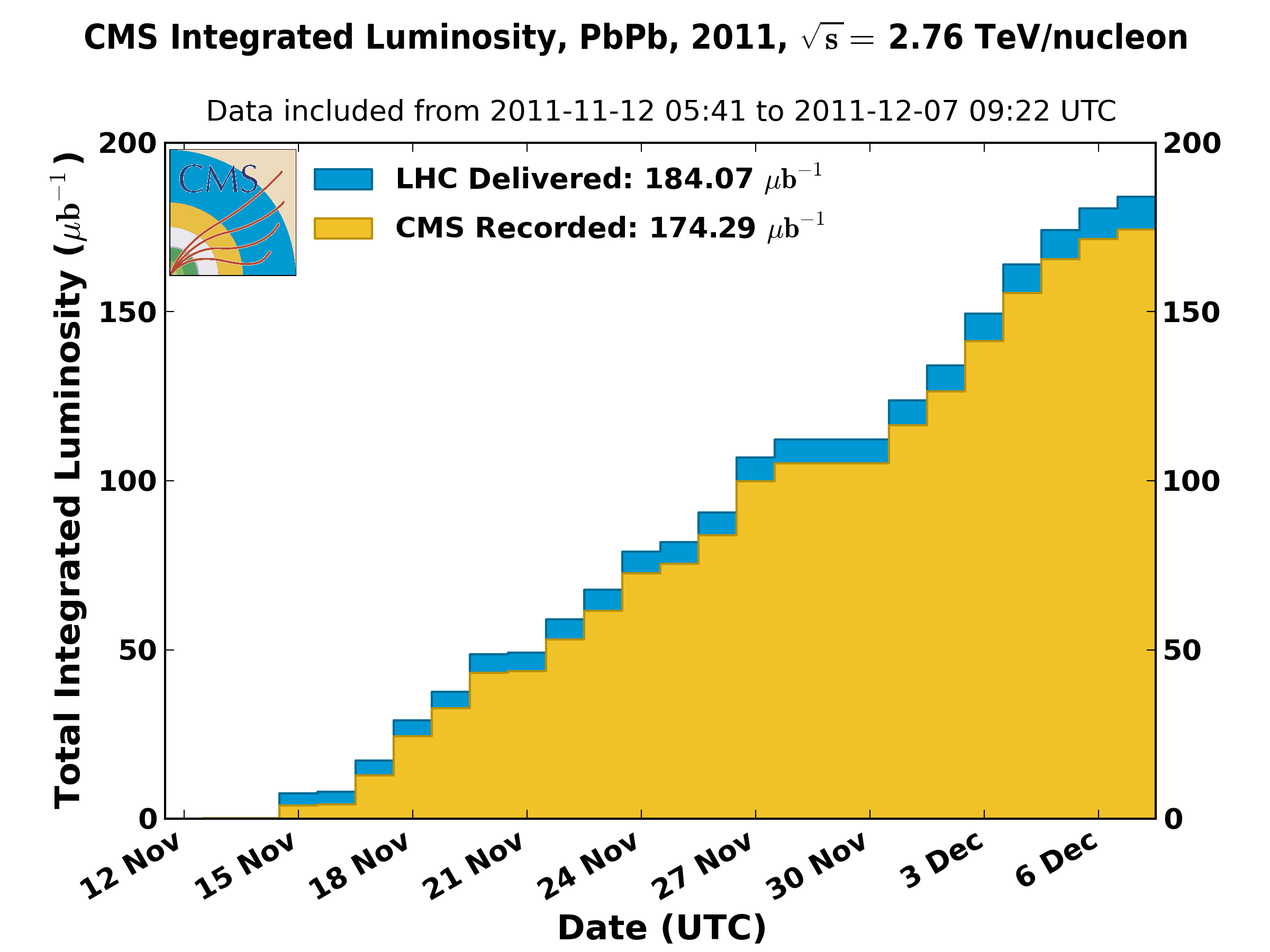}
\includegraphics[width=0.45\textwidth]{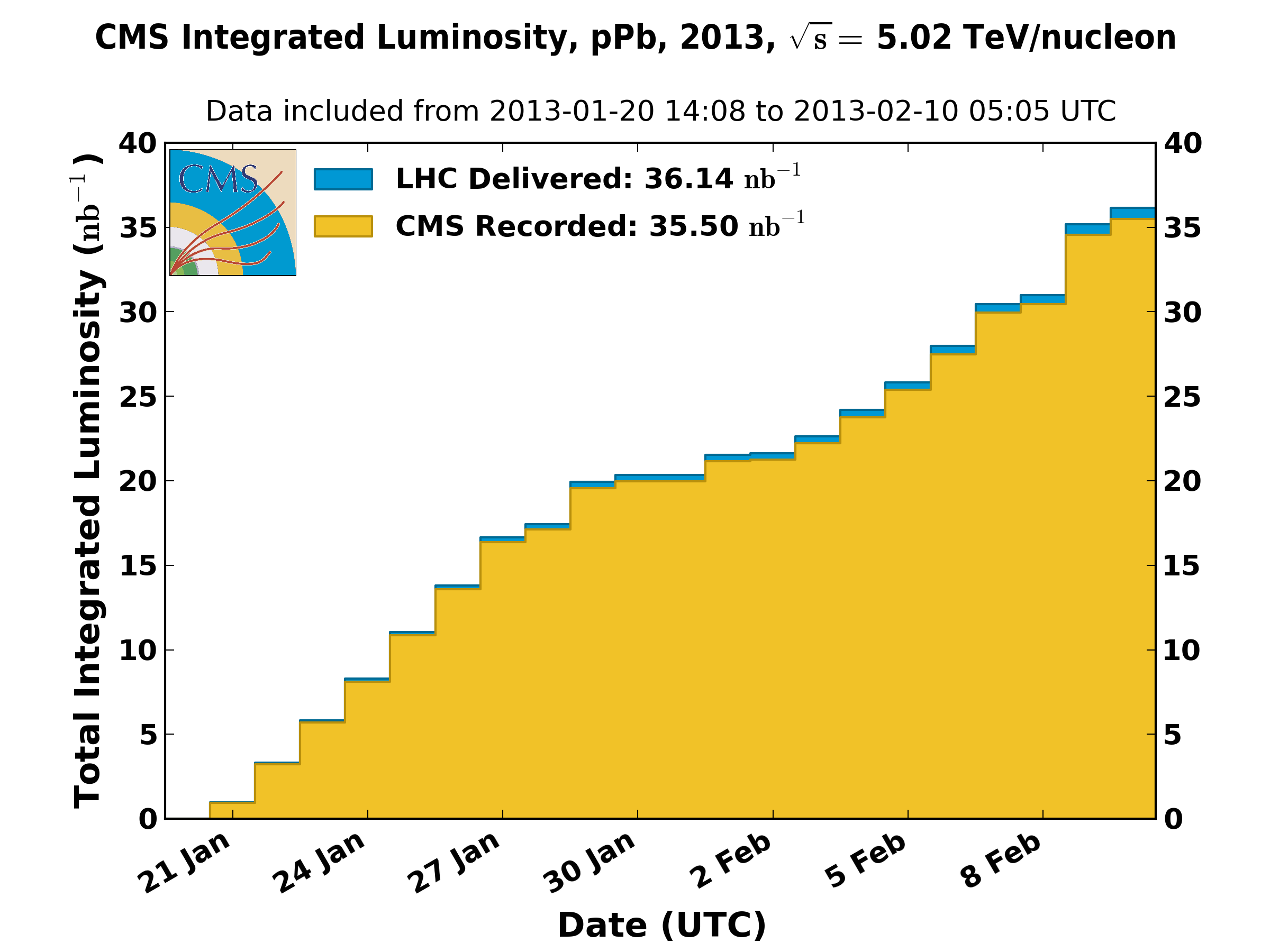}
\caption{Top: Cumulative luminosity delivered by the LHC to CMS during stable $pp$-collision beams shown for different data-taking years. 
Bottom: CMS integrated luminosity for PbPb collisions (left) and $p$Pb collisions (right).~\cite{lumiTwiki}} 
\label{CMSlumiInt}
\end{figure}

\begin{figure}[tbh!]
\centering
~\;\;\;
\includegraphics[width=0.54\textwidth]{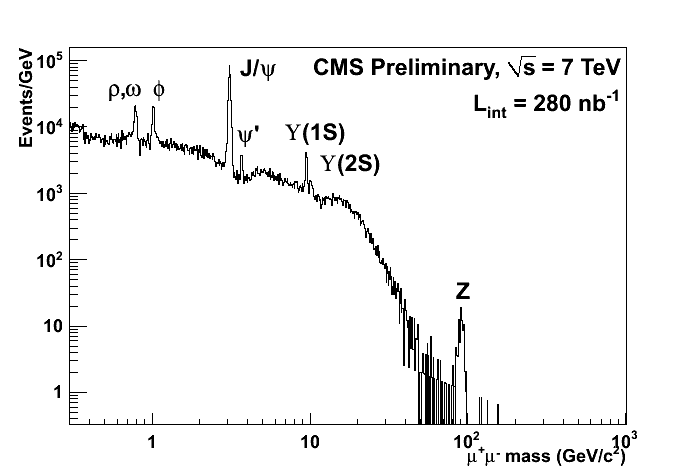} 
\includegraphics[width=0.36\textwidth]{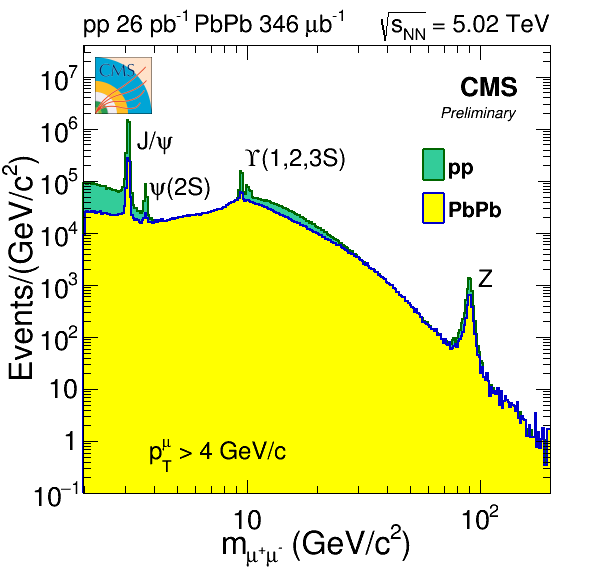}
\includegraphics[width=0.55\textwidth]{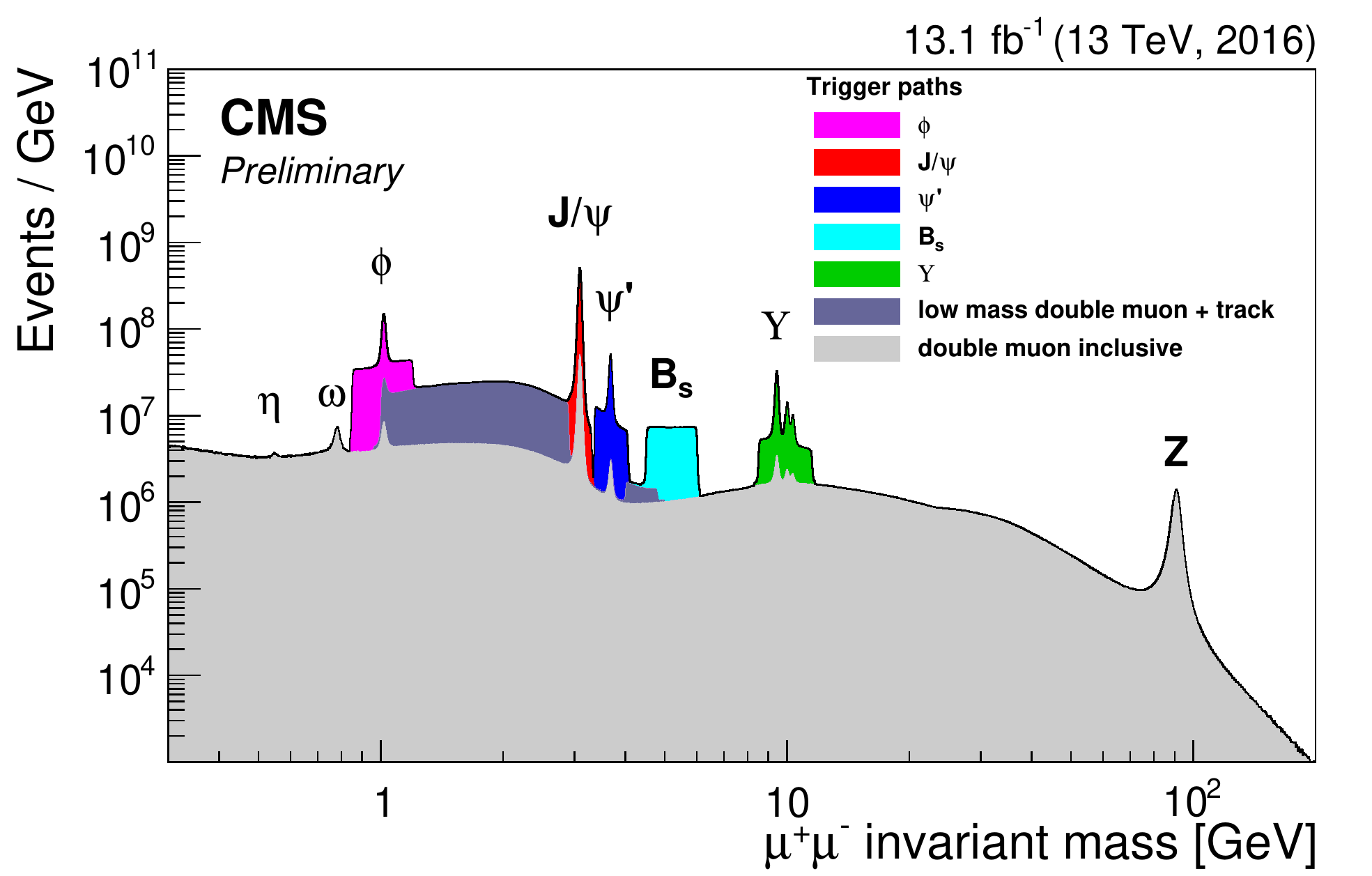}
\includegraphics[width=0.35\textwidth]{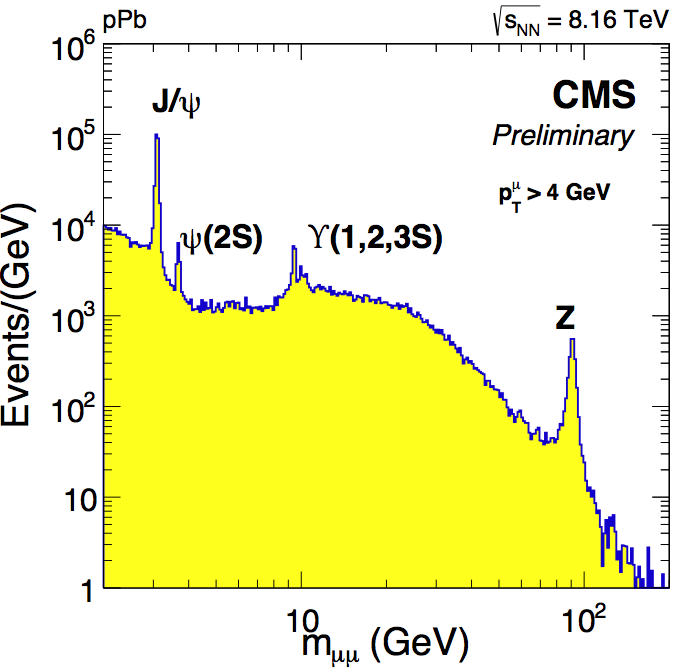}
\\
\caption{Dimuon invariant mass spectrum at CMS. Top left~\cite{ichep2010}: first version of the spectrum plot, 
produced with the very first LHC data, collected in 2010 and corresponding to 280 nb$^{-1}$ in $pp$ collisions 
at $\sqrt{s}=7$\,\TeV. Bottom left~\cite{dimuonPlot}: the most recent version, produced with a data set 
of 13.1 fb$^{-1}$ $\sqrt{s}=13$\,\TeV collected in 2016, which also highlights the dedicated heavy-flavor triggers 
in use by the experiment. Top right: the dimuon invariant mass spectrum in PbPb collisions. 
Bottom right: the dimuon invariant mass spectrum in $p$Pb collisions.}  
\label{dimuonMass}
\end{figure}

The Compact Muon Solenoid (CMS) is a general purpose detector at the
LHC. It explores both $pp$ collisions and collisions involving heavy
ions. The charged particles produced in the collisions leave
trajectories in the CMS all-silicon Tracker. The particle momentum is
precisely measured in the 3.8 T axial magnetic field. Since the start of its operation, several large data sets
have been delivered by the LHC. The integrated luminosity recorded by CMS
is shown in Fig.~\ref{CMSlumiInt}~\cite{lumiTwiki}. Early into the start
of the LHC data taking in 2010, a full spectrum of the dimuon invariant
mass was reconstructed by CMS. As evidenced in Fig.~\ref{dimuonMass},
even with an initial data set corresponding to a few hundred nb$^{-1}$, CMS was able to observe many
SM particles, in a wide mass range from the $\omega$ to the $Z$ boson.
The figure essentially combines many particle physics discoveries of the
past 50 years into a single picture.  

Taking advantage of its good momentum resolution, CMS is able to
distinguish the three lowest lying $\Upsilon$ resonances in the dimuon
decay channel. None of the heavy ion experiments at RHIC were able to
achieve this, and ALICE and ATLAS at the LHC are only able to resolve
the $\Upsilon$ resonances with difficulty. Compared to LHCb and the
Tevatron experiments, CMS has both complementary and wider kinematic
acceptance coverage. These advantages make CMS uniquely suited for
bottomonium measurements.

In this review, measurements of $S$-wave bottomonium $\Upsilon(nS)$
production in $pp$, PbPb, and $p$Pb collisions are summarized. The relative production of $P$-wave bottomonia $\chi_b(nP)$ is
presented. Several effective models of QCD have been compared to and challenged by
these measurements. The QCD predicted deconfined medium has been probed by
comparing the measurements from different types of collisions. The first
observation of bottomonium pair production and searches for new states
are also discussed.

\section{Production in proton-proton collisions}
\label{sec-Yinpp}

Historically, quarkonium production has been poorly understood. A
number of effective QCD approaches attempt to describe the
nonperturbative evolution of the $q\bar{q}$ pair into a color neutral bound
state, including the color-singlet model (CSM)~\cite{CSM-1, CSM-2}, the
color-octet mechanism (COM)~\cite{COM-1, COM-2} dominant in NRQCD, and
the color-evaporation model (CEM)~\cite{CEM}. The models have been
proposed to explain hadroproduction measurements. The similarities and
differences among them are discussed in detail
elsewhere~\cite{Quarkonium}. These models provide different predictions
for the quarkonium production cross sections and polarizations.
Bottomonium states are heavier and more non-relativistic than charmonium
states, so the comparison between experiment and theory is expected to
be more reliable for bottomonium~\cite{ReviewOfBottomonium}. A
precise measurement of bottomonium production is thus crucial for distinguishing among the models.

In recent years, cross section and polarization results of
$\Upsilon$ production in hadron collisions have been reported by
CDF~\cite{CDF,PolarizationCDF}, D0~\cite{D0}, CMS, ATLAS~\cite{xsecATLAS,xsecATLAS2}, 
and LHCb~\cite{xsecLHCb,xsecLHCb2}. However, none of the current
theoretical models predict both the rate and the polarization as observed 
in any of these experiments. This situation is referred to as the ``quarkonium
puzzle.'' In this section, we summarize the $\Upsilon$ production
measurements for the different center-of-mass regimes probed by CMS.

The  $\Upsilon(nS)$ states are typically reconstructed through the dimuon decay
channel, $\Upsilon(nS)\to\mu\mu$. While the branching fractions for
these decays are only about 2\%, Fig.~\ref{dimuonMass} shows that CMS is
able to collect a large number of $\Upsilon$ events even in a limited
integrated luminosity data set and to cleanly resolve the lightest three
$\Upsilon$ states.

\subsection{$\Upsilon$ cross section}	
\label{sec-xsec}

The first preliminary results on quarkonium production cross sections at the LHC
were reported~\cite{ichep2010} only a few months after the start of data
taking, in summer 2010, based on less than 0.3\,pb$^{-1}$ collected by
CMS. The first publication~\cite{xsecPRD} of the cross sections of the
individual  $\Upsilon(nS)$ states in $pp$ collisions at $\sqrt{s}=
7$\,\TeV followed based on a 10 times larger data set, corresponding to an
integrated luminosity of 3.1\,pb$^{-1}$. 

\begin{figure} 
\begin{center}
\includegraphics[width=0.5\textwidth]{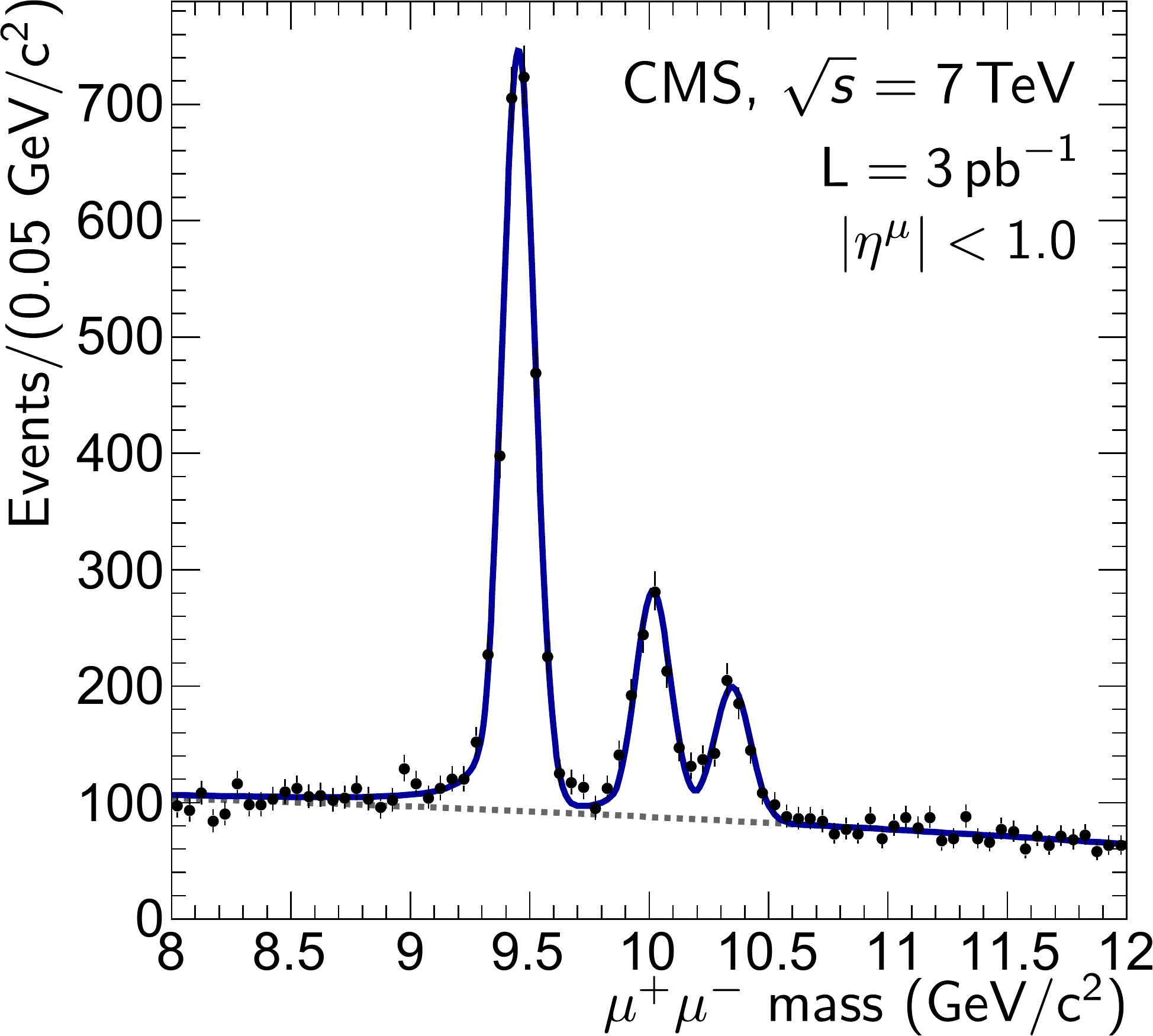}
\caption{The dimuon invariant mass distribution between 8 and 12\,GeV,
reconstructed using $pp$ data corresponding to 3 pb$^{-1}$ collected by
CMS in 2010 ($\sqrt{s}=7$\,\TeV)~\cite{xsecPRD}. The $\Upsilon(1S)$,
$\Upsilon(2S)$, and $\Upsilon(3S)$ resonances are clearly visible and
fully resolved.} 
\label{UpsMass} 
\end{center} 
\end{figure}

The  $\Upsilon(nS)$  differential cross section
is measured as~\cite{xsecPRD}:  
\begin{equation}
\frac{d^2\sigma(pp\rightarrow\Upsilon(nS)X)}{d\pt dy}\cdot B(\Upsilon(nS)\rightarrow\mu^+\mu^-)=\frac{N_{\Upsilon(nS)}(A,\epsilon)}{\mathcal{L}\cdot\Delta \pt\cdot\Delta y}, 
\label{xsecdef}
\end{equation}
where $B$ is the dimuon branching fraction, $N$ is the signal yield
corrected for the event weights given by the inverse product of the detector acceptance 
$A$ and the combined trigger and reconstruction efficiencies $\epsilon$, $\mathcal{L}$ is
the integrated luminosity, and $\Delta \pt$ and $\Delta y$ are the 
widths of the bins in transverse momentum (\pt) and rapidity ($y$). 
The symbol $X$ in Eq.~(\ref{xsecdef}) is used to indicate that the measurements
include feed-down contributions originating from decays of higher-mass bottomonia as indicated in Fig.~\ref{Bottomonium}. 

The signal yield before correction is determined using an extended unbinned
maximum-likelihood fit to the dimuon invariant mass spectrum, as shown
in Fig.~\ref{UpsMass}. The dimuon candidates are
required to satisfy $|y|<2$, while the individual muon candidates must
satisfy kinematic thresholds that depend on pseudorapidity $\eta$, to
ensure that the trigger and reconstruction efficiencies are high and
stable~\cite{xsecPRD}:
\begin{eqnarray}
\begin{array}{ll}
\pt^{\mu} > 3.5\,\GeV \text{, if }|\eta^{\mu}|<1.6, \\
\pt^{\mu} > 2.5\,\GeV \text{, if }1.6<|\eta^{\mu}|<2.4. 
\end{array}
\label{acceptanceCut}
\end{eqnarray}
Such a detector acceptance region defined by the \pt, $\eta$, and $y$ requirements is referred to as the fiducial region. 
\begin{figure}[h]
\begin{center}
\includegraphics[width=0.37\textwidth]{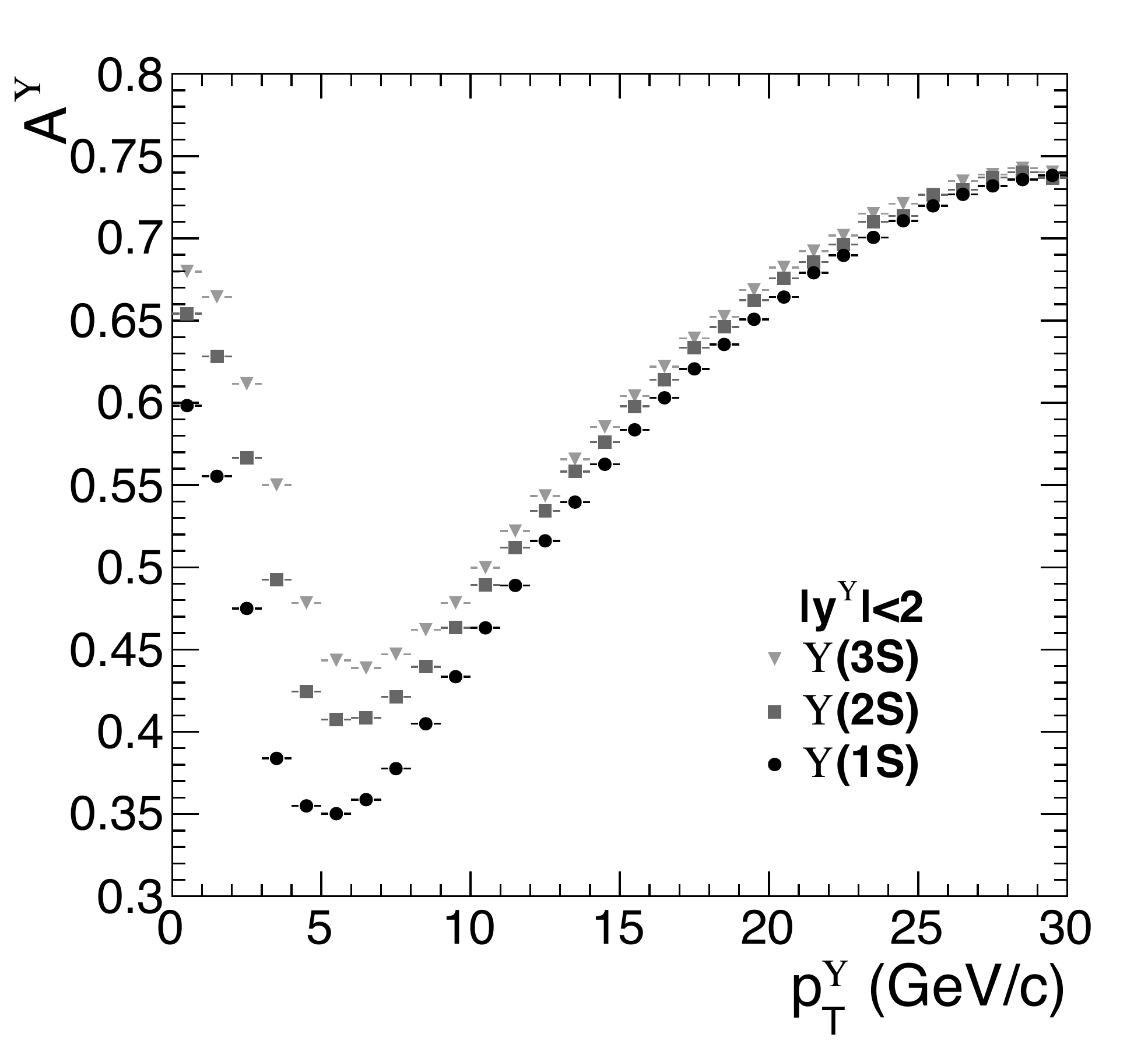}
\includegraphics[width=0.53\textwidth]{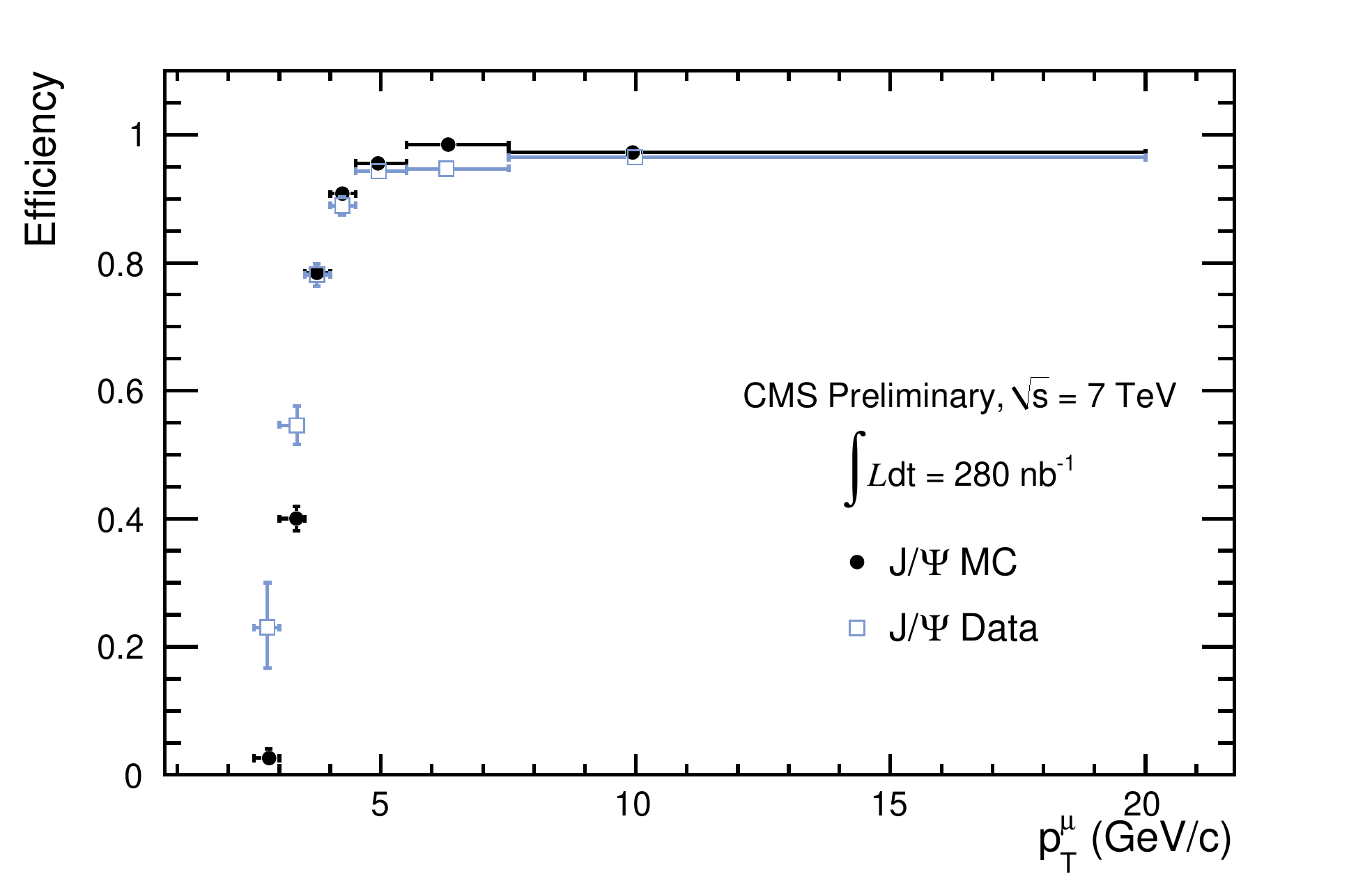}
\caption{Left: the unpolarized  $\Upsilon(nS)$  acceptances integrated over rapidity as a function of $\Upsilon$ \pt; 
Right: single muon trigger efficiency, as a function of muon $\pt$, shown for the central region 
$|\eta^{\mu}|<1.2$, evaluated with a data-driven method (T\&P). \label{AandE}}
\end{center}
\end{figure}
The $\Upsilon\rightarrow\mu^+\mu^-$ signal acceptance $A$ is
defined as the fraction of $\Upsilon$ mesons with $|y|<2$ and 
with both muons in the fiducial volume of the detector, 
and is determined from Monte Carlo simulation.
The acceptance as a function of $\Upsilon$ \pt
is shown in Fig.~\ref{AandE}~(left)~\cite{xsecPRD}. 
The seemingly peculiar form 
of this graph shows that, when the $\Upsilon$ is produced at rest, both muons are likely to reach 
the muon detector and pass the $\pt^{\mu}$ threshold. 
When the $\Upsilon$ acquires a small boost, one of the muons may be below 
the muon detection threshold defined in Eq.~(\ref{acceptanceCut}), 
and the acceptance drops. This is due to the large mass of the $\Upsilon$ compared to the $\pt^{\mu}$ threshold. 
For an $\Upsilon$ momentum above about half the $\Upsilon$ mass (in natural units), 
the strongly boosted $\Upsilon$ passes large momentum to the two daughter muons and 
the acceptance starts to increase with $\Upsilon$ \pt. The trigger and reconstruction efficiency 
of the accepted muon is determined with a data-based
tag-and-probe (T\&P) technique~\cite{xsecPRD,tnp}. An example efficiency turn-on curve 
as a function of the muon \pt is
displayed in Fig.~\ref{AandE} (right)~\cite{xsecPRD}. 
Other selection criteria, such as the separation between the two muons, and the vertex quality of the dimuon candidate, 
are also applied, and accounted for in the total efficiency. 

Assuming unpolarized $\Upsilon(nS)$ production, the product of the $\Upsilon(nS)$ cross section 
and dimuon branching fraction, in $pp$ collisions at 7\,\TeV, in the rapidity region $|y|<2$ is~\cite{xsecPRD}: 
\begin{eqnarray}
\begin{array}{rcl}
\sigma(pp &\rightarrow& \Upsilon(1S)X)\cdot B(\Upsilon(1S)\rightarrow\mu^+\mu^-) \\
&=& 7.37\pm0.13(\text{stat.})^{+0.61}_{-0.42}(\text{syst.})\pm0.81(\text{lumi.}) \text{nb, } \\
\sigma(pp &\rightarrow& \Upsilon(2S)X)\cdot B(\Upsilon(2S)\rightarrow\mu^+\mu^-) \\
&=& 1.90\pm0.08(\text{stat.})^{+0.20}_{-0.14}(\text{syst.})\pm0.21(\text{lumi.}) \text{nb, } \\
\sigma(pp &\rightarrow& \Upsilon(3S)X)\cdot B(\Upsilon(3S)\rightarrow\mu^+\mu^-) \\
&=& 1.02\pm0.07(\text{stat.})^{+0.11}_{-0.08}(\text{syst.})\pm0.11(\text{lumi.}) \text{nb. } 
\end{array}
\label{xsec3pb}
\end{eqnarray}

\begin{figure}[t]
\centering
\includegraphics[width=0.37\textwidth]{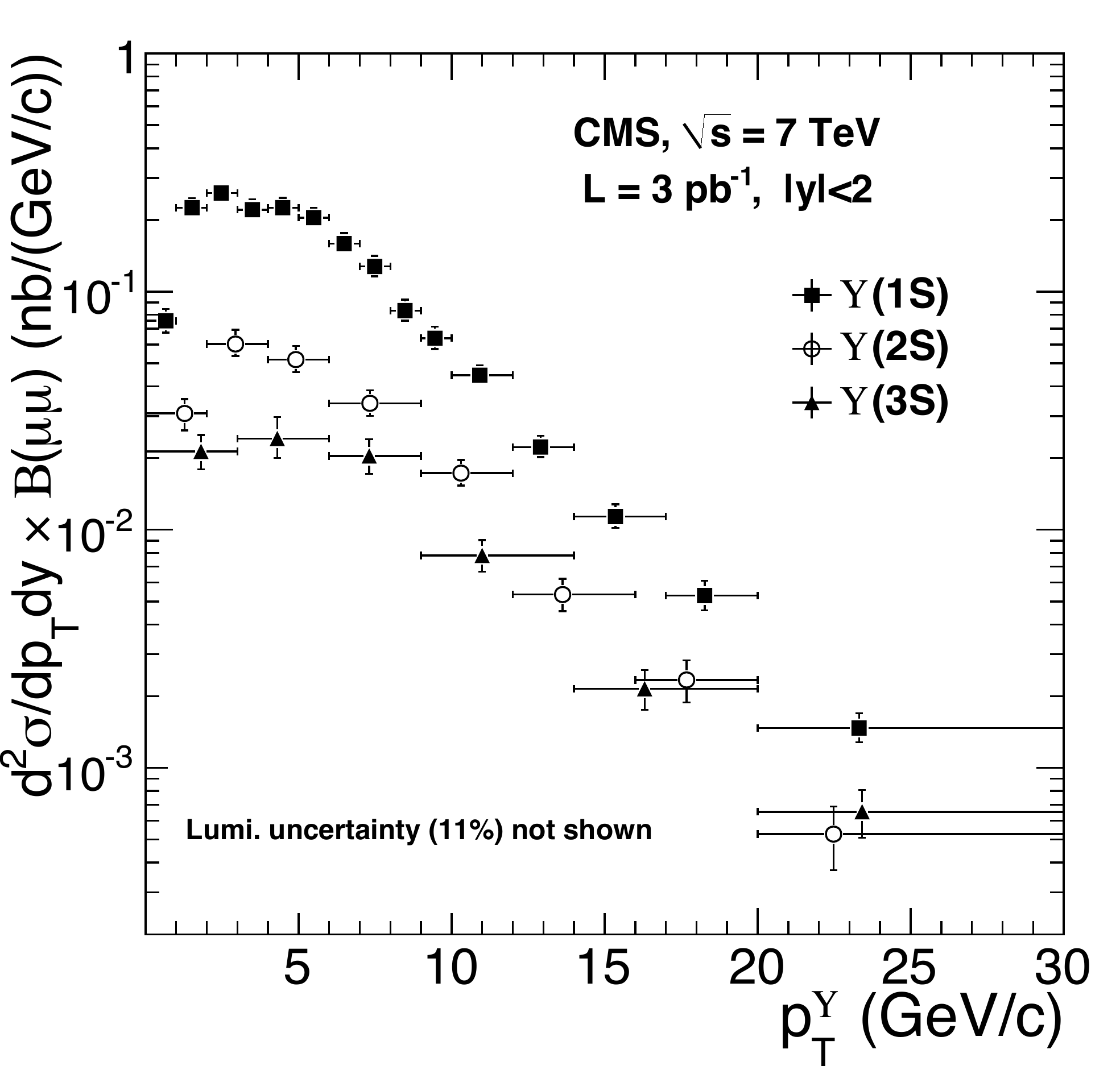}
\includegraphics[width=0.53\textwidth]{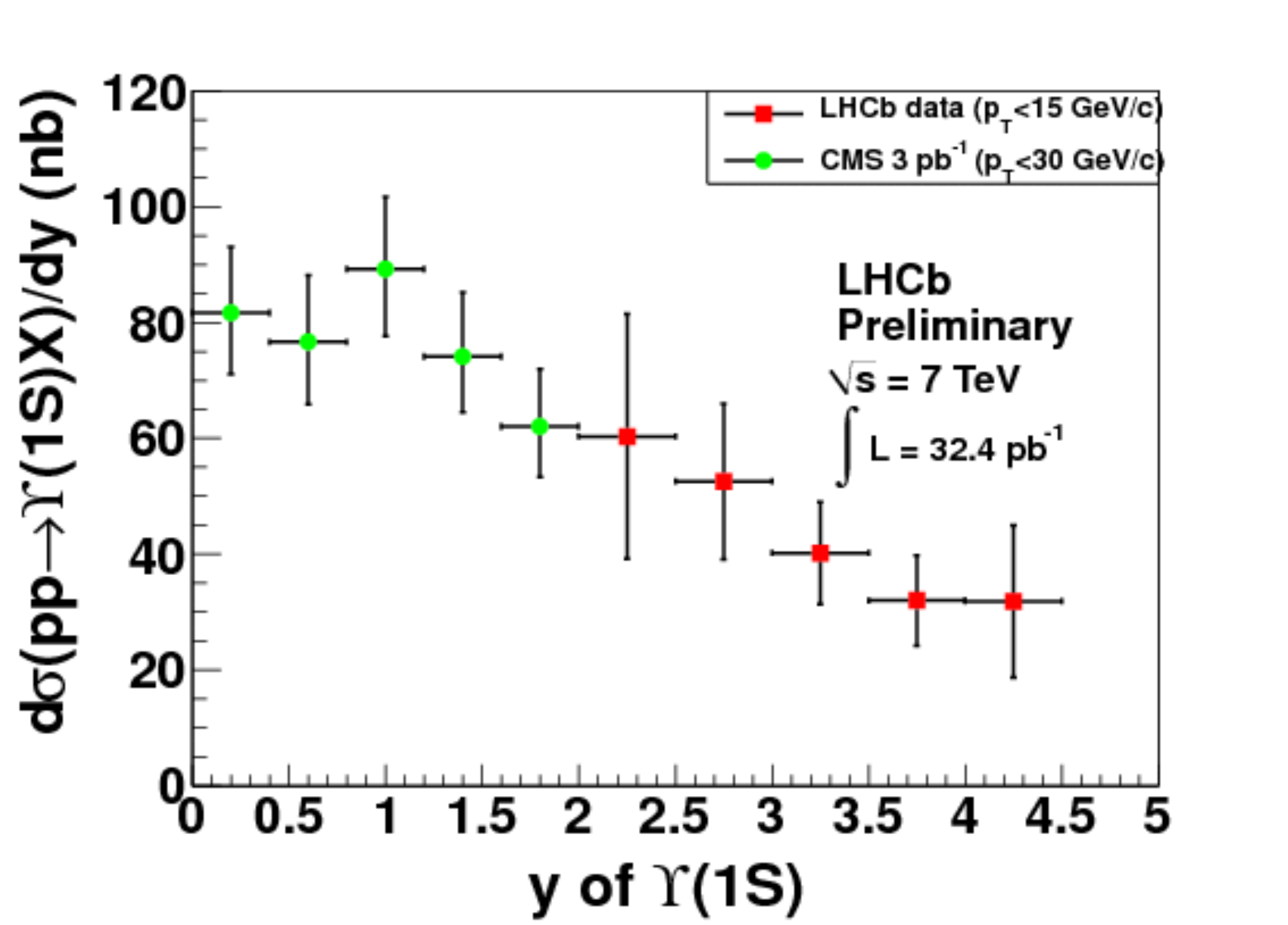}
\caption{Left: $\Upsilon(nS)$ double differential cross sections as a function of $\Upsilon$ \pt 
in the rapidity region $|y|<2$; Right: $\Upsilon(1S)$ differential cross section as a function of $\Upsilon$ rapidity, where the extended and complementary coverage achieved by the LHC experiments is illustrated.  
\label{Y-xsec-3pb}}
\end{figure}

The differential $\Upsilon(nS)$ cross sections as functions of $\Upsilon$ \pt and
rapidity are shown in Fig.~\ref{Y-xsec-3pb}~\cite{xsecPRD}. 
The measurements are performed down to zero $\Upsilon$ \pt. 
The complementarity of the CMS and LHCb~\cite{xsecLHCb} detectors in terms of phase-space coverage is also illustrated.
Heralding the start of the LHC era, the first bottomonium measurement
performed at the LHC covers about the same $\Upsilon$ \pt range as the
Tevatron experiments had reached at the end of their
lifetime~\cite{CDF,D0}. The measured cross section is about three times
larger than that measured at the Tevatron. This increase in the cross
section values is expected, given the increased center-of-mass energy at
the LHC compared to the Tevatron. 

\begin{figure}[t]
\centering
\includegraphics[width=0.32\textwidth]{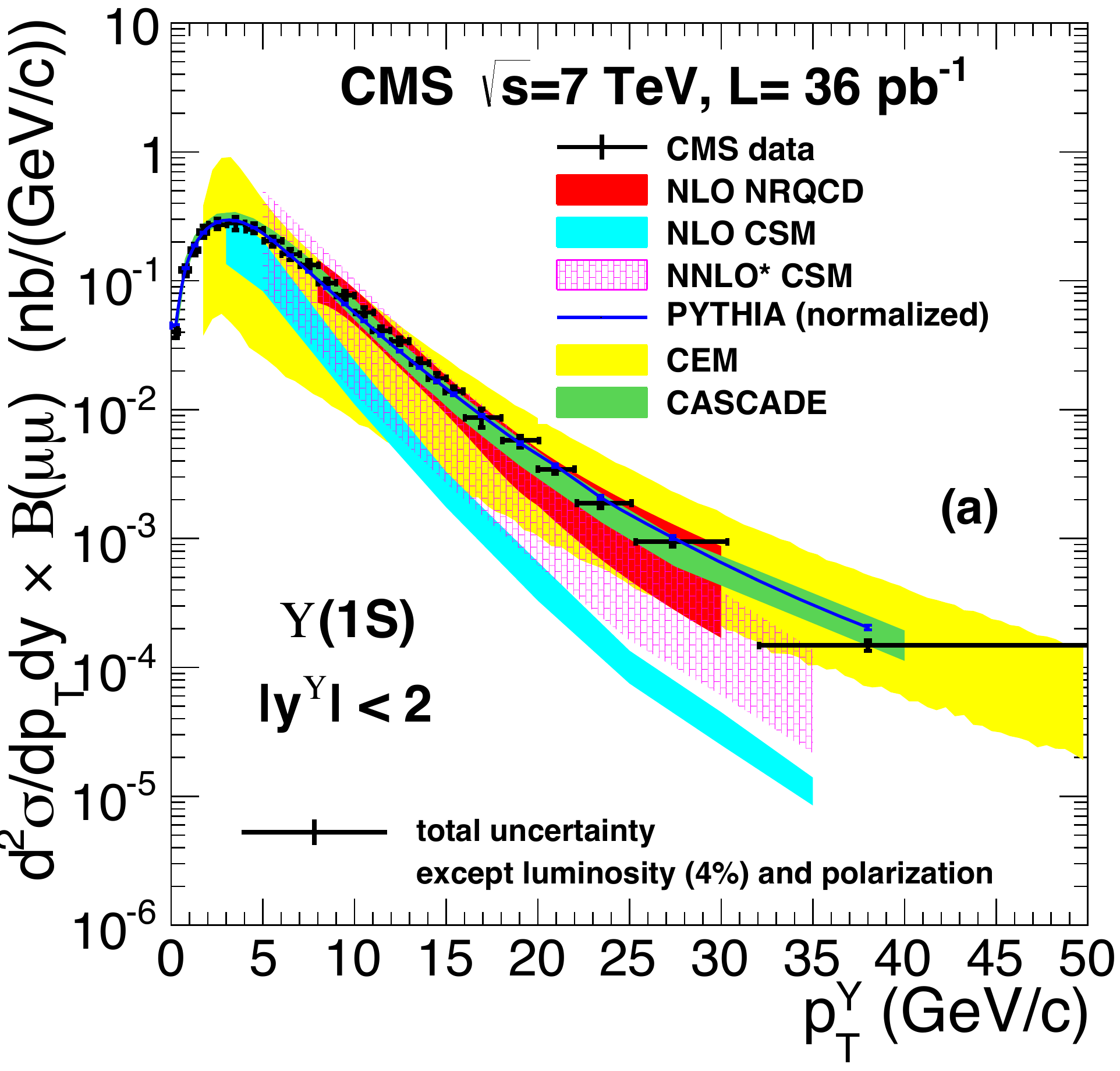}
\includegraphics[width=0.32\textwidth]{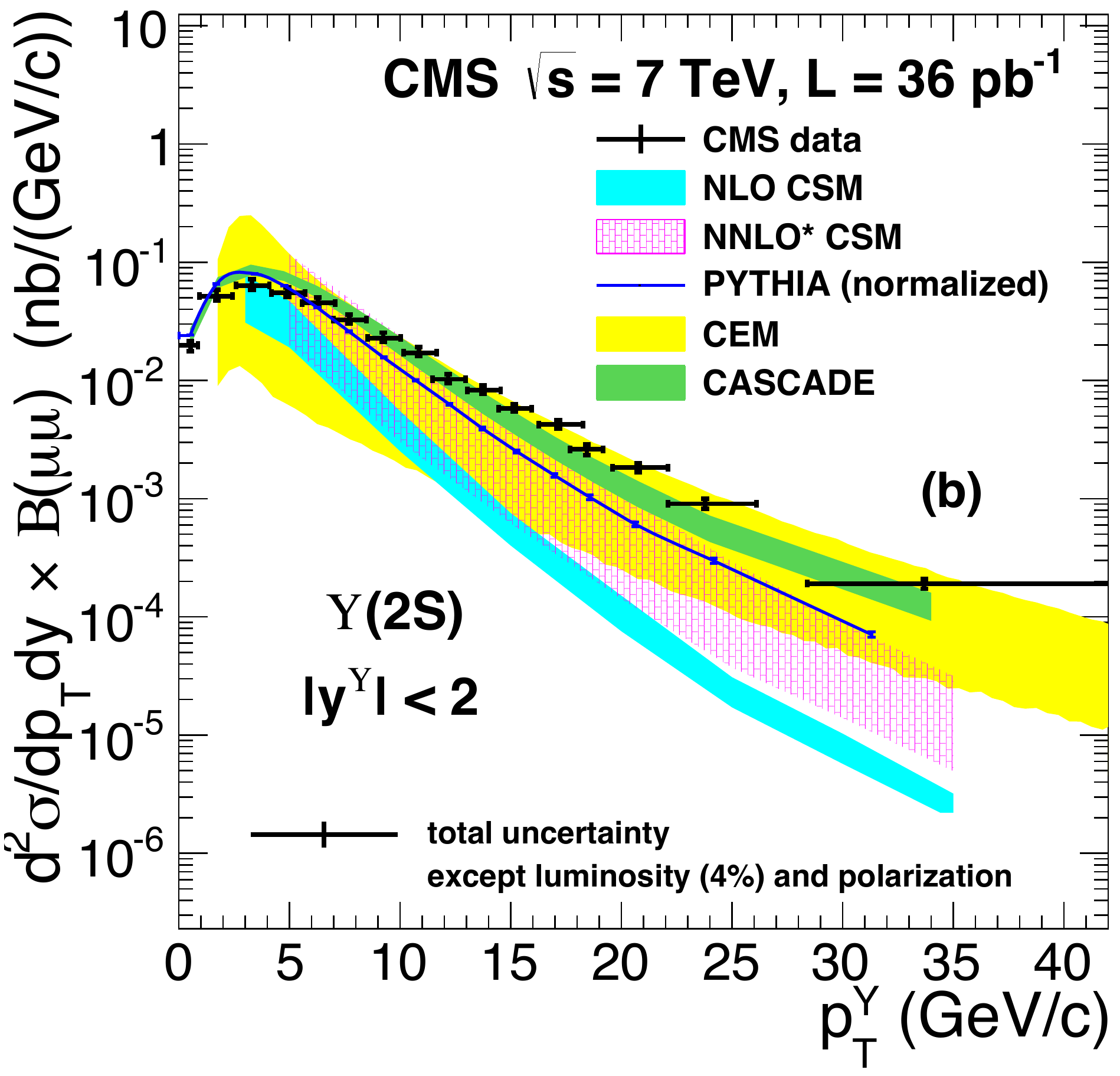} 
\includegraphics[width=0.32\textwidth]{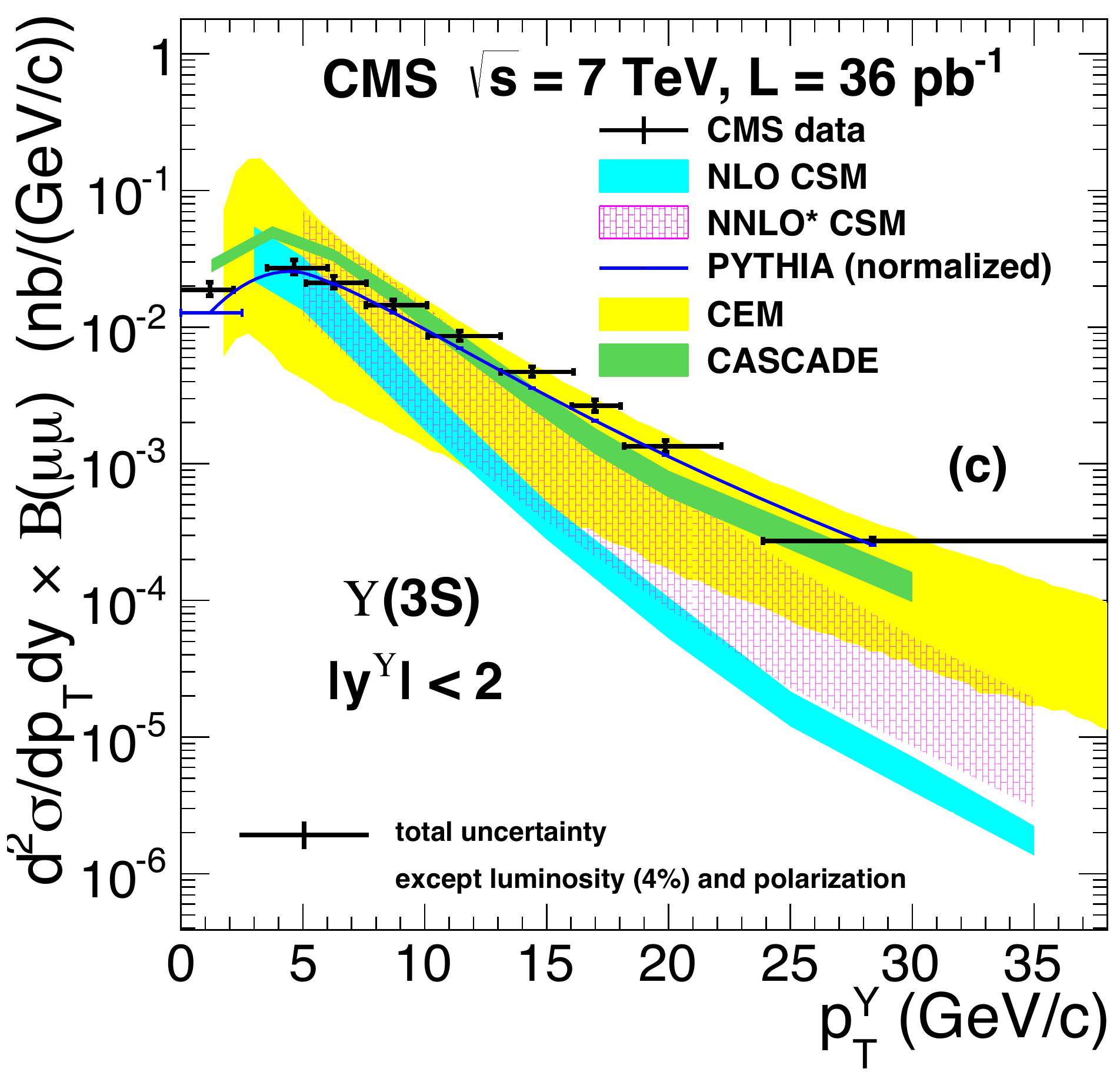} \\
\includegraphics[width=0.445\textwidth]{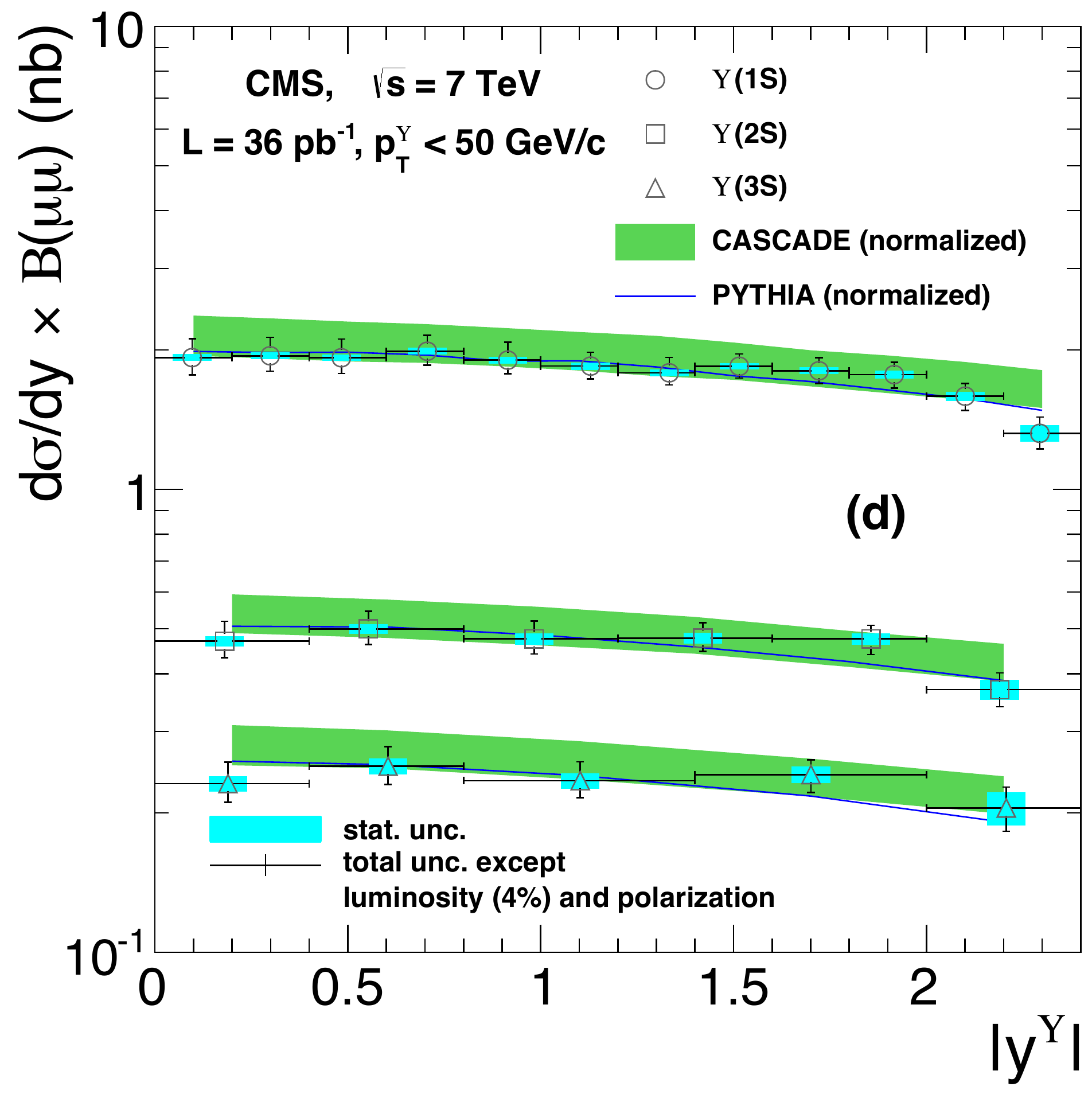} 
\includegraphics[width=0.455\textwidth]{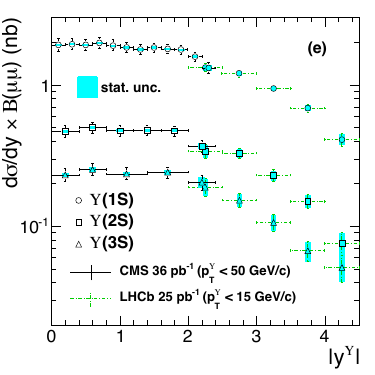} 
\caption{The differential cross sections of $\Upsilon(1S)$ (a), $\Upsilon(2S)$ (b), 
and $\Upsilon(3S)$ (c) in $pp$ collisions at 7\,\TeV as a function of $\Upsilon$ \pt 
in the rapidity region $|y|<2$, and as a function of $\Upsilon$ rapidity (d). 
The complementarity of the CMS and LHCb~\cite{xsecLHCb} measurements 
in terms of phase-space coverage is illustrated in (e). \label{Y-xsec-40pb}}
\end{figure}

To measure the $\Upsilon$ differential cross section more precisely,
especially in the high \pt region, the analysis was subsequently
extended to a larger data set, also collected in 2010~\cite{xsecPLB},
corresponding to an integrated luminosity of about 36 pb$^{-1}$. In a
wider $\Upsilon$ rapidity and transverse momentum range, $|y|<2.4$ and 
$\pt < 50$\,\GeV, the product of $\Upsilon$ production cross section and
dimuon branching fraction at 7\,\TeV is found to be~\cite{xsecPLB}: 
\begin{eqnarray}
\begin{array}{rcl}
\sigma(pp &\rightarrow& \Upsilon(1S)X)\cdot B(\Upsilon(1S)\rightarrow\mu^+\mu^-) \\ 
&=& 8.55\pm0.05(\text{stat.})^{+0.56}_{-0.50}(\text{syst.})\pm0.34(\text{lumi.})\,\text{nb, } \\
\sigma(pp &\rightarrow& \Upsilon(2S)X)\cdot B(\Upsilon(2S)\rightarrow\mu^+\mu^-) \\ 
&=& 2.21\pm0.03(\text{stat.})^{+0.16}_{-0.14}(\text{syst.})\pm0.09(\text{lumi.})\,\text{nb, } \\
\sigma(pp &\rightarrow& \Upsilon(3S)X)\cdot B(\Upsilon(3S)\rightarrow\mu^+\mu^-) \\ 
&=& 1.11\pm0.02(\text{stat.})^{+0.10}_{-0.08}(\text{syst.})\pm0.04(\text{lumi.})\,\text{nb. } 
\end{array}
\label{xsec40ipb}
\end{eqnarray}
To facilitate the comparison with the results given in Eq.~(\ref{xsec3pb}), 
a matching rapidity requirement was applied. 
The results agree within uncertainties in the $|y|<2$ region; 
for example,  
$\sigma(pp \rightarrow \Upsilon(1S)X)\cdot B(\Upsilon(1S)\rightarrow\mu^+\mu^-) = 7.50 \pm 0.05 \text{\,(stat.)\,nb}.$
is statistically compatible with the result in Eq.~(\ref{xsec3pb}).
The differential cross sections, corrected for acceptance and efficiency, as a function of
$\Upsilon$ \pt and $y$ are shown in Fig.~\ref{Y-xsec-40pb}~\cite{xsecPLB}. 
The comparisons of the measured results with the
CASCADE~\cite{cascade} MC generator, the CEM model without
feed-down~\cite{CEM}, NRQCD COM at next-to-leading order (NLO) including
feed-down~\cite{COM-2}, CSM to NLO, and NNLO*~\cite{CSM-2}, are also
shown in the figure. 

The cross sections reported above assume unpolarized
production. Polarization affects the spatial distribution of the two
muons from the $\Upsilon$ decay, and therefore the fraction of the
signal that is in the detector acceptance region. For extreme
assumptions for the polarization, i.e. either fully transverse or fully
longitudinal, the cross section results vary by about $\pm20$\%.
A fiducial cross section may be defined as the $\Upsilon$
yields corrected only by the efficiency
but not by the geometric acceptance, which can be used to compare
with theoretical models without making any assumption for the polarization.
The product of the fiducial $\Upsilon(nS)$ cross section and dimuon branching fraction is~\cite{xsecPLB}: 
\begin{eqnarray}
\begin{array}{rcl}
\sigma(pp &\rightarrow& \Upsilon(1S)X)\cdot B(\Upsilon(1S)\rightarrow\mu^+\mu^-) \\ 
&=& 3.06\pm0.02(\text{stat.})^{+0.20}_{-0.18}(\text{syst.})\pm0.12(\text{lumi.}) \text{nb, } \\
\sigma(pp &\rightarrow& \Upsilon(2S)X)\cdot B(\Upsilon(2S)\rightarrow\mu^+\mu^-) \\
&=& 0.910\pm0.011(\text{stat.})^{+0.055}_{-0.046}(\text{syst.})\pm0.036(\text{lumi.}) \text{nb, } \\
\sigma(pp &\rightarrow& \Upsilon(3S)X)\cdot B(\Upsilon(3S)\rightarrow\mu^+\mu^-) \\
&=& 0.490\pm0.010(\text{stat.})^{+0.029}_{-0.029}(\text{syst.})\pm0.020(\text{lumi.}) \text{nb. } 
\end{array}
\end{eqnarray}

As \pt increases, higher-order corrections become more significant in
several of the theoretical models. Therefore, cross section measurements
in high \pt regions are important for distinguishing among the models. 
The CMS analysis was further expanded using the full 4.9 fb$^{-1}$ $pp$
collision data taken at $\sqrt{s}=7$\,\TeV in 2011 with the \pt coverage
increased to 100\,\GeV~\cite{xsecPLB2015}. 

\begin{figure}[t]
\begin{center}
\includegraphics[width=0.48\textwidth]{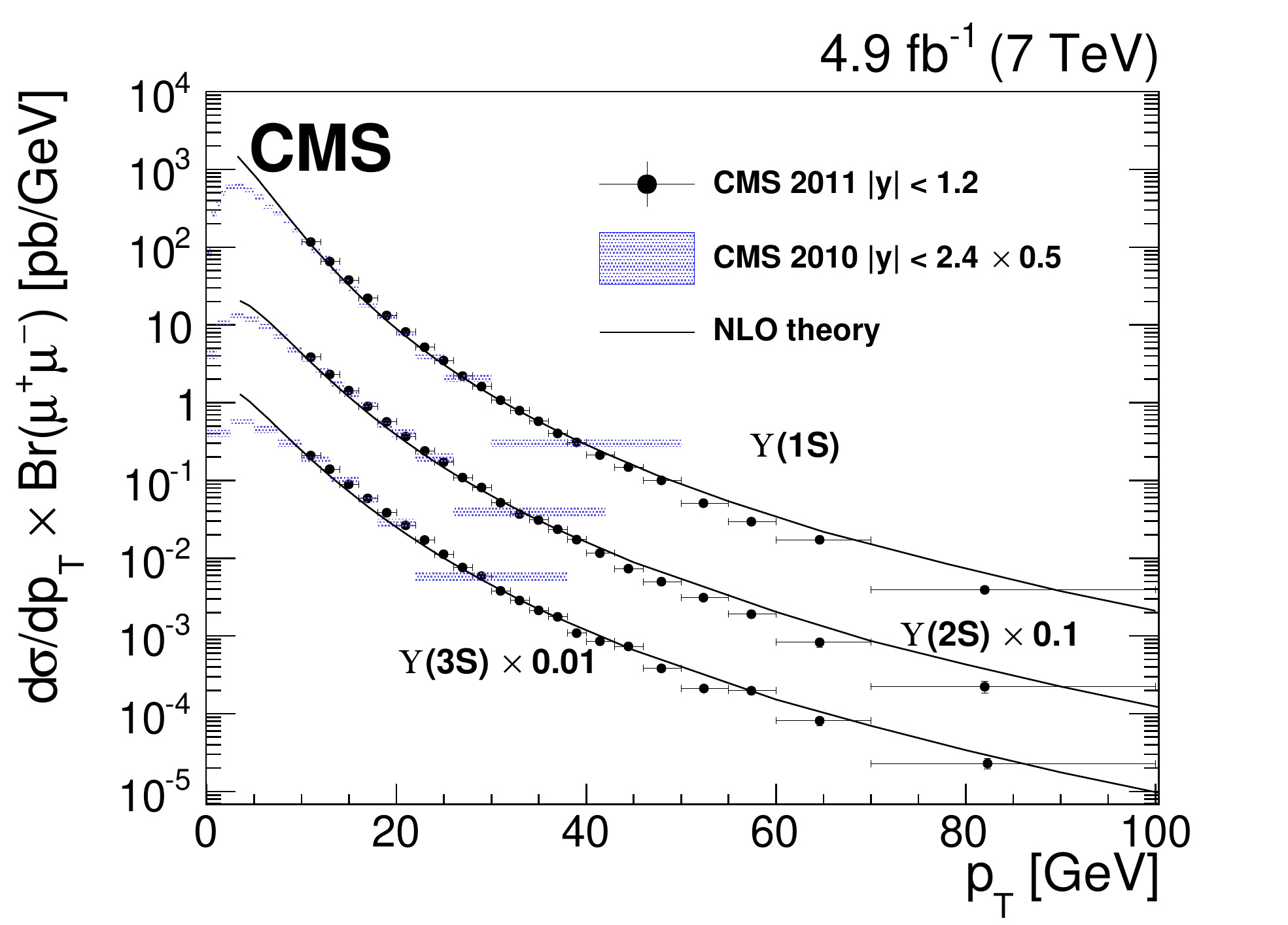}
\includegraphics[width=0.48\textwidth]{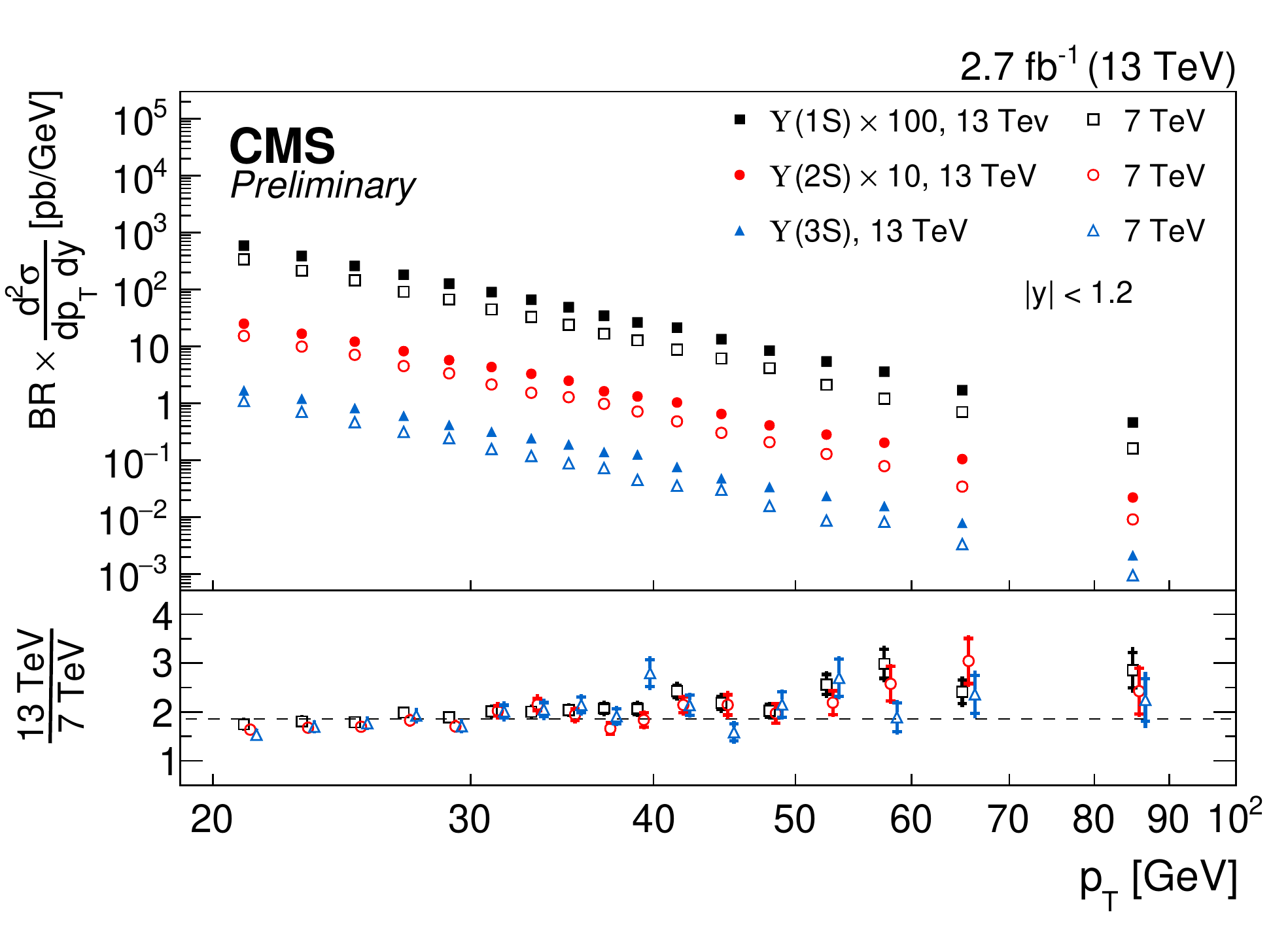} 
\caption{
The $\Upsilon(nS)$ \pt differential cross sections times branching fractions for 7\,\TeV  (left) and 13\,\TeV (right).
The uncertainty in the luminosity measurements is not included.
Comparisons with theoretical expectations are shown in the left plot, while the right plot compares measurements performed at the two different center-of-mass energies. 
}
\label{Y-xsec-fb}
\end{center}
\end{figure}

Figure~\ref{Y-xsec-fb}~(left)~\cite{xsecPLB2015} shows the differential
cross section as a function of \pt from the two groups of data, one
integrated over the rapidity range of $|y|<1.2$ (CMS 2011, black dots)
and the other scaled to the same range (CMS 2010, cross-hatched areas).
The solid lines are the NLO NRQCD color-octet calculations from
Ref.~\citen{extendTheory}, with the range of \pt further extended to
$\pt<100$\,\GeV by the corresponding authors, which describe the trends
of the data points for all the three $\Upsilon$ states. This significant
increase of the \pt range is also useful in studying the \pt dependence
in other models.

A similar measurement was repeated using a data set corresponding to 2.7\,fb$^{-1}$, collected in 2015, at the beginning of the second LHC 
data-taking period (Run 2).  This was the first result of the
$\Upsilon$ cross section at $\sqrt{s} = 13$\,\TeV reported by
CMS~\cite{BPH-15-005}. An analysis strategy similar to the one
previously described was used: dimuons were selected in the central
rapidity region, $|y|<1.2$, and tighter muon \pt acceptance thresholds
were adopted, namely $p^{\mu}_{T}>4.5~(4.0)$\,\GeV for
$|\eta^{\mu}|<0.3~(0.3<|\eta^{\mu}|<1.4)$. 
The \pt reach attained with this measurement based on the initial data
is already  100\,\GeV, and will increase with enlarged Run 2 datasets.  
While the focus is in the high \pt region, a lower threshold 
is effectively imposed from tightening trigger rate requirements.
This measurement is reported from an $\Upsilon$ \pt of 20\,\GeV.
The results are shown in Fig.~\ref{Y-xsec-fb}~(right)~\cite{BPH-15-005}, where a comparison of the
$\Upsilon(nS)$ differential cross sections for the 7 and 13\,\TeV data
sets is also given.
The cross sections for all bottomonium states at 13\,\TeV are larger
than the corresponding cross sections at 7\,\TeV by a factor of two to
three. This increase is expected from the $\sqrt{s}$-induced variation
in the parton distribution functions, and is confirmed by simulations
using the PYTHIA\,8 description. The extensions of NRQCD and other
theoretical models at 13\,\TeV are currently in progress.

\begin{figure}[th!]
\begin{center}
~\;
\includegraphics[width=0.4\textwidth]{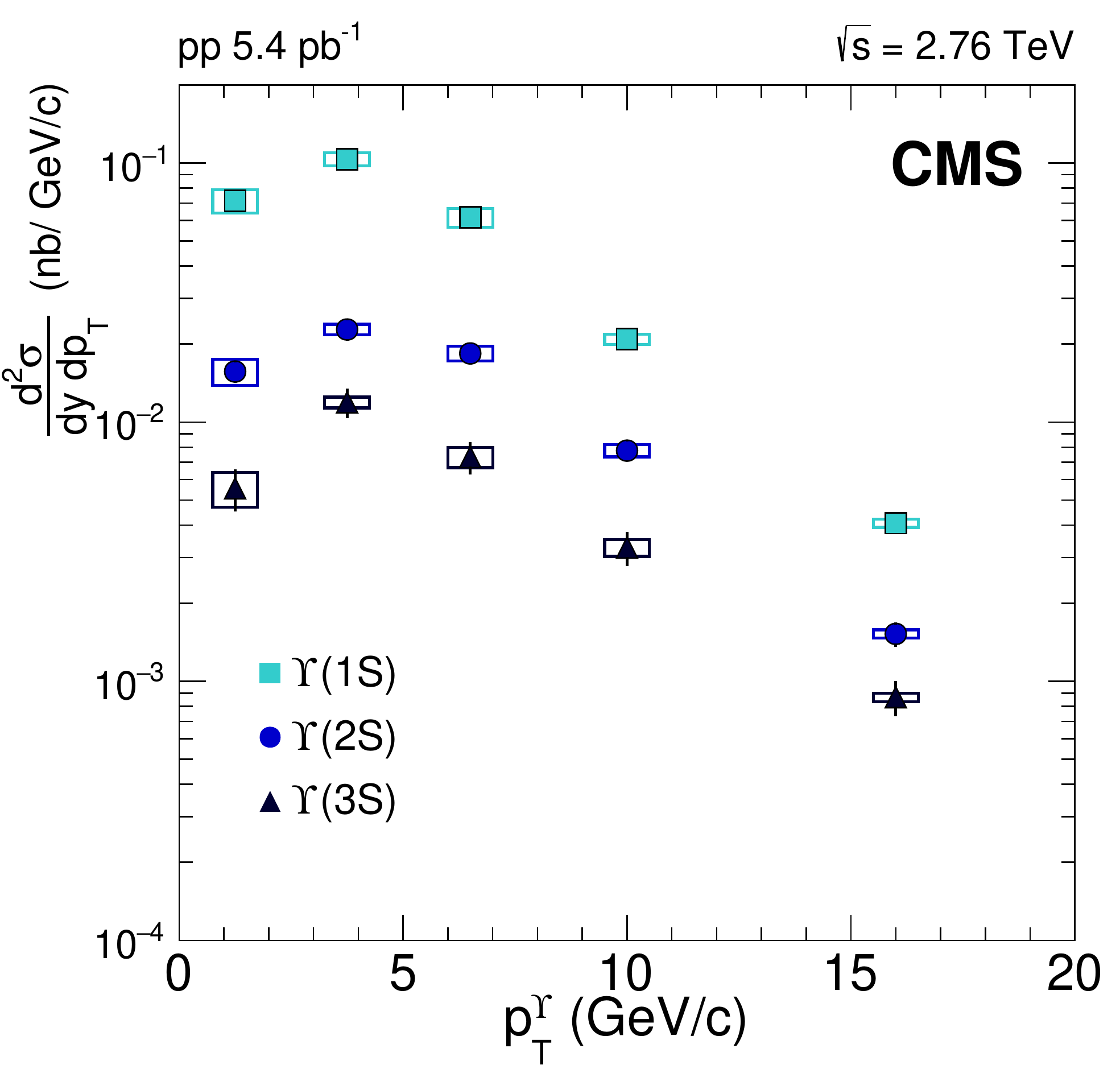}~\;\;\;\;
\includegraphics[width=0.4\textwidth]{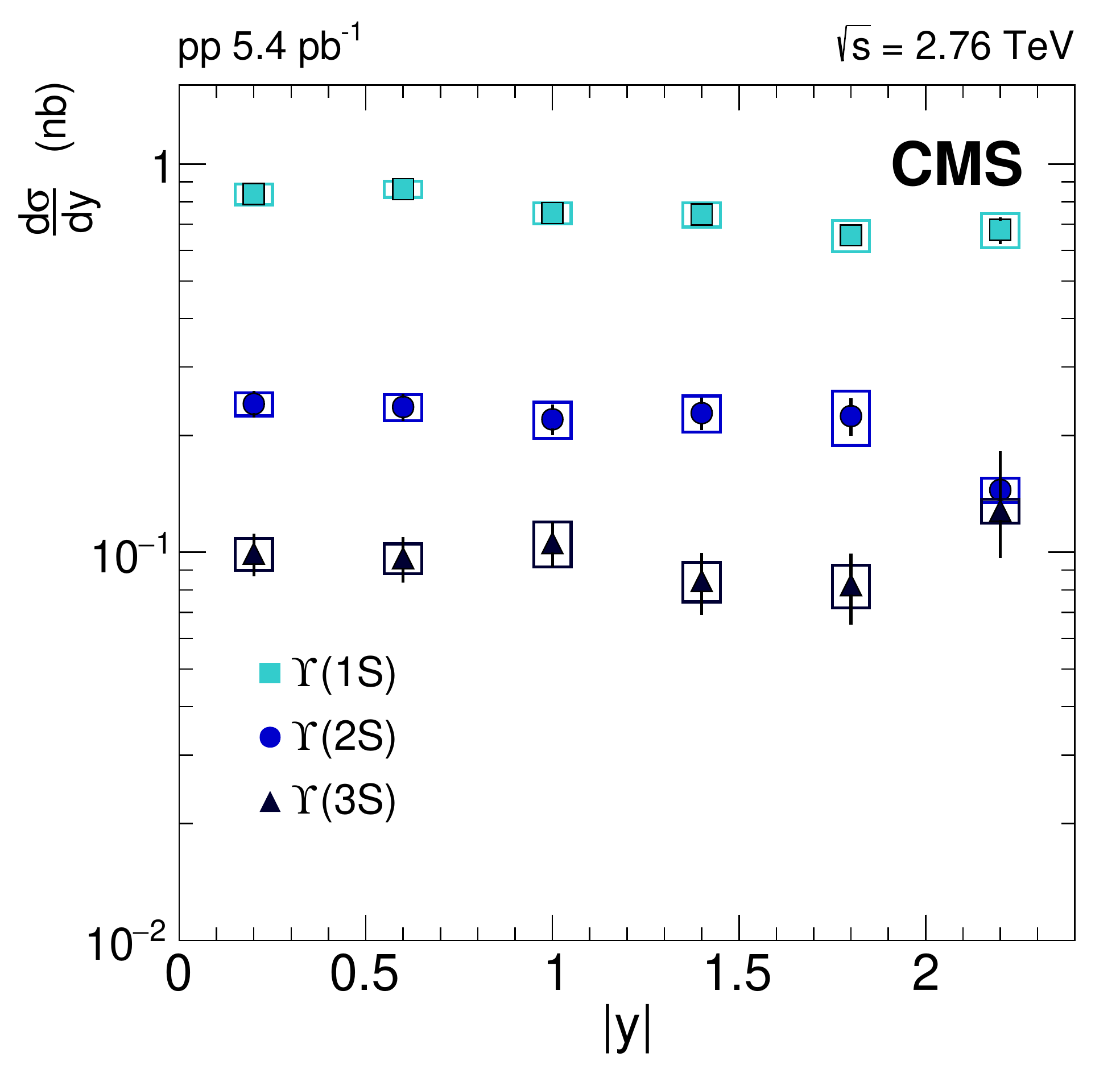}
\includegraphics[width=0.45\textwidth]{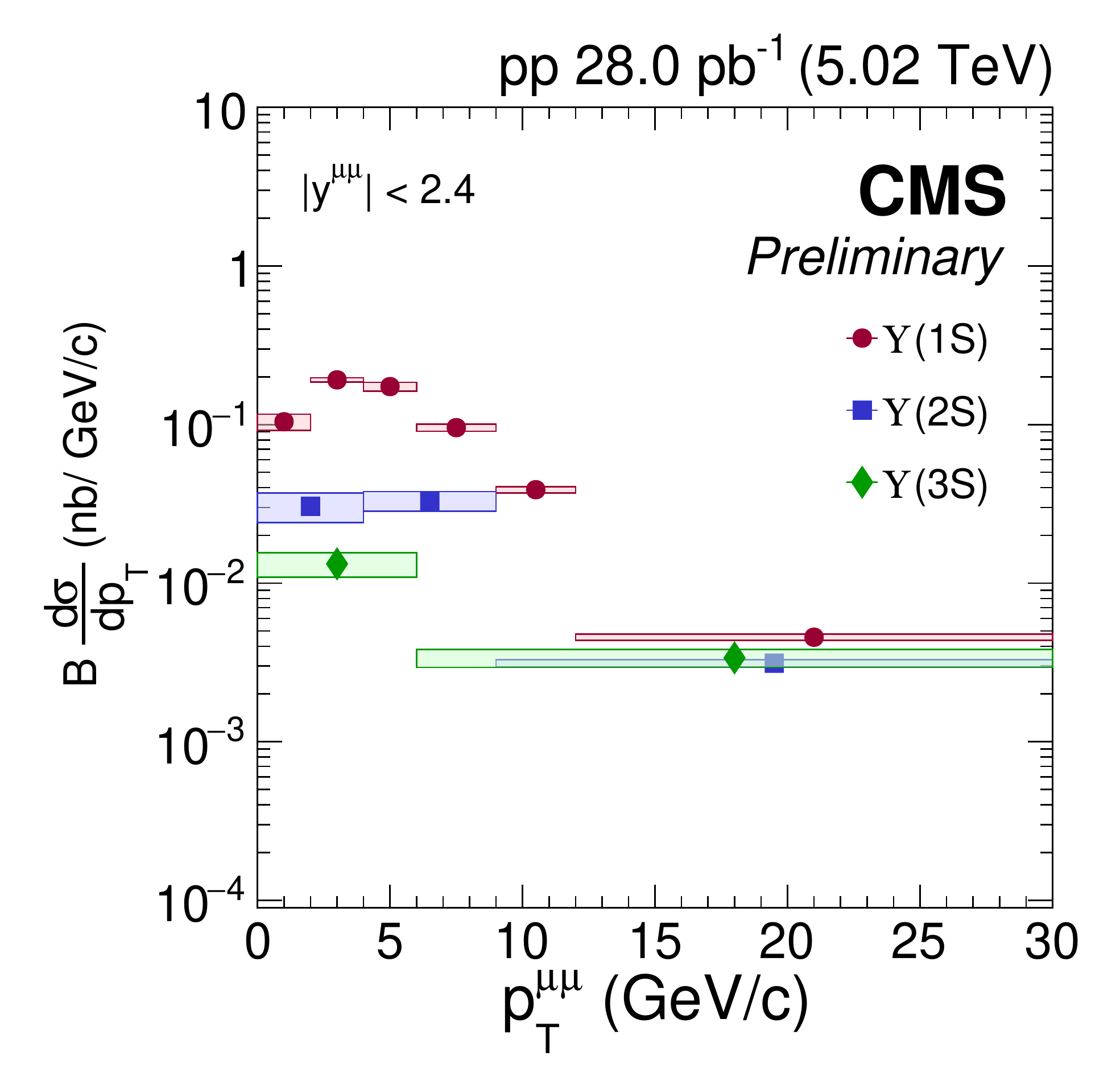}
\includegraphics[width=0.45\textwidth]{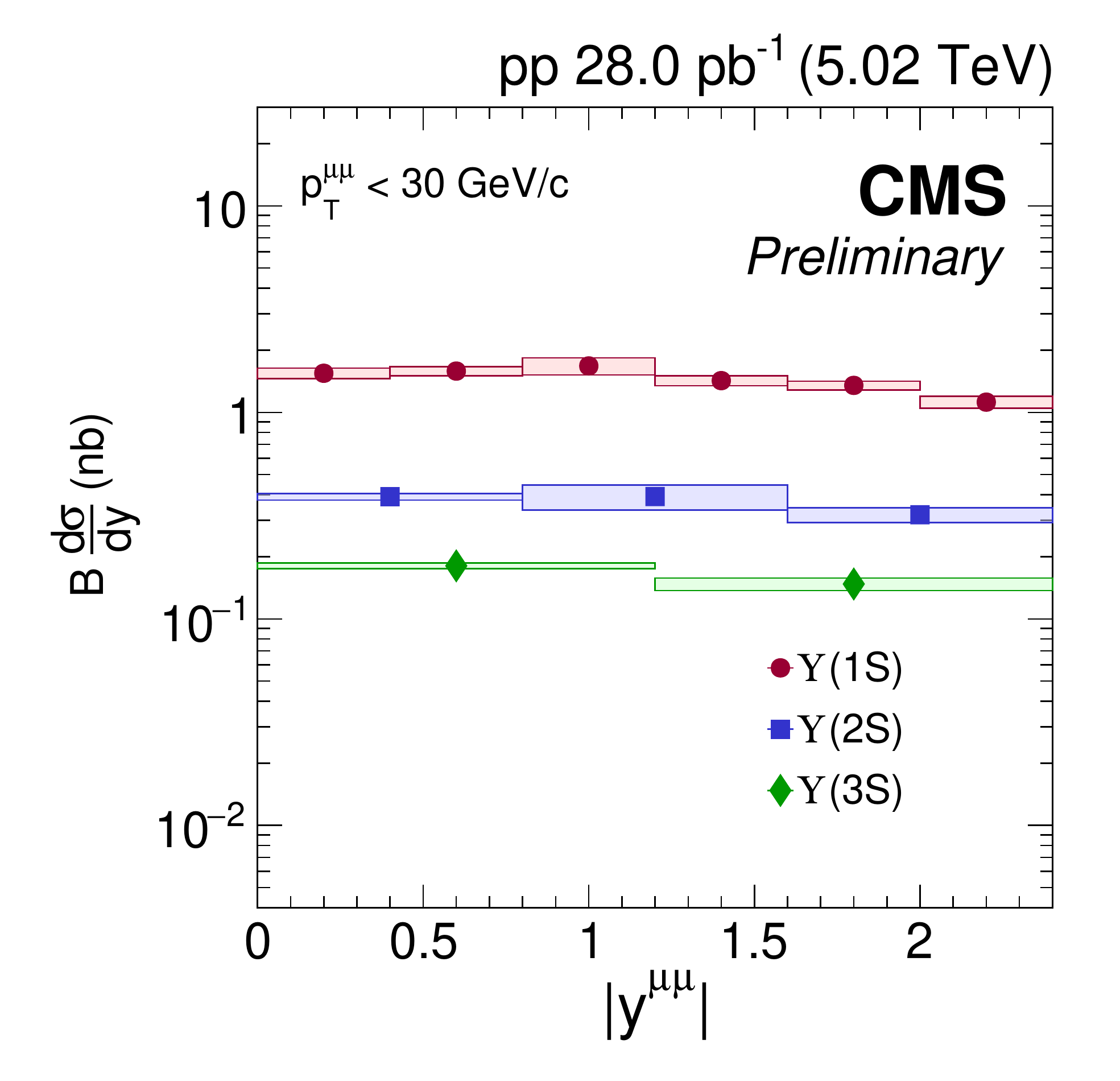}
\caption{$\Upsilon(nS)$ differential cross sections times branching fractions 
as a function of \pt (left) and rapidity (right), in $pp$ collisions taken 
at 2.76\,\TeV (top) and 5.02\,\TeV (bottom).  
}
\label{Y-xsec-2p76-5p02}
\end{center}
\end{figure}

Reduced event number data sets accumulated in special $pp$ runs at different center-of-mass
collision energies,  of $\sqrt{s} = 2.76$\,\TeV (Ref.~\citen{HIN-15-001}) and $\sqrt{s}=5.02$\,\TeV (Ref.~\citen{HIN-16-023}),  
corresponding to 5.4 and 28\,pb$^{-1}$, respectively, 
have been also explored. 
The differential cross section measurements of the individual $\Upsilon(nS)$ states as a function of \pt and rapidity are shown in Fig.~\ref{Y-xsec-2p76-5p02}.
Despite the modest statistical power and reduced \pt reach, because of the small size of the data sets accumulated, these measurements add useful information for characterizing the production $\sqrt{s}$ dependence. 

\begin{figure}[th!] \begin{center}
\includegraphics[width=0.29\textwidth]{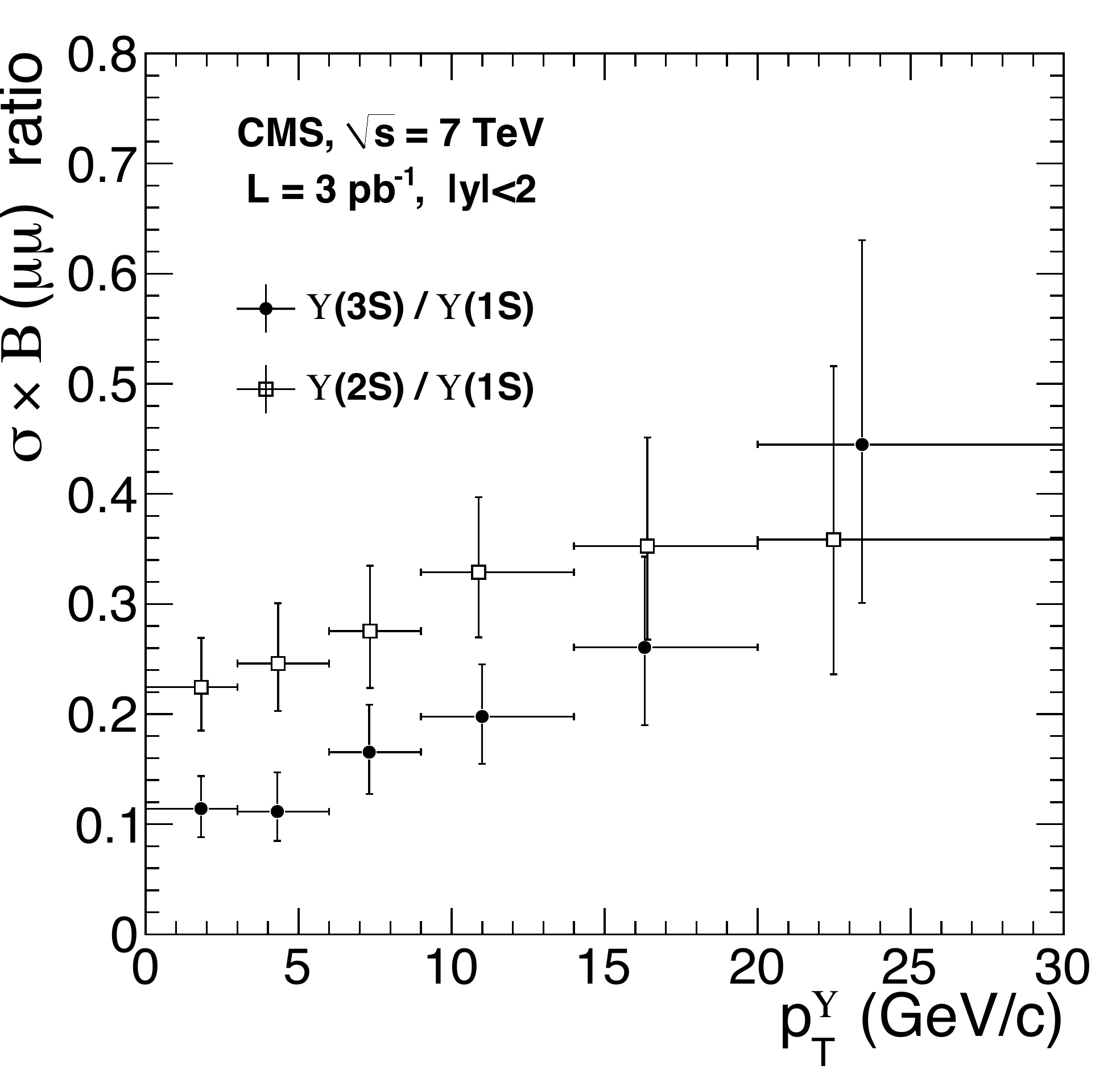}
\includegraphics[width=0.28\textwidth]{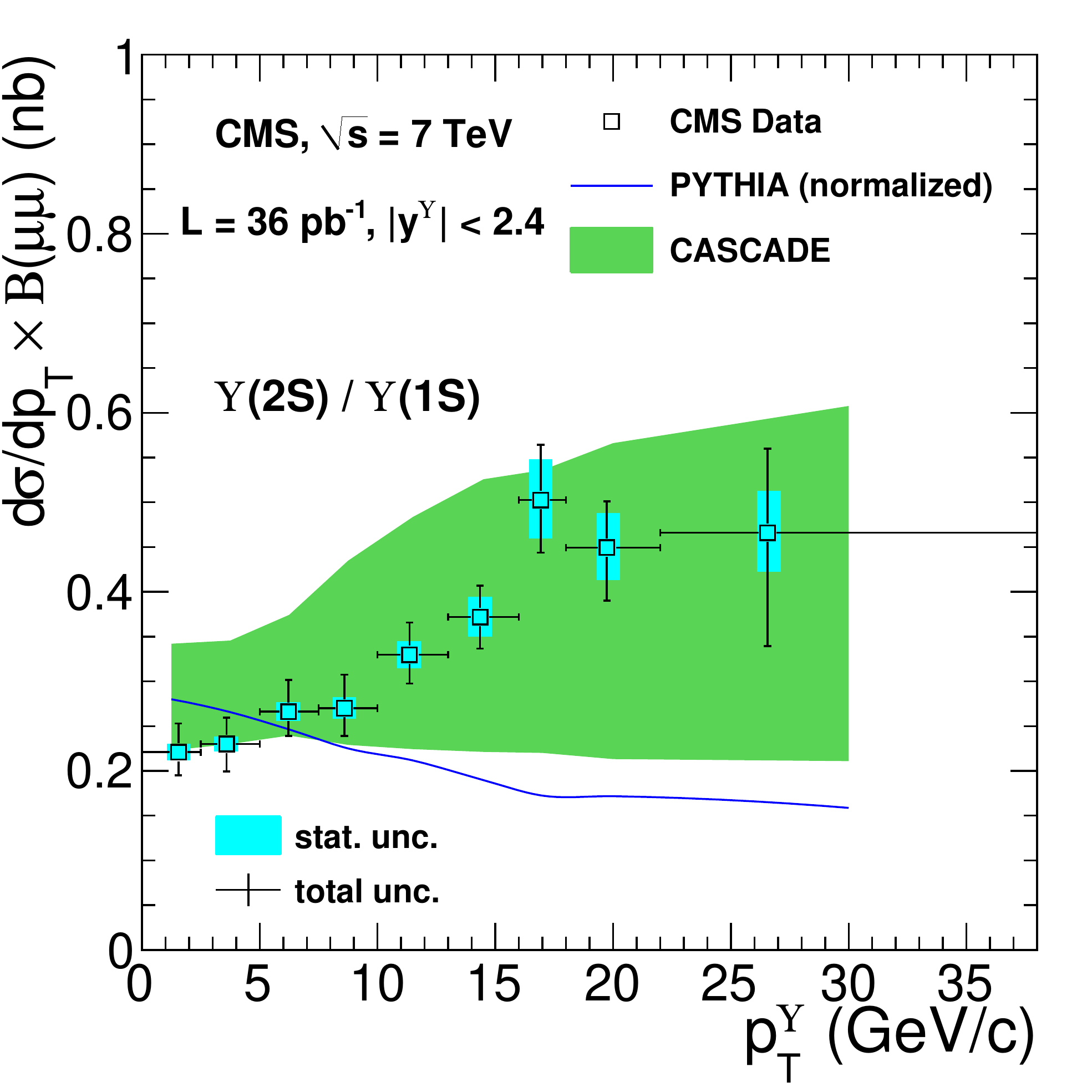}
\includegraphics[width=0.4\textwidth]{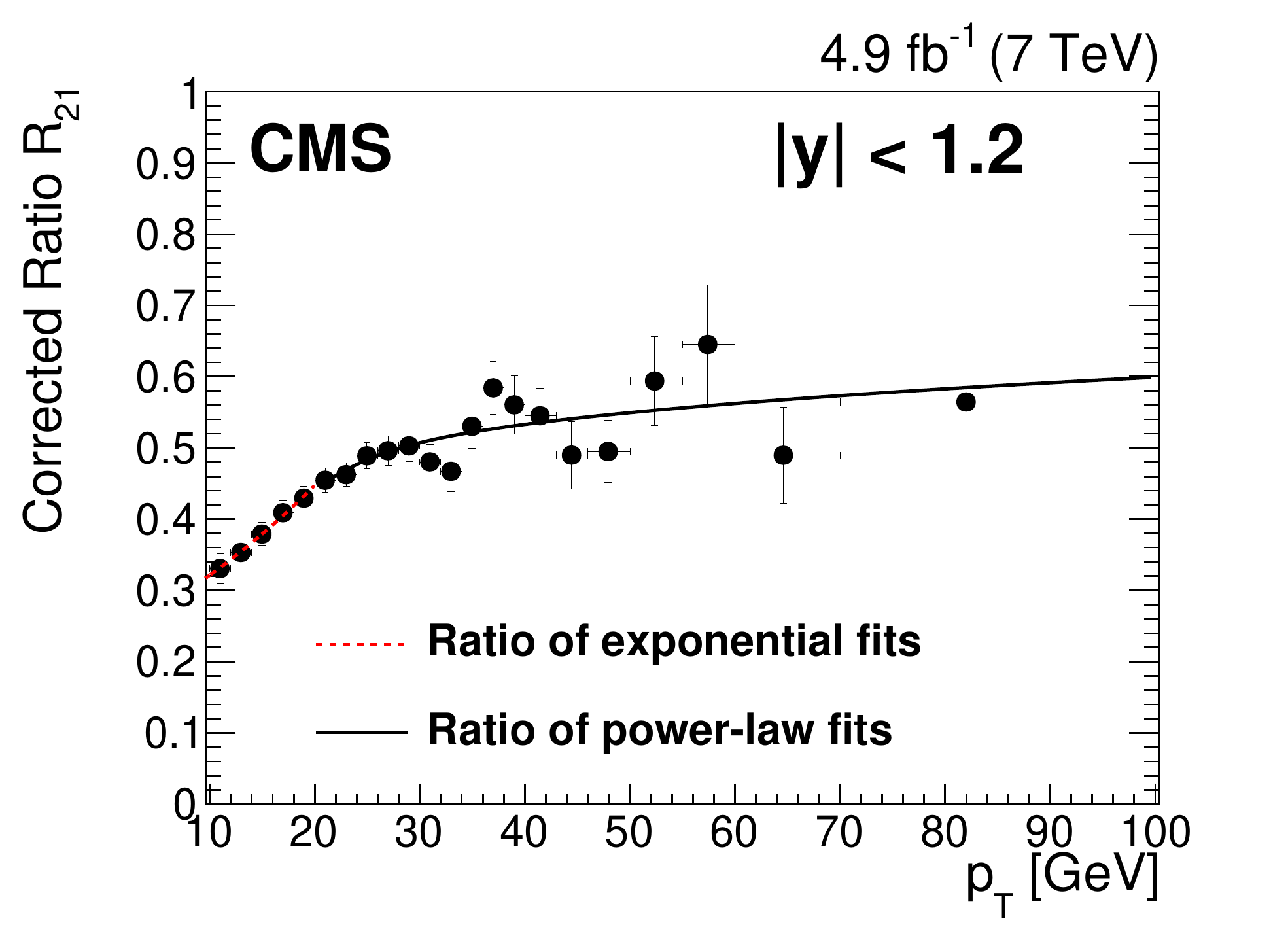} 
\caption{The ratio of $\Upsilon(nS)$ differential cross section times
branching fractions as a function of \pt, measured in data sets
corresponding to 3, 36, and 4900 pb$^{-1}$. In the middle plot, the
ratio is compared with predictions from PYTHIA (line) and CASCADE
(band). In the right-hand plot, the $\Upsilon(1S)$ and $\Upsilon(2S)$
differential cross sections are fitted with exponential ($\pt < 20$\,\GeV)
and power-law ($\pt > 20$\,\GeV) functions, and the curves on the plot
are the ratios of the fitted functions. }
\label{Y-xsec-ratios}
\end{center}
\end{figure}

The relative production cross sections provide further information, 
from both experimental and theoretical 
perspectives. For this reason CMS has provided cross section ratios
since the earliest measurements~\cite{xsecPRD, xsecPLB, xsecPLB2015},
and these are summarized in Fig.~\ref{Y-xsec-ratios}. They illustrate a
clear increase in the production ratios, $\Upsilon(nS)/\Upsilon(1S)$,
particularly at low \pt values. As larger data sets probe the higher
\pt region, a saturation threshold appears at about three
times the $\Upsilon$ mass, as displayed in the right-side panel of
Fig.~\ref{Y-xsec-ratios}.

\subsection{$\Upsilon$ polarization}	
\label{sec-pol} 

Measurements of quarkonium polarization provide important information
about the production mechanisms, and complement the cross section
measurements. As was discussed, the latter also have a sizable
dependence on the former. The strategy adopted by CMS involves the measurements
of the assumed unpolarized cross sections and separately of the polarizations. The
two form the proper inputs to carry out global fits for disentangling
and characterizing the competing production mechanisms. 

Although the $\psi(nS)$ and $\Upsilon(nS)$ cross sections measured at
Tevatron~\cite{CDF, CDFJpsi, D0} and LHC~\cite{xsecPRD, xsecLHCb,
xsecPLB, CMSJpsi} can be reproduced by NRQCD COM calculations, the
corresponding predictions~\cite{BGong2} for ``strong'' transverse
polarizations are in stark contrast with the negligible polarizations
observed by the experiments~\cite{PolarizationCDF-Jpsi}. And, although
heavy quarkonia from color singlets are expected to be produced with
longitudinal polarization~\cite{CSM-1, ReviewOfBottomonium}, this is
also not observed.   

The polarization can be measured through the analysis of the angular distribution of the
two leptons produced in the $\Upsilon\rightarrow\mu^{+}\mu^{-}$
decay~\cite{pol1,pol2}: 
\begin{equation} W(\theta,
\phi)\propto\frac{1}{3+\lambda_{\theta}}(1+\lambda_{\theta}\cos^{2}\theta
+\lambda_{\phi}\sin^{2}\theta\cos2\phi+\lambda_{\theta\phi}\sin2\theta\cos\phi) \,, 
\end{equation} 
where $\phi$ and $\theta$ are the azimuthal and polar angles of the
outgoing leptons with respect to the quantization axis ($z$-axis) of the
chosen polarization frame. $\lambda$ are the set of polarization parameters;
the parameter $\lambda_{\theta}$ is 0 (1) for fully longitudinal
(transverse) polarization. 
The polarization parameters depend on the reference frame in which they are measured.
The three most commonly used reference frames are the helicity frame (HX), where the $z$-axis coincides
with the $\Upsilon$ momentum direction in the collision center-of-mass
frame; the Collins--Soper (CS) frame~\cite{CS}, where the $z$-axis is
chosen as the bisector of the two beam directions in the $\Upsilon$ rest
frame; and the perpendicular helicity (PX) frame~\cite{PX}, which is
orthogonal to the CS frame. The $y$-axis is always taken along the
direction of the vector product of the two beam directions in the $\Upsilon$ rest frame.
In addition, one can define some observables that do not depend on the frame in which they are measured. 

The bottomonium states are heavier and satisfy the non-relativistic
approximation better than charmonium states. Measurements of the
$\Upsilon$ states, especially in the high \pt region, are thus expected
to provide more robust tests of NRQCD. Earlier measurements from 
CDF~\cite{PolarizationCDF} and D0~\cite{PolarizationD0} collaborations
were found to be in disagreement with the theoretical predictions, and
also disagreed to some degree between the two experiments. In these
earlier measurements only the $\lambda_\theta$ parameter was extracted,
in a single polarization frame.  
CMS measured all the polarization parameters $\lambda_{\theta}$, 
$\lambda_{\phi}$, and  $\lambda_{\theta\phi}$, in the complementary
polarization frames mentioned above, as well as the frame-invariant
quantity
$\tilde{\lambda}=(\lambda_{\theta}+3\lambda_{\phi})/(1-\lambda_{\phi})$,
using the $pp$ data sample at $\sqrt{s}=7$\,\TeV corresponding to an
integrated luminosity of 4.9 fb$^{-1}$~\cite{PolarizationCMS}. 

\begin{figure}[t]
\centerline{\includegraphics[width=0.9\textwidth]{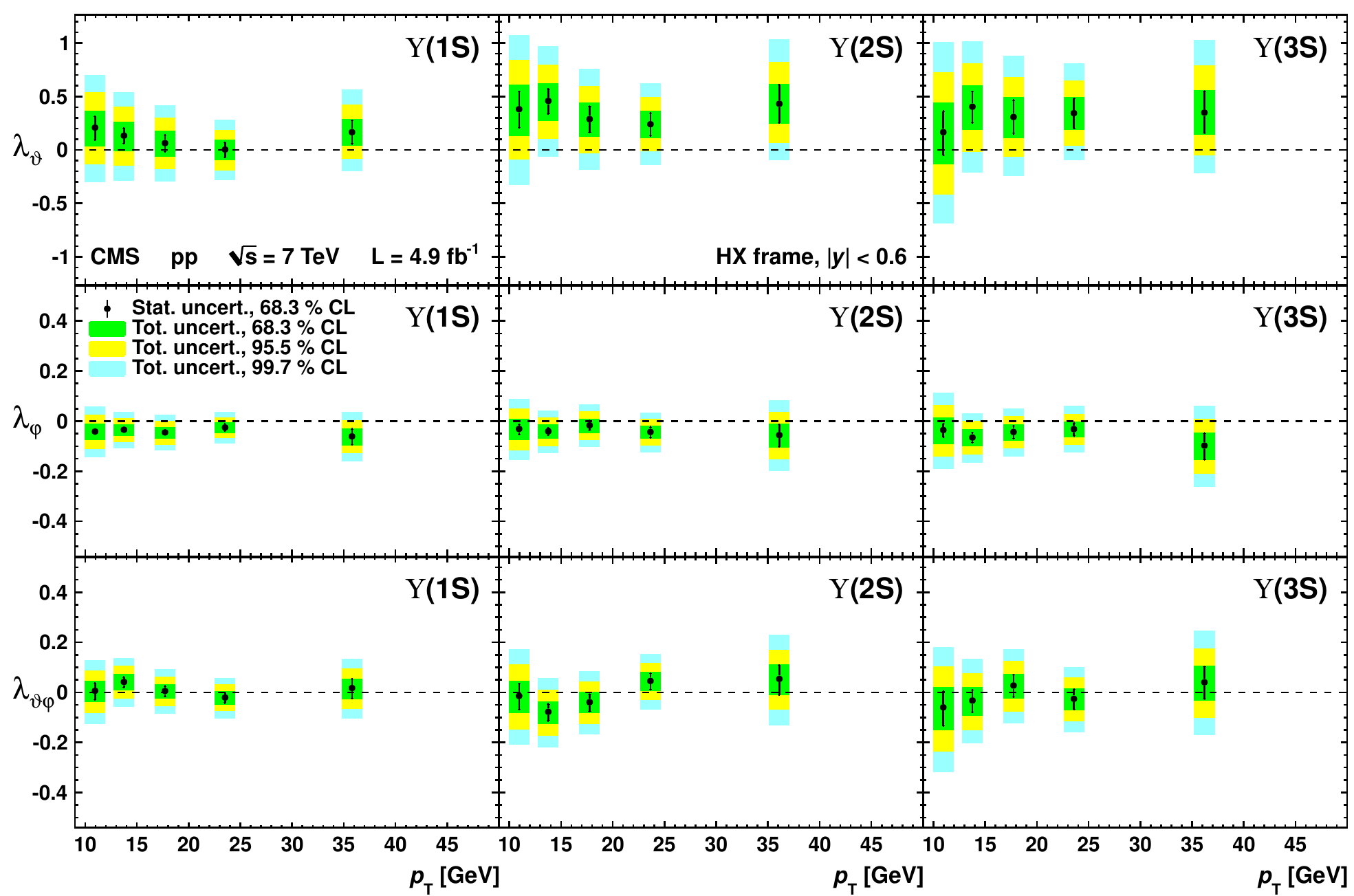}}
\caption{The measured $\Upsilon(nS)$ polarization parameters; shown as a function 
of $\Upsilon$ \pt for $|y|<0.6$ in the helicity frame. The error bars display the statistical uncertainty, 
while the bands show the total uncertainty at the 1$\sigma, 2\sigma, \rm{and~} 3\sigma$ levels.
}
\label{Y-pol-5fb} 
\end{figure}

Figure~\ref{Y-pol-5fb} shows the $\lambda_{\theta}$, 
$\lambda_{\phi}$, and  $\lambda_{\theta\phi}$ measurements as function of \pt~\cite{PolarizationCMS}, for the $\Upsilon(nS)$ states in
the HX frame, in the rapidity range $|y|<0.6$. Measurements in
complementary kinematic regions and reference frames were also
performed. The frame-invariant parameters $\tilde{\lambda}$ for the
three $\Upsilon$ states were also studied as a function of the
$\Upsilon(nS)$ \pt, and are in good agreement in the HX, CS, and PX
frames~\cite{PolarizationCMS}. All  polarization parameters are
compatible with zero or small values in the three polarization frames. 

\begin{figure}[t]
\centerline{\includegraphics[width=0.7\textwidth]{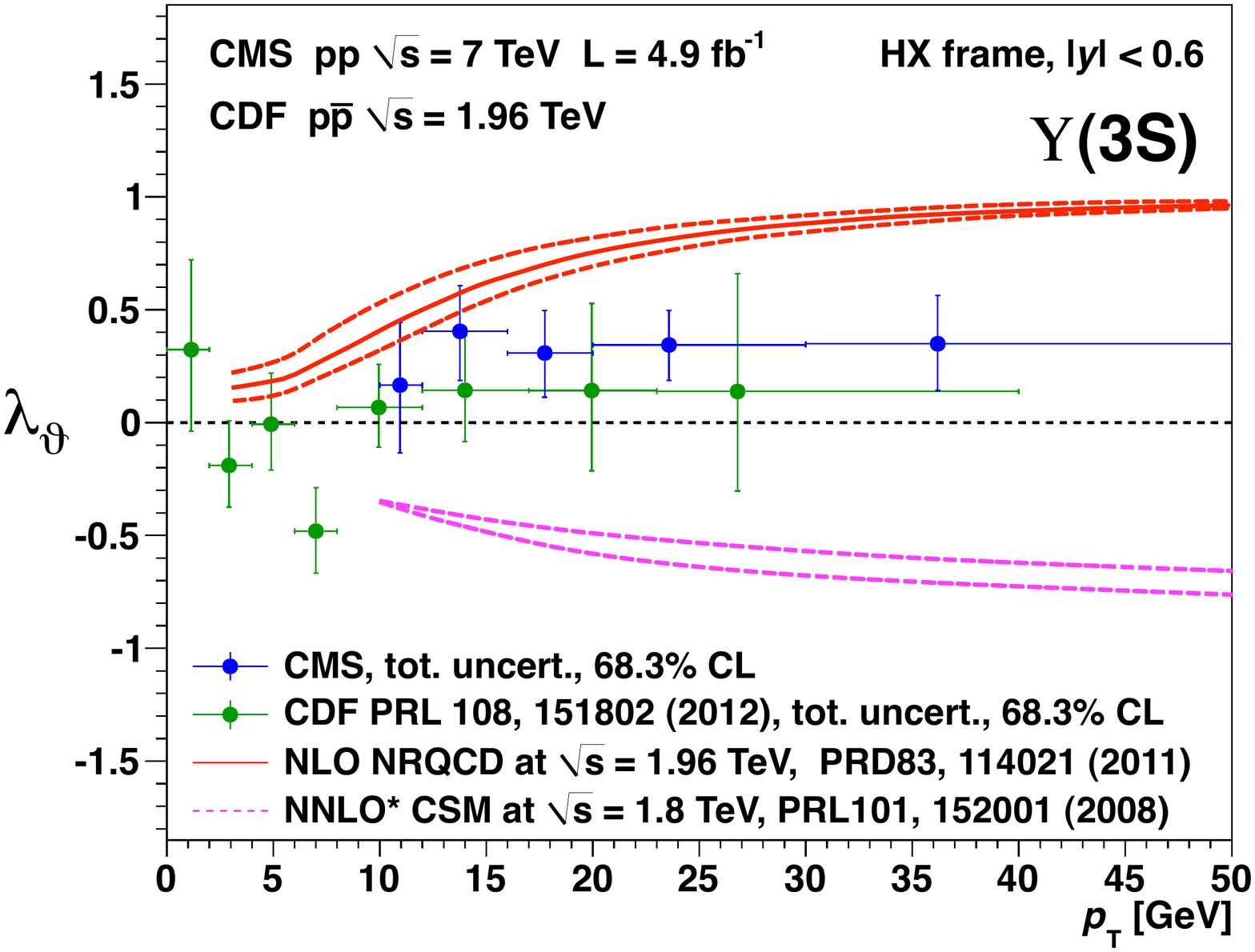}}
\caption{
Comparison of the CMS and CDF polarization measurements with model expectations, 
shown for the $\Upsilon(3S)$ state in the helicity frame in the central rapidity region $|\eta|<0.6$.~\cite{PolarizationCMS,BPH-11-023}. 
}
\label{Y-pol-theory} 
\end{figure}

In summary, the measurements exclude large longitudinal and transverse
polarizations for $\Upsilon(nS)$, in extended \pt and $y$ ranges
compared with previous experiments. This result is in disagreement with
the theoretical predictions for high-energy hadron collisions, as shown
in Fig.~\ref{Y-pol-theory}~\cite{PolarizationCMS,BPH-11-023}. 

The measurements of $\Upsilon$ cross section and polarization have led to 
new theoretical interpretations of the quarkonium puzzle. For instance,
in a recent study~\cite{PFaccioli}, global fits using both cross section and 
polarization measurements are performed
 to determine the nonperturbative parameters of
bound-state formation. This study reveals unexpected
hierarchies in the phenomenological long-distance parameters, which
brings a new understanding of the bound-state formation mechanism in
QCD. 


\section{Suppression in heavy ion collisions}
\label{sec-suppression}

\subsection{$\Upsilon$ suppression in PbPb}
\label{sec-PbPb}

Quarks and gluons are normally bound together to form composite
particles. However, QCD allows for strongly interacting matter to
undergo a phase transition to an unbound (deconfined) state at
sufficiently high temperature and density. The unique medium of quarks
and gluons in this deconfined state where the partons are no longer
confined to hadrons is referred to as the quark-gluon plasma.  
This medium can be produced in heavy ion collisions, where once the
heavy quarkonium states are formed they are expected to unbind due to
the strong interactions with partons in the medium 
through a QCD Debye screening mechanism~\cite{Satz}. Above a certain temperature, the weaker
bound states, such as $\Upsilon(3S)$, are expected to unbind more
completely compared to the more strongly bound states, e.g. $\Upsilon(1S)$. 
At even higher temperatures, more of the weakly bound states
are expected to dissolve. In the experiment, this sequential unbinding
(also referred to as melting) of quarkonium states is expected to be
observed as a sequential suppression of their yields. 
The suppression of heavy quarkonium states was accordingly proposed as
the smoking-gun signature of the phase transition, and its sequential
pattern as a probe of the medium temperature~\cite{Satz,PRL107,Phys5-132}.

The NA50~\cite{NA50-1,NA50-2,NA50-3} and NA60~\cite{NA60} experiments 
at the CERN Super Proton Synchrotron, and the PHENIX~\cite{PHENIX} 
and STAR~\cite{STAR} experiments at BNL RHIC had measured suppression of
$J/\psi$ and $\psi(2S)$ yields in heavy ion collisions. However, these experiments 
were not able to carry out quantitative studies of the $\Upsilon(nS)$ states. 
Bottomonium states are regarded as better probes because
recombination effects are believed to be much less significant than in
the charmonia case~\cite{ZhuangYRHIC}. Although the
recombination effect is expected to increase for bottomonia from RHIC to LHC energies, 
it is predicted to remain small~\cite{ZhuangYLHC}. 

\begin{figure}[t]
\centerline{\includegraphics[width=0.7\textwidth]{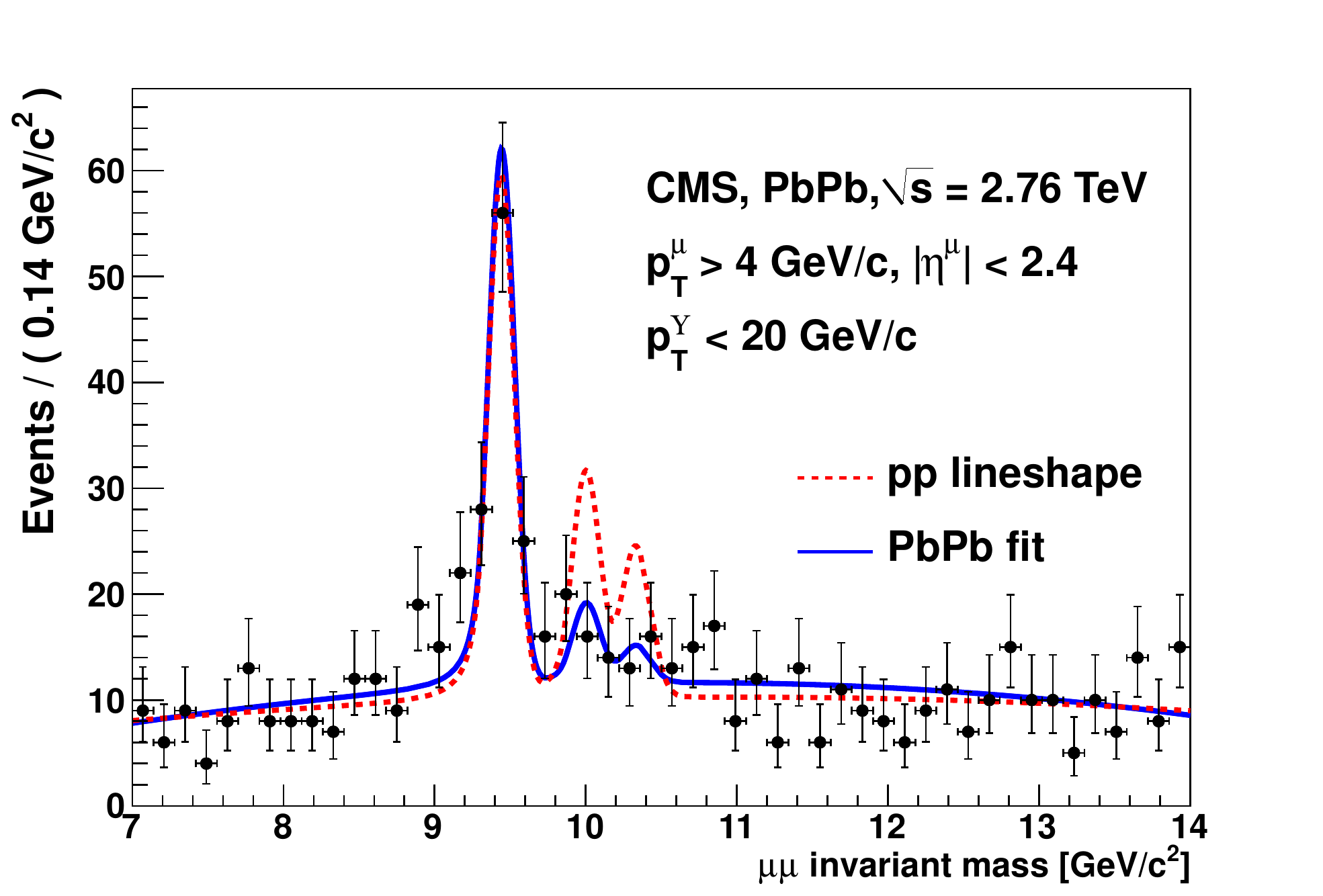}}
\caption{Dimuon invariant mass distribution in the vicinity of the $\Upsilon(nS)$ states, 
from a PbPb data set corresponding to  7.3 $\mu$b$^{-1}$. 
The solid (blue) line shows the fit to the PbPb data; the dashed (red)
line shows the shape obtained from the fit to the $pp$ data, after normalizing to the $\Upsilon$(1S) yield in PbPb.} 
\label{overlay2011}
\end{figure}

The first indication of $\Upsilon$ suppression in heavy ion collisions
was reported by CMS in 2011~\cite{PRL107, QM2011, JHEP}. This result is
based on data collected during the first LHC PbPb run in 2010 
and a special $pp$ run in 2011, at the same 2.76\,\TeV center-of-mass collision energy 
per nucleon pair ($\sqrt{s_{NN}}$). The PbPb and $pp$ data sets correspond to integrated
luminosities of 7.3 $\mu$b$^{-1}$ and 230\,nb$^{-1}$, respectively.
Thanks to the good momentum resolution of the CMS detector and the large
event samples, the three $\Upsilon$ resonances observed in the dimuon
mass spectrum were well separated for both PbPb and $pp$. Similar
techniques were applied to the two data sets to extract yields and
calculate cross sections. The results are shown in
Fig.~\ref{overlay2011}~\cite{HIN-11-007}, where the solid (blue) line
represents the fit to the mass spectrum in PbPb, and the dashed (red) line
represents the line-shape from the fit to the $pp$ data. In this figure,
all three $\Upsilon(nS)$ peaks are normalized such that the
$\Upsilon(1S)$ yield in the $pp$ line-shape matches the $\Upsilon(1S)$
yield in the PbPb fit. The $\Upsilon(2S)$ and $\Upsilon(3S)$ resonances
in PbPb collisions are clearly more strongly suppressed than the
$\Upsilon(1S)$, compared with the $pp$ result. The statistical
significance of the effect has been evaluated to be 2.4 $\sigma$. 
The double ratio, $[\Upsilon(2S+3S)/\Upsilon(1S)]_{\text{PbPb}}/[\Upsilon(2S+3S)/\Upsilon(1S)]_{pp}$, 
measured in the kinematic region defined by $\pt^{\mu}>4$\,\GeV and
$|\eta^{\mu}|<2.4$, is found to be $0.31^{+0.19}_{-0.15}(\text{stat.})\pm0.03(\text{syst.})$~\cite{PRL107}, 
significantly smaller than unity. 

Besides the hot-nuclear-matter (HNM), corresponding to the QGP, the
suppression of $\Upsilon$ production can also be caused by
cold-nuclear-matter (CNM) effects~\cite{CNM}. But the CNM effects cancel
to first order in the $\Upsilon$ double ratio measurements. For
example, one of the initial-state effects, ``shadowing,''~\cite{CNM} is expected to
suppress all three $\Upsilon$ resonances by almost the same factor, so
it has a small impact on their ratio. One of the final-state effects,
``nuclear absorption,''~\cite{CNM} is expected to be less important at LHC
collision energies. (To further probe CNM effects,
measurements in proton-lead collisions are conducted 
as described in Sec. ~\ref{sec-pPb}.) 
Other effects that could affect the suppression measurements, 
such as differences in the detector acceptance and efficiency, 
are similar for the different $\Upsilon$ states and 
largely cancel in the double ratio analysis.

\begin{figure}[t]
\centerline{\includegraphics[width=0.5\textwidth]{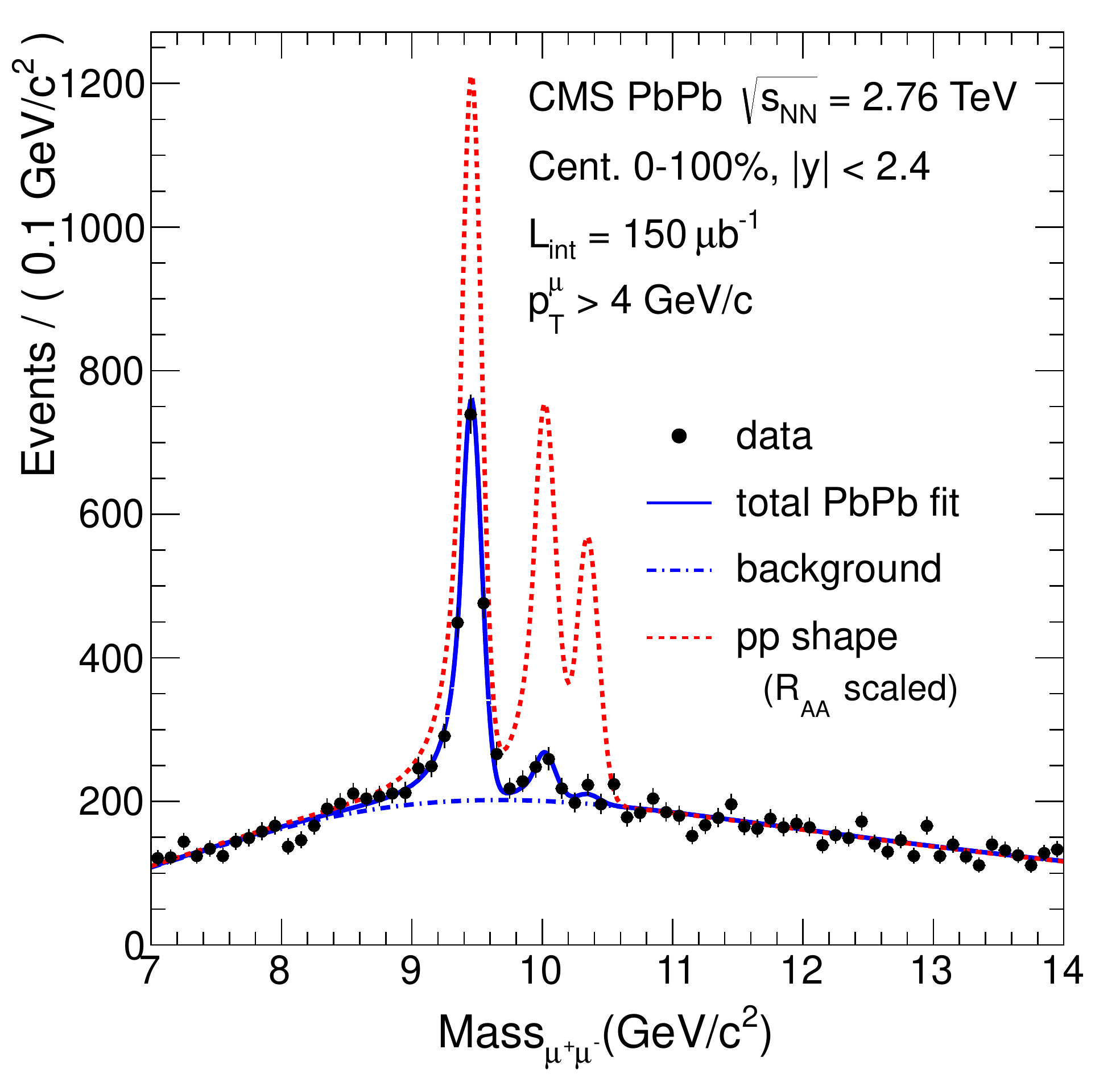}}
\caption{Dimuon invariant mass distribution from the 150 $\mu$b$^{-1}$
PbPb data. The solid (blue) line shows the fit to the PbPb data; the 
dashed (red) line illustrates the corresponding signals in $pp$ data, scaled
by the $R_{AA}$ values. }
\label{overlay2012}
\end{figure}

The integrated luminosity of the second LHC PbPb run exceeded 150
$\mu$b$^{-1}$ at the end of 2011, which is approximately 20 times larger
than the 2010 integrated luminosity. With this large data set, 
the relative suppression of excited $\Upsilon$ states
with respect to the  $\Upsilon(1S)$ ground state in PbPb was observed
with a significance exceeding 5$\sigma$~\cite{PRL109}. 
A comparison of dimuon mass spectra in PbPb and $pp$ is shown in
Fig.~\ref{overlay2012}~\cite{HIN-11-011}. The $\Upsilon(1S)$ state is
clearly suppressed in PbPb relative to $pp$, while the $\Upsilon(2S)$
and $\Upsilon(3S)$ states are suppressed to an even greater degree. 
The double ratios for $\Upsilon(2S)$ and $\Upsilon(3S)$ are measured as:~\cite{PRL109}
\begin{eqnarray}
\begin{array}{ll}
\displaystyle \frac{[\Upsilon(2S)/\Upsilon(1S)]_{\text{PbPb}}}{[\Upsilon(2S)/\Upsilon(1S)]_{pp}}=0.21\pm0.07(\text{stat.})\pm0.02(\text{syst.}), \\[8pt]
\displaystyle \frac{[\Upsilon(3S)/\Upsilon(1S)]_{\text{PbPb}}}{[\Upsilon(3S)/\Upsilon(1S)]_{pp}}=0.06\pm0.06(\text{stat.})\pm0.06(\text{syst.})<0.17(95\%\text{CL}).  
\end{array}
\label{doubleRatio}
\end{eqnarray}

In addition to the relative suppression of the two excited $\Upsilon$
states with respect to the ground state, the absolute suppression of all
three individual $\Upsilon$ states was also measured. This is quantified
by the nuclear modification factor, $R_{AA}$, that is defined as the
ratio of the yield per nucleon-nucleon collision in PbPb relative to
that in $pp$, corrected for efficiencies and normalized by luminosities.
It is measured as
\begin{equation}
R_{AA}=\frac{\mathcal{L}_{pp}}{T_{AA}N_{MB}}\frac{\Upsilon(nS)|_{\text{PbPb}}}{\Upsilon(nS)|_{pp}}\frac{\epsilon_{pp}}{\epsilon_{\text{PbPb}}}, 
\label{raaequ}
\end{equation}
where $T_{AA}$ is the nuclear overlap function~\cite{JHEP}, 
and $N_{MB}$ is the number of minimum-bias events sampled by the event selection. 
When $R_{AA} < 1$, suppression in PbPb is observed; otherwise, there is no indication of 
medium effects. 
A more detailed explanation of the $R_{AA}$ observable can be found in Ref.~\citen{PRL109}. 
The absolute suppression of the three individual $\Upsilon$
resonances is measured to be:~\cite{PRL109} 
\begin{eqnarray} 
\begin{array}{ll}
R_{AA}(\Upsilon(1S))=0.56\pm0.08(\text{stat.})\pm0.07(\text{syst.}), \\
R_{AA}(\Upsilon(2S))=0.12\pm0.04(\text{stat.})\pm0.02(\text{syst.}), \\
R_{AA}(\Upsilon(3S))=0.03\pm0.04(\text{stat.})\pm0.01(\text{syst.})<0.10
(95\%\text{CL}). 
\end{array} 
\label{raa} 
\end{eqnarray}

\begin{figure}[h]
\centering
\includegraphics[width=0.62\textwidth]{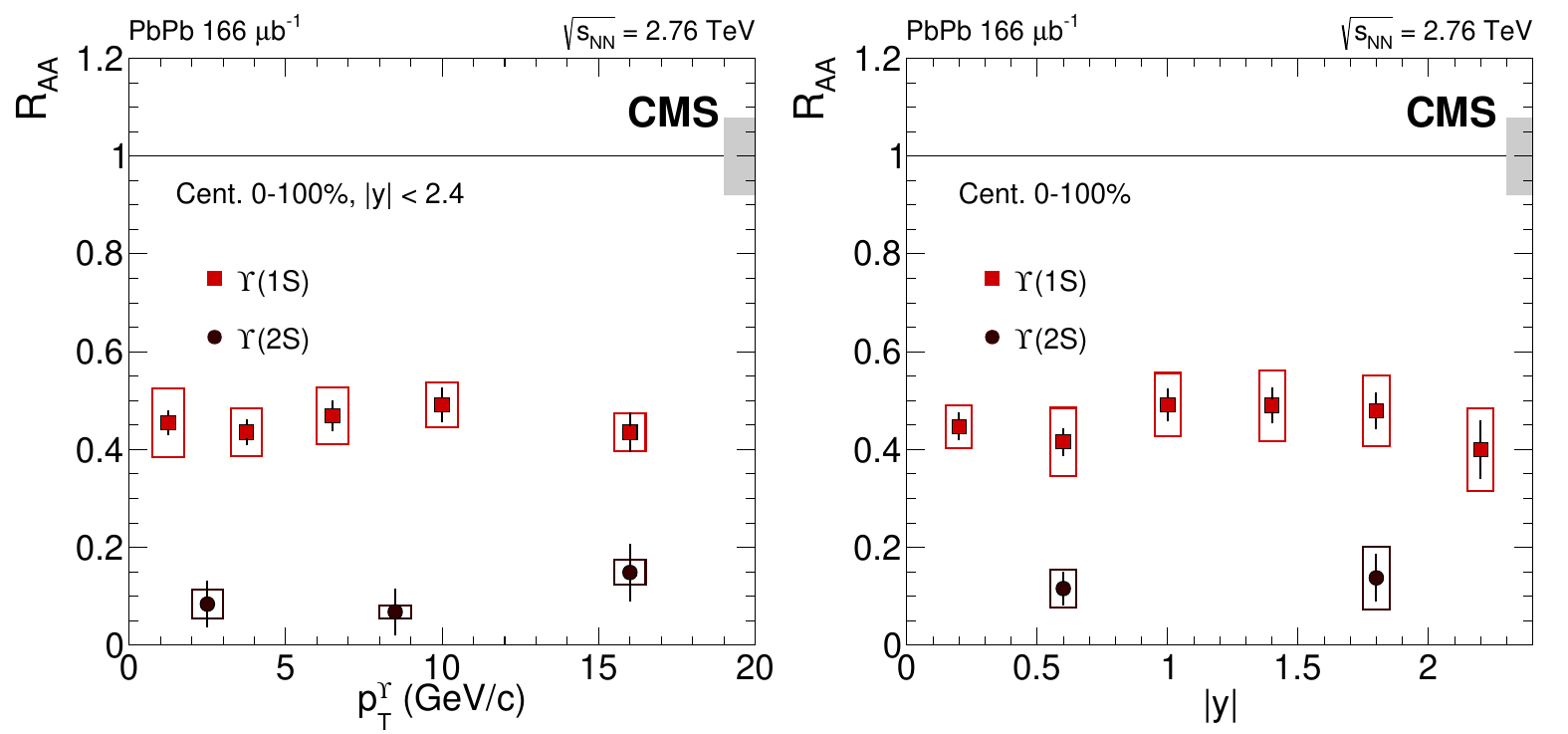}
\includegraphics[width=0.33\textwidth]{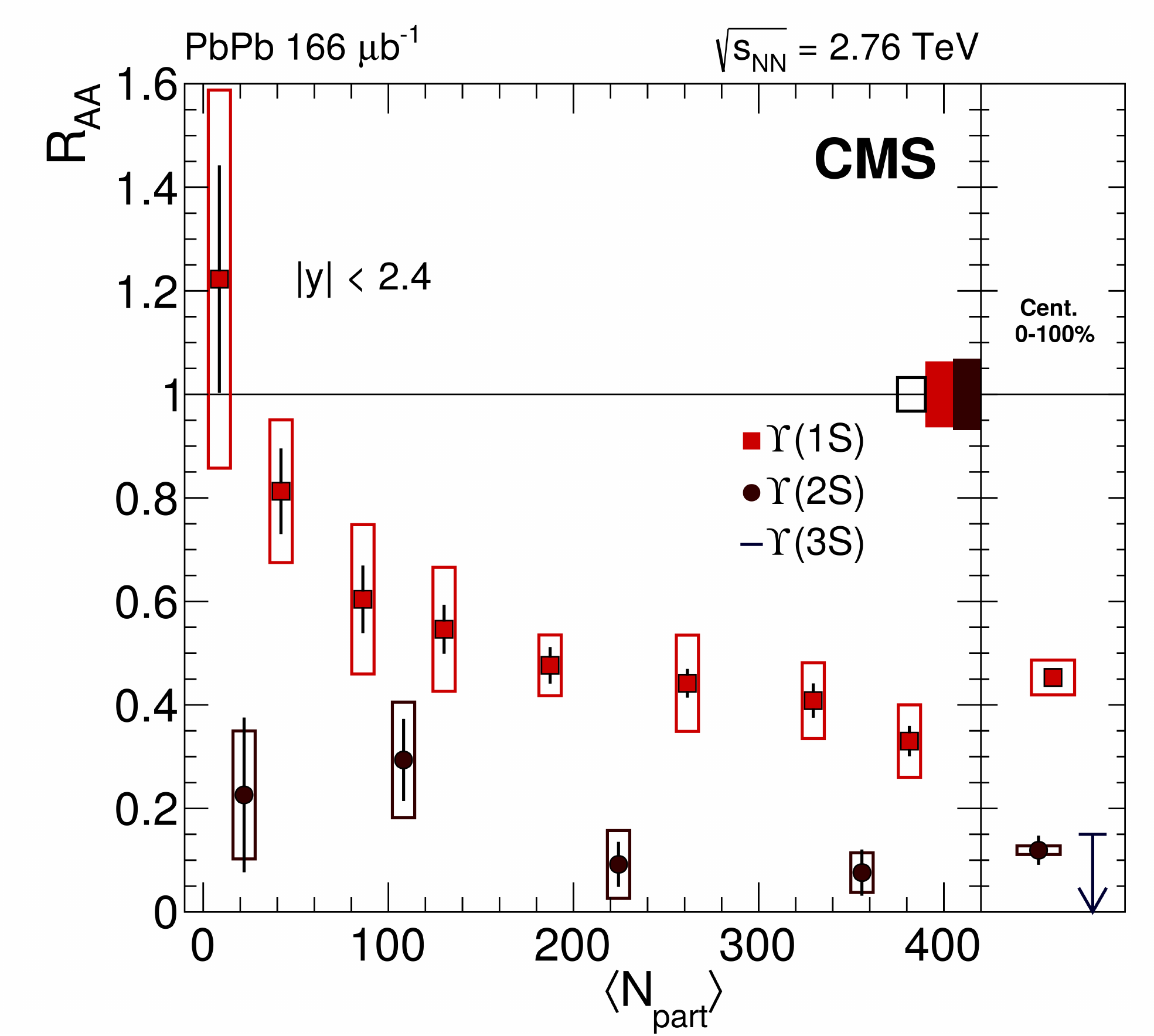}
\caption{The suppression factor, $R_{AA}$, for $\Upsilon(1S)$ and $\Upsilon(2S)$ 
as a function of \pt (left), $|y|$ (middle), and centrality (right).
Centrality is represented as the average number of participating
nucleons $N_{\text{part}}$. The upper limit of the $\Upsilon(3S)$
$R_{AA}$ is displayed with an arrow in the centrality integrated panel.
Systematic (statistical) uncertainties are drawn as error boxes (bars),
while the global (fully correlated) uncertainties are shown at unity. } 
\label{RAA-pT-y}
\end{figure}

Subsequently, the $pp$ event sample was increased by about 20 times in
2013 (5.4 pb$^{-1}$) ~\cite{HIN-15-001}, which allowed for a better
differential $R_{AA}$ study as function of the $\Upsilon$ rapidity,
transverse momentum, and centrality. Figure~\ref{RAA-pT-y} shows the
$\Upsilon(1S)$ and $\Upsilon(2S)$ $R_{AA}$ versus rapidity (left plot)
and transverse momentum (middle plot).  

Centrality is an important parameter in QGP
matter studies because it is directly related to the overlap region of the
colliding nuclei. It is determined from the energy deposits in the forward calorimeter~\cite{HF}, 
starting from 0\% for the most central collisions. 
With Glauber model calculations~\cite{Npart,Glauber}, the centrality variable can be 
expressed in terms of the number of nucleons participating in the collisions, $N_{\text{part}}$. 
In PbPb collisions, $N_{\text{part}}$ is the number of
nucleons (at most 208 for a Pb nucleus) that collide at least once with nucleons
in the other Pb nucleus. In Fig.~\ref{RAA-pT-y} (right), the $R_{AA}$  values
for $\Upsilon(1S)$ and $\Upsilon(2S)$ are shown as functions of
$N_{\text{part}}$~\cite{HIN-15-001}. For $\Upsilon(1S)$, the suppression
was observed to increase with the centrality of the collisions.
Comparisons with theoretical models are given in
Ref.~\citen{proceeding}. 

\begin{figure}[h!]
\centering
\includegraphics[width=0.32\textwidth]{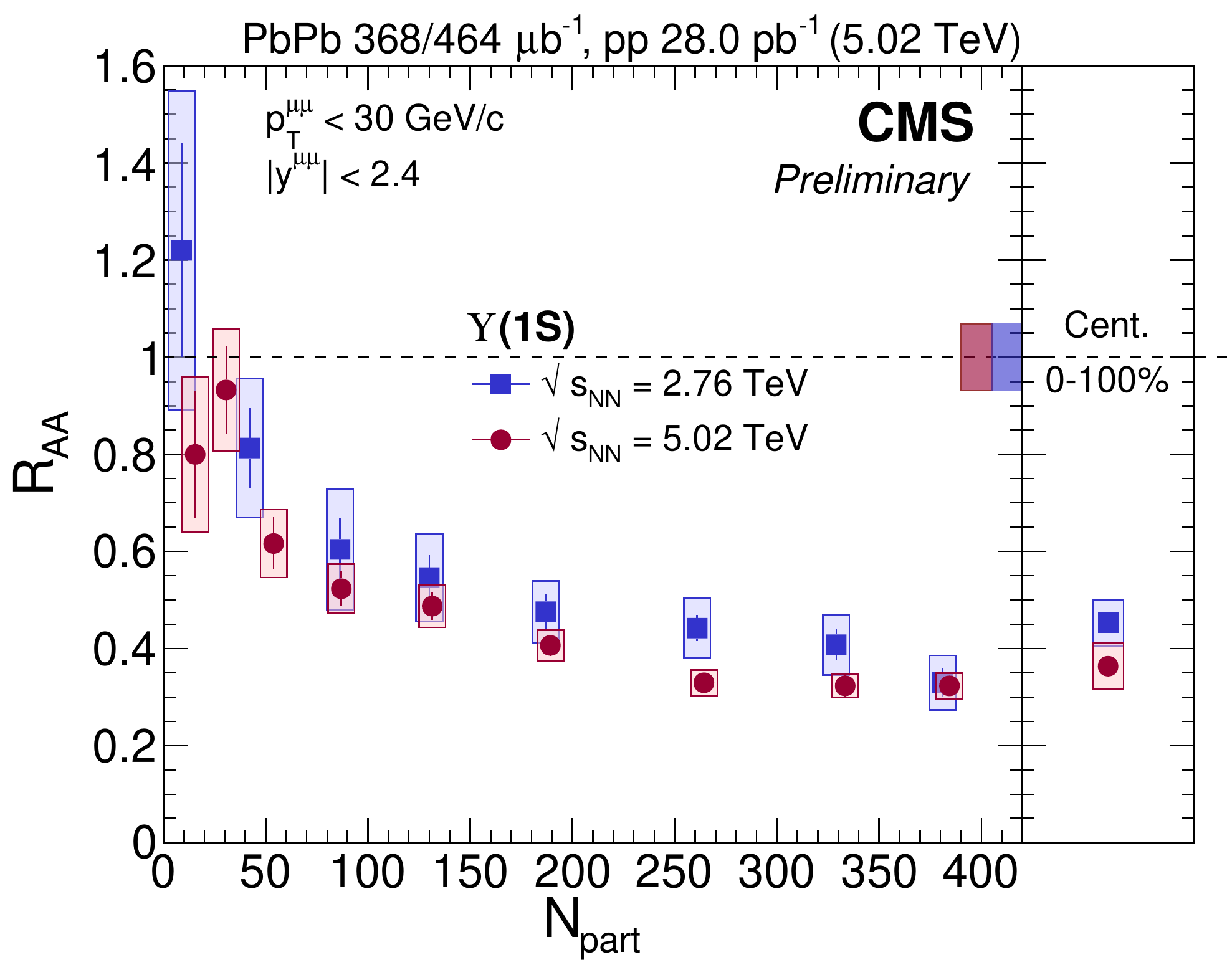}
\includegraphics[width=0.32\textwidth]{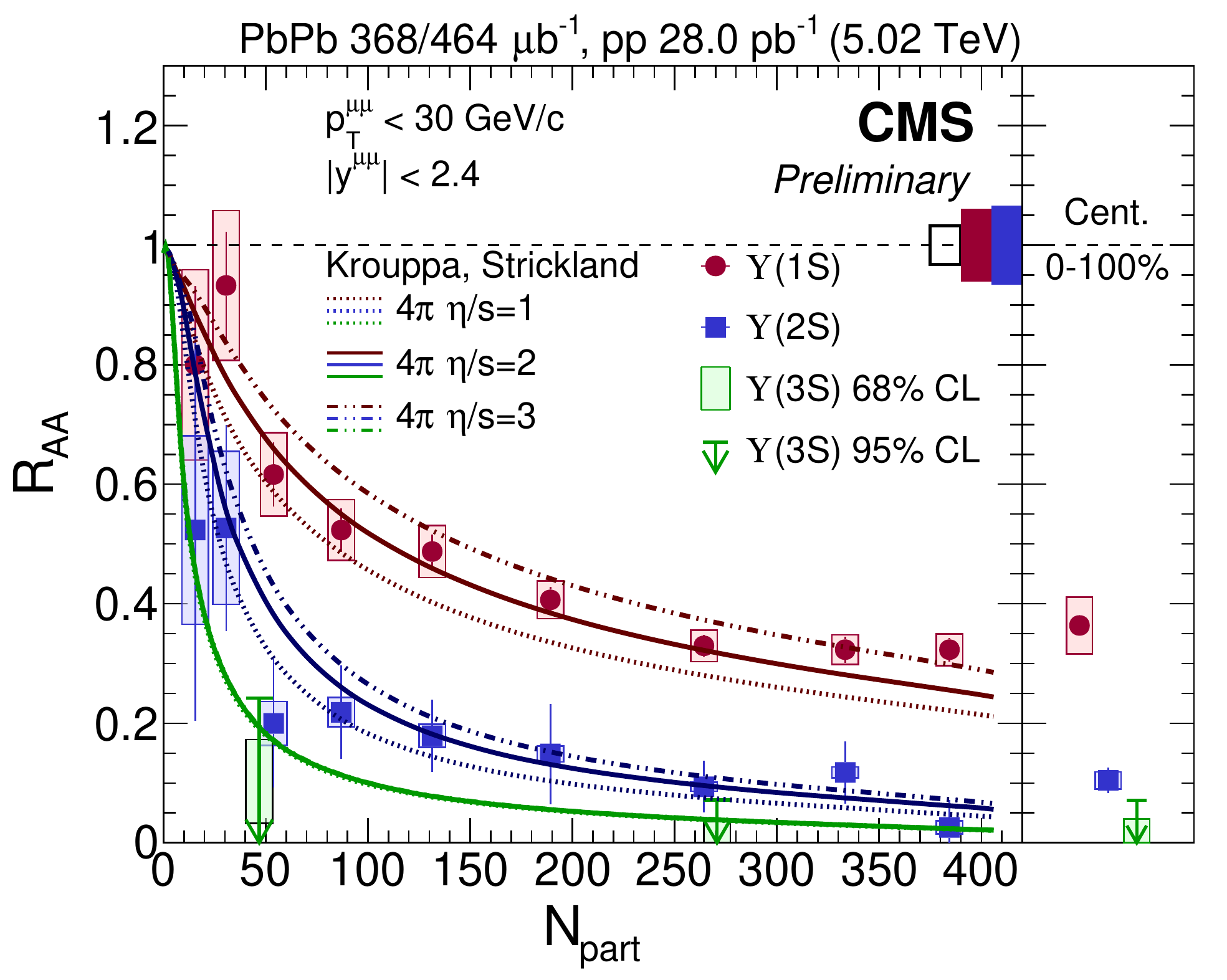} 
\includegraphics[width=0.32\textwidth]{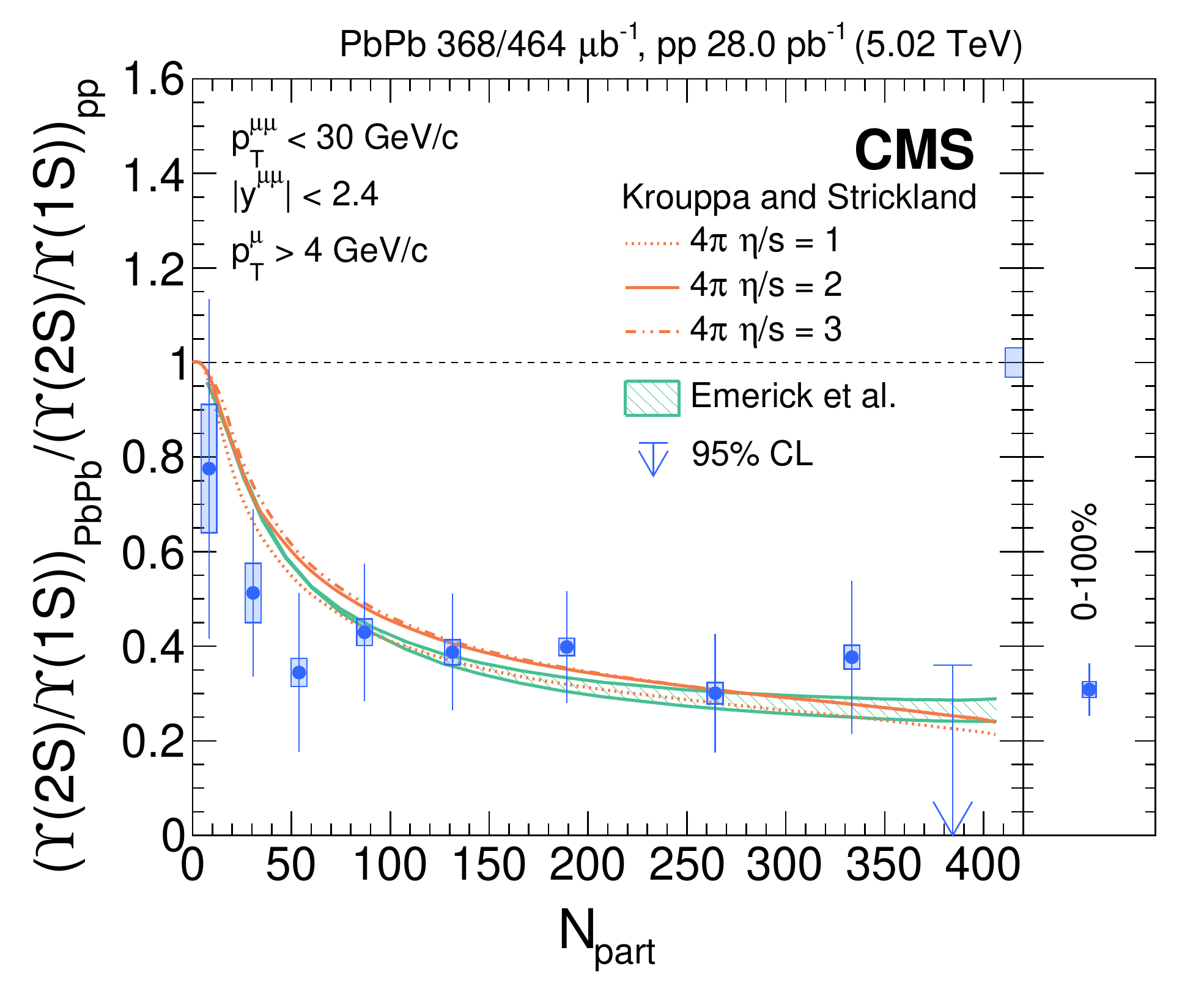}
\caption{Left: $R_{AA}$ versus centrality for $\Upsilon(1S)$ at 2.76\,\TeV
(blue) and 5.02\,\TeV (red).\cite{QM2017, HIN-16-023} Middle: $R_{AA}$ versus centrality for
$\Upsilon(1S)$, $\Upsilon(2S)$, and $\Upsilon(3S)$ compared with a
hydrodynamic model.\cite{QM2017, HIN-16-023, Strickland}. 
Right: Double ratio versus centrality for $\Upsilon(2S)$ at 5.02\,\TeV.~\cite{HIN-16-008}  }
\label{raa5tev}
\end{figure}

The suppression of $\Upsilon$ states in PbPb collisions at
$\sqrt{s_{NN}} = 5.02$\,\TeV is expected to be stronger than that measured
at 2.76\,\TeV, because the temperature of the medium is higher owing to the
higher collision energy. CMS results 
reported at Quark Matter 2017~\cite{QM2017, HIN-16-023, HIN-16-008} support this expectation,
as shown in Fig.~\ref{raa5tev} (left). In Fig.~\ref{raa5tev} (middle),
the $\Upsilon(nS)$ $R_{AA}$ versus centrality is compared with model expectations~\cite{QM2017, Strickland}, 
which contain bottomonia placed in an anisotropic hydrodynamic model. 
Other comparisons of $R_{AA}$ as functions of $\Upsilon$ rapidity and
transverse momentum can be found in Refs.~\citen{QM2017} and~\citen{HIN-16-023}. 
The centrality-integrated double ratio for $\Upsilon(2S)$ is measured 
as $0.308\pm0.055(\text{stat.})\pm0.019(\text{syst.})$.~\cite{HIN-16-008} The $\Upsilon(2S)$ double ratio 
as a function of centrality is shown in Fig.~\ref{raa5tev} (right). 
The centrality-integrated double ratio for $\Upsilon(3S)$ is less than 0.29 at 95\% CL.~\cite{HIN-16-008}

\begin{figure}[b!]
\centering
\includegraphics[width=0.45\textwidth]{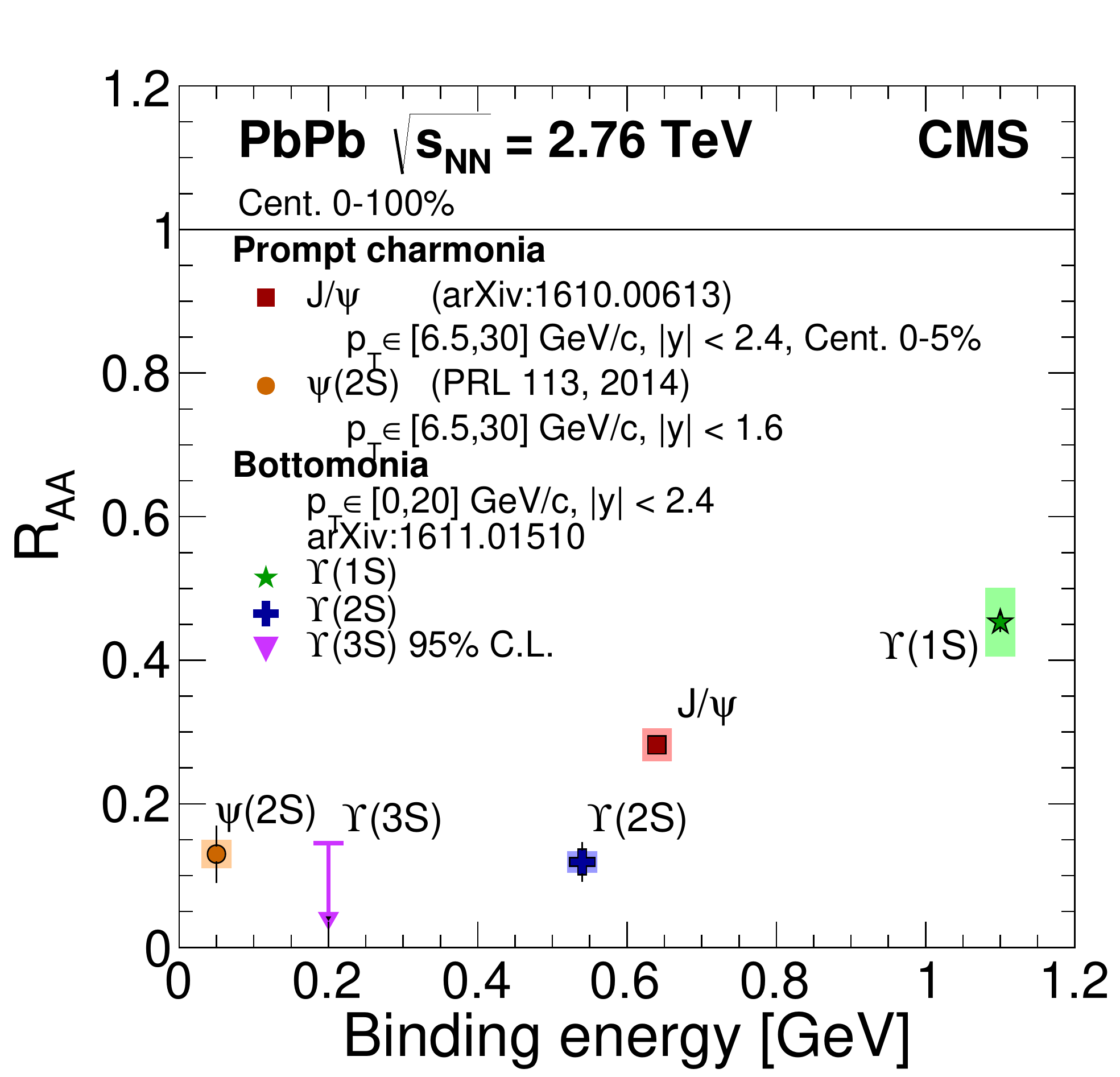}
\includegraphics[width=0.45\textwidth]{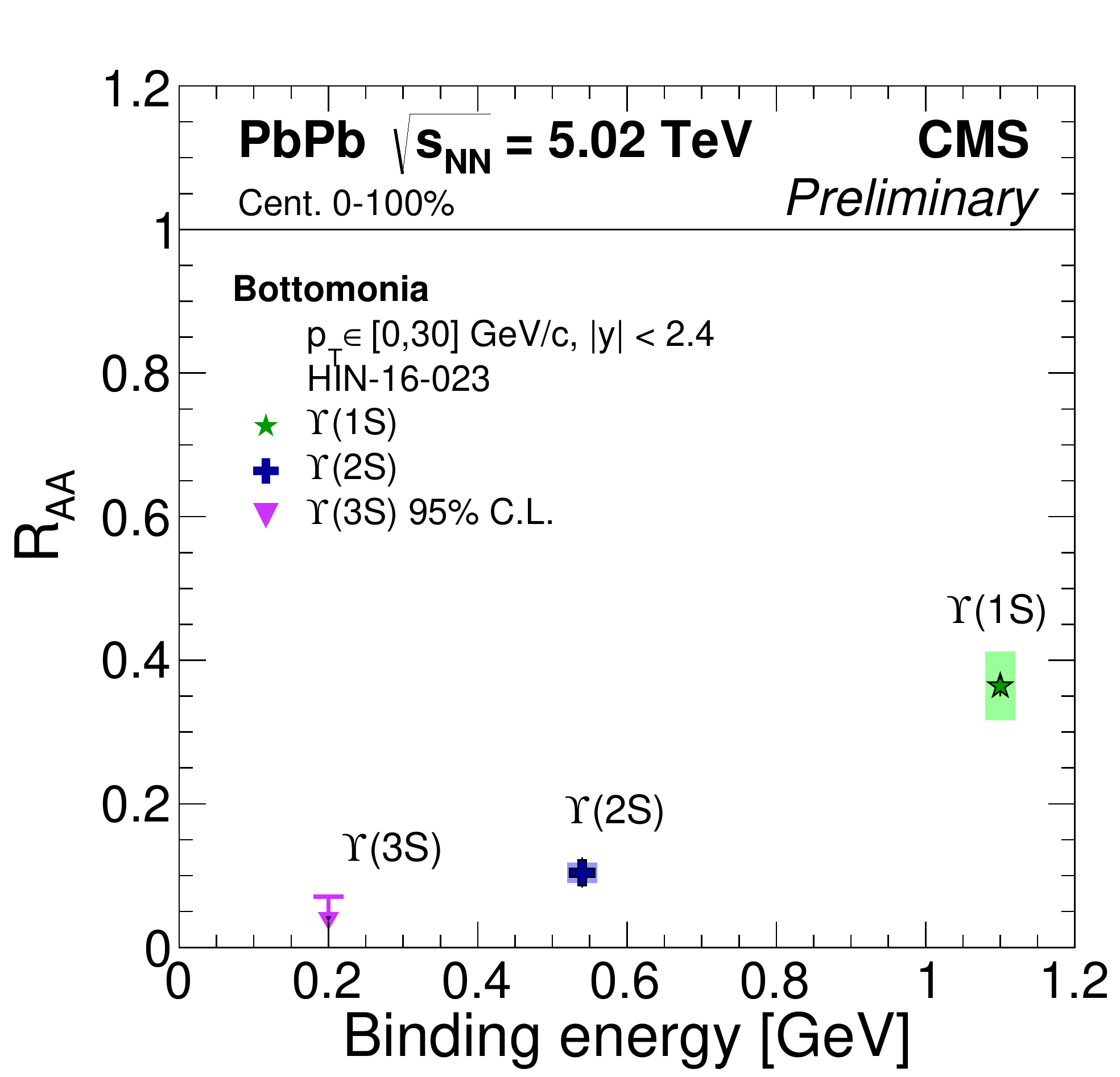}
\caption{The sequential suppression pattern of $S$-wave quarkonium states, shown as the measured 
$R_{AA}$ observable as a function of the binding energy of the individual quarkonium states, 
at 2.76\,\TeV (left)~\cite{HIN-12-007} and at 5.02\,\TeV (right)~\cite{QM2017}.}
\label{raaVSdE}
\end{figure}

The sequential suppression behavior may be best illustrated by displaying $R_{AA}$ as a function of the binding energy of the studied quarkonium states. This is represented in Fig.~\ref{raaVSdE}~\cite{HIN-12-007,QM2017}, for all $S$-wave quarkonium states. 
In this way CMS has experimentally established the sequential pattern of quarkonium suppression.


\subsection{$\Upsilon$ suppression effects in $p$Pb and in $pp$}
\label{sec-pPb}

\begin{figure}[t]
\centerline{\includegraphics[width=0.5\textwidth]{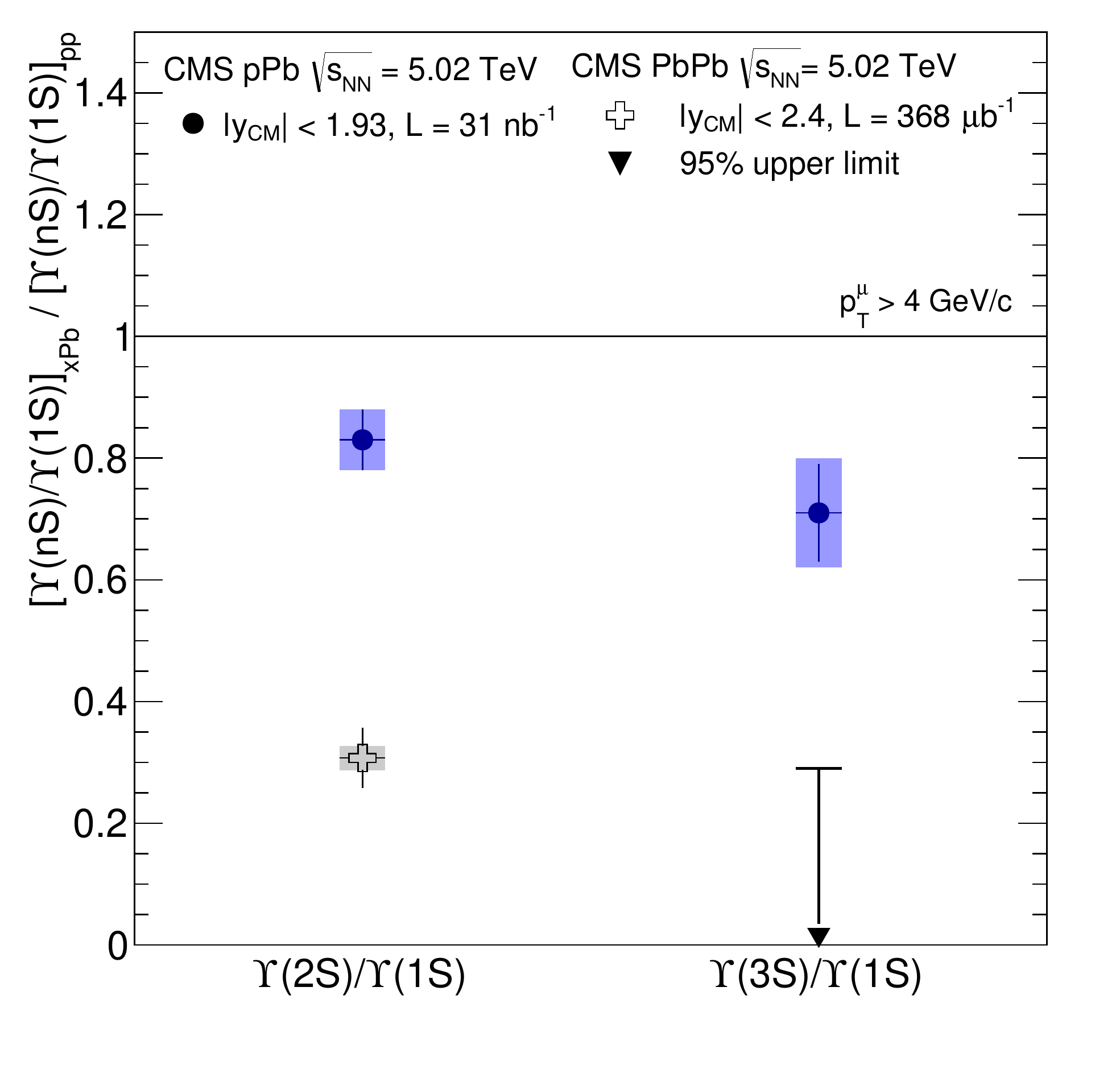}}
\caption{Double ratios of the $\Upsilon(2S)$ and $\Upsilon(3S)$ states
to the $\Upsilon(1S)$ ground state in $p$Pb collisions at 5.02\,\TeV with
respect to $pp$ collisions at 2.76\,\TeV (circles) and compared with
similar ratios for PbPb collisions at 5.02\,\TeV, which used a 5.02\,\TeV $pp$ 
data set for the reference. }
\label{DR-PbPb-pPb}
\end{figure}

As discussed above, cold-nuclear-matter effects will influence the
formation of bottomonium bound states. Some CNM effects could even cause
a sequential suppression dependent on the binding energy of the $b\bar{b}$ 
pairs. The suppression observed in PbPb collisions is a combined
effect caused by both CNM and HNM, while a possible suppression in
proton-lead ($p$Pb) collisions would be attributed to CNM only. 
Consequently, it is essential to study $\Upsilon$ production
in the $p$Pb reference system, as it is representative of the possible non-HNM 
suppression effects. This knowledge can be extrapolated
into the PbPb system so that the fraction of suppression due to HNM in
PbPb collisions can be understood.

The $p$Pb data set collected by CMS in 2013 corresponds to an integrated
luminosity of 31 nb$^{-1}$. The $p$Pb double ratios were measured to
be~\cite{pPb}:
\begin{eqnarray}
\begin{array}{ll}
\displaystyle \frac{[\Upsilon(2S)/\Upsilon(1S)]_{p\text{Pb}}}{[\Upsilon(2S)/\Upsilon(1S)]_{pp}}=0.83\pm0.05(\text{stat.})\pm0.05(\text{syst.}), \\
\displaystyle \frac{[\Upsilon(3S)/\Upsilon(1S)]_{p\text{Pb}}}{[\Upsilon(3S)/\Upsilon(1S)]_{pp}}=0.71\pm0.08(\text{stat.})\pm0.09(\text{syst.}).  
\end{array}
\label{doubleRatio-pPb}
\end{eqnarray}

The results lie slightly below unit, pointing to the presence of possible non-HNM effects. 
Such contributions reflect only as small corrections to the magnitude of the sequential suppression in PbPb, 
which is highly significant and associated to HNM.  A 
comparison between $p$Pb/$pp$ and PbPb/$pp$ double ratios is provided in Fig.~\ref{DR-PbPb-pPb}~\cite{pPb}. 

\begin{figure}[t]
\centering
\includegraphics[width=0.43\textwidth]{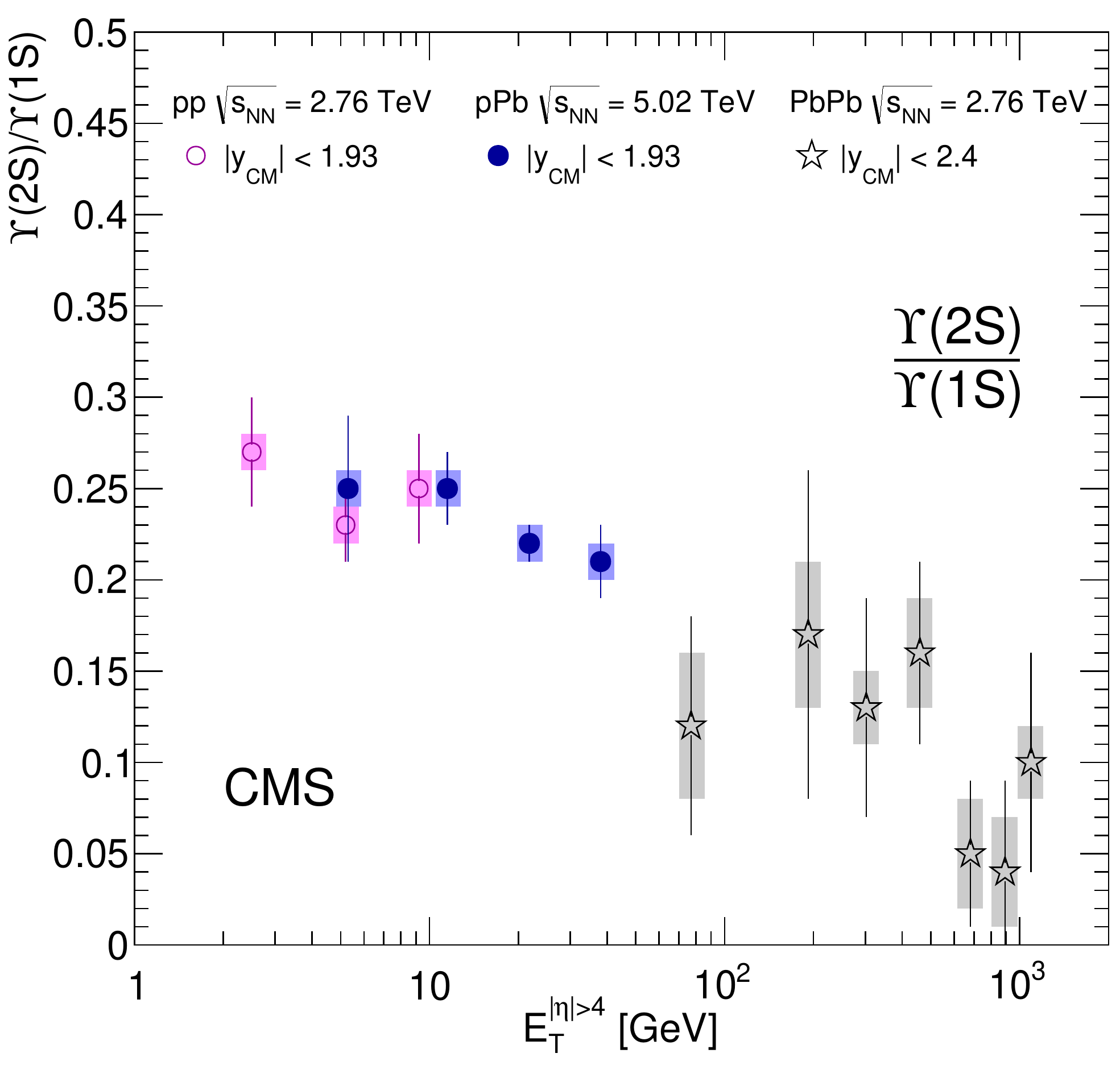} \, 
\includegraphics[width=0.43\textwidth]{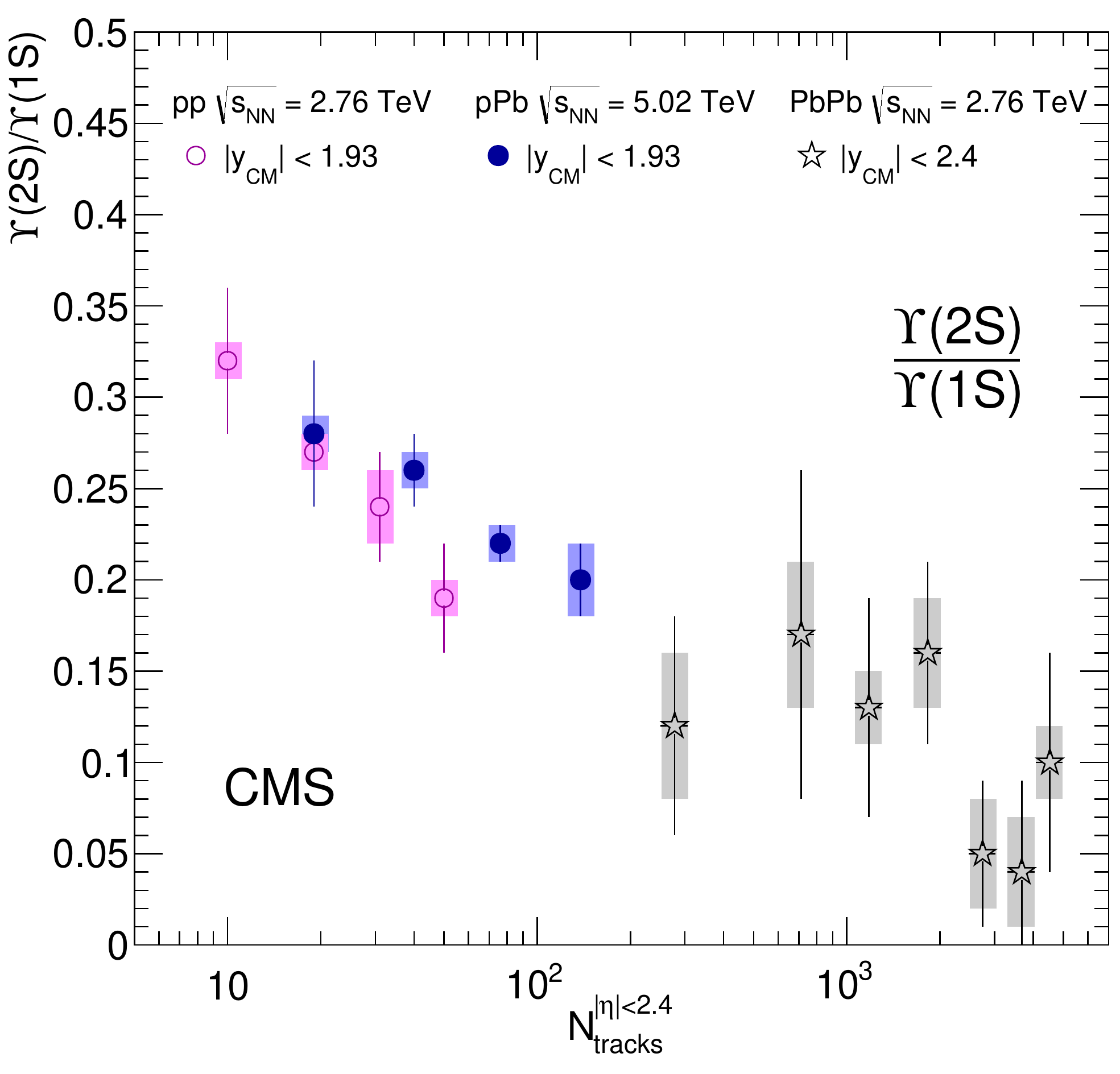} 
\caption{The $\Upsilon(2S)$/$\Upsilon(1S)$ ratio as function of the event activity variables 
in $pp$, $p$Pb, and PbPb collisions.  \label{evtActivity}}
\end{figure}

The $\Upsilon(nS)$/$\Upsilon(1S)$ single ratios were also studied in the three collision systems, 
along with their dependency on event activity variables. In the left
panel of Fig.~\ref{evtActivity}~\cite{pPb}, the ratios are plotted with
respect to the raw transverse energy deposited in the most forward part
of the hadron calorimeters at $4.0<|\eta|<5.2$, while in the right
panel, the ratios are plotted with respect to the number of charged
particle tracks, excluding the two muons, with $\pt>400$\,\MeV and at
$|\eta|<2.4$. The tracks are required to originate from the same vertex
as the muon pairs.

\begin{figure}[h!]
\centering
\includegraphics[width=0.45\textwidth]{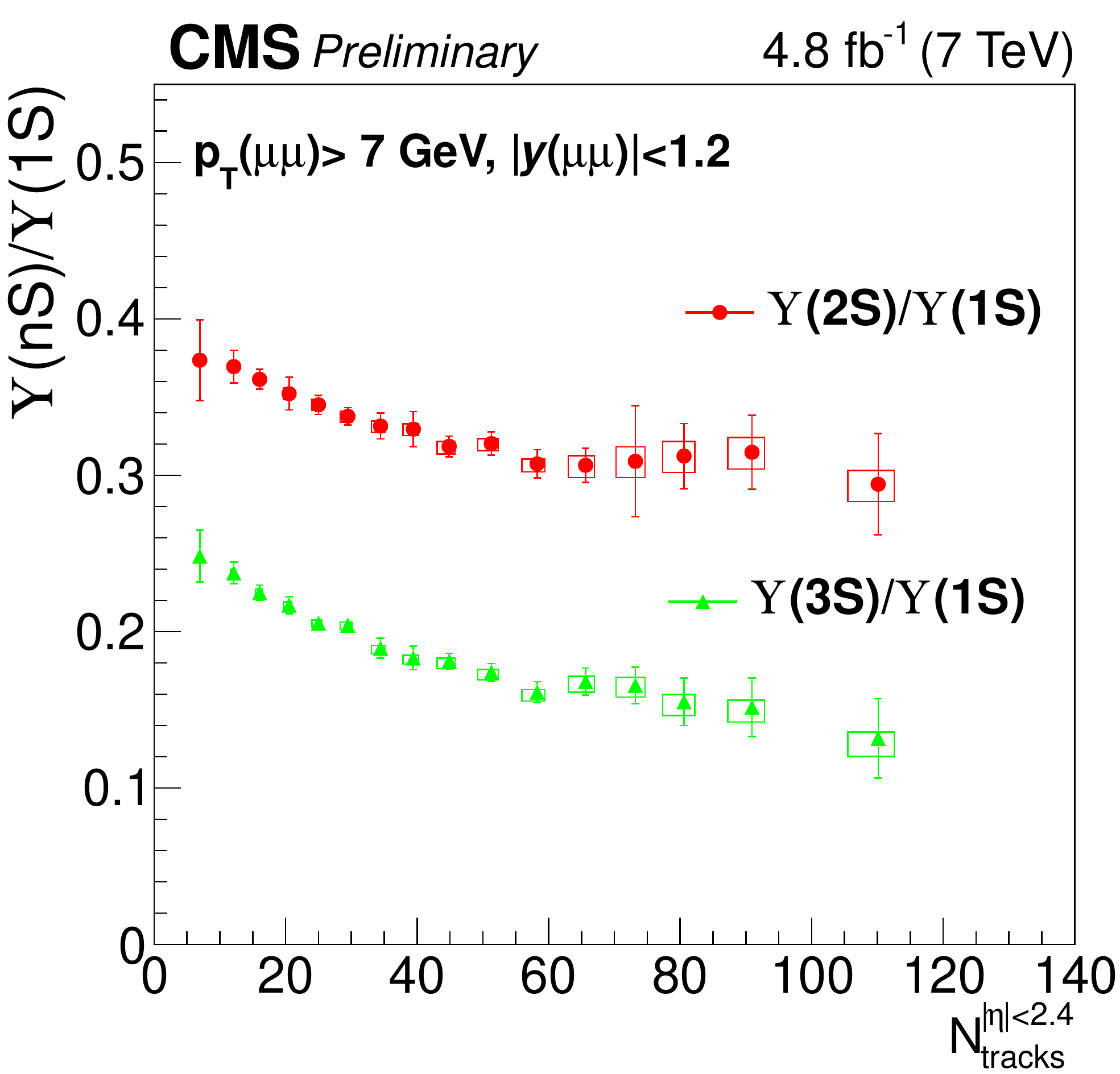}
\includegraphics[width=0.45\textwidth]{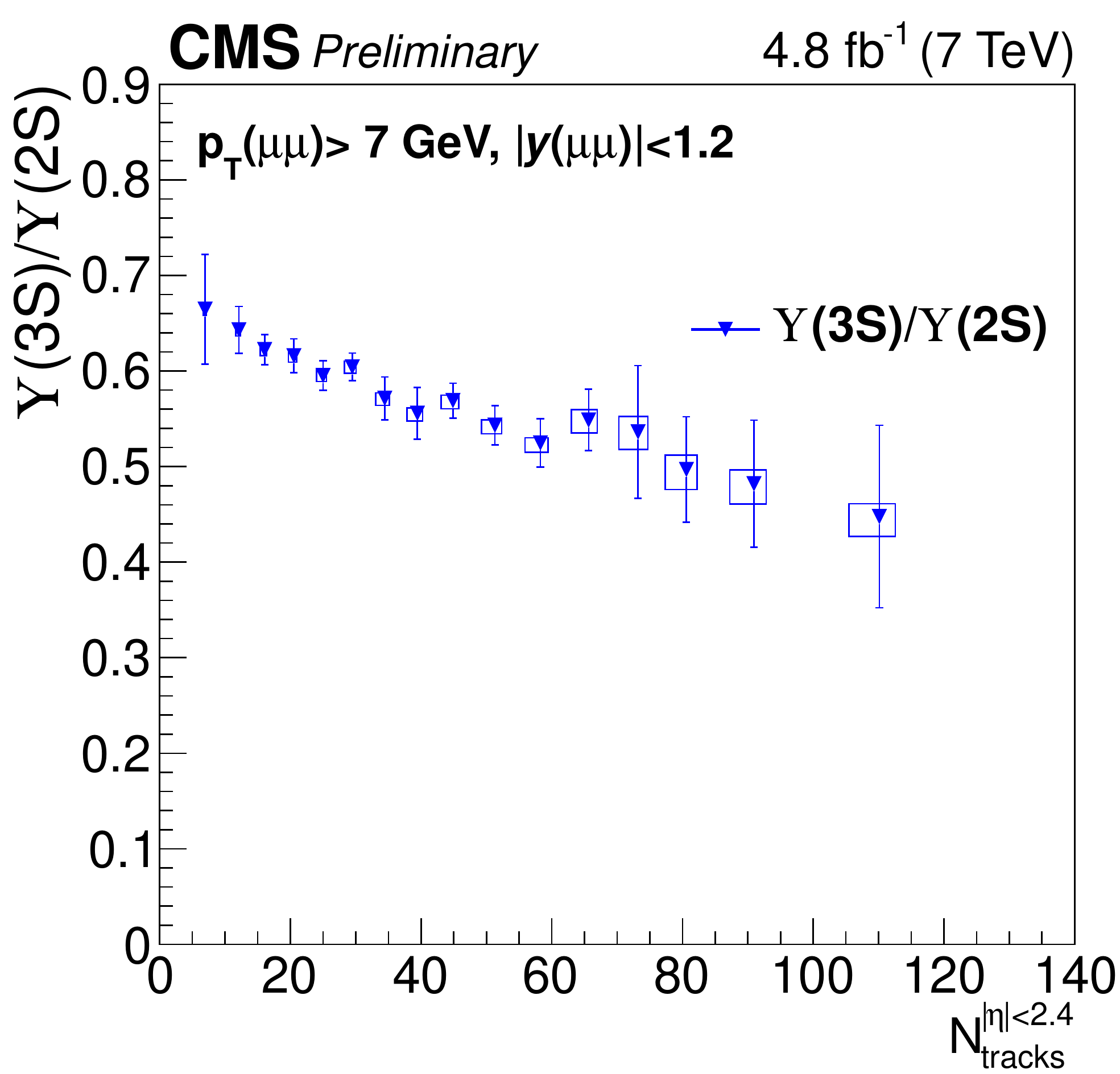}
\caption{Production ratios for $\Upsilon(nS)/\Upsilon(1S)$ (left) and $\Upsilon(3S)/\Upsilon(2S)$ (right) 
as a function of the number of charged particles, $N^{|\eta|<2.4}_{\mathrm{tracks}}$. 
The $\Upsilon(nS)$ mesons are required to satisfy $\pt> 7$\,\GeV and $|y|< 1.2$. 
Error bars represent statistical uncertainties, while empty squares show the systematic uncertainties.~\cite{BPH-14-009} } 
\label{ratios-multiplicity}
\end{figure}

\begin{figure}[h]
\centering
\includegraphics[width=0.32\textwidth]{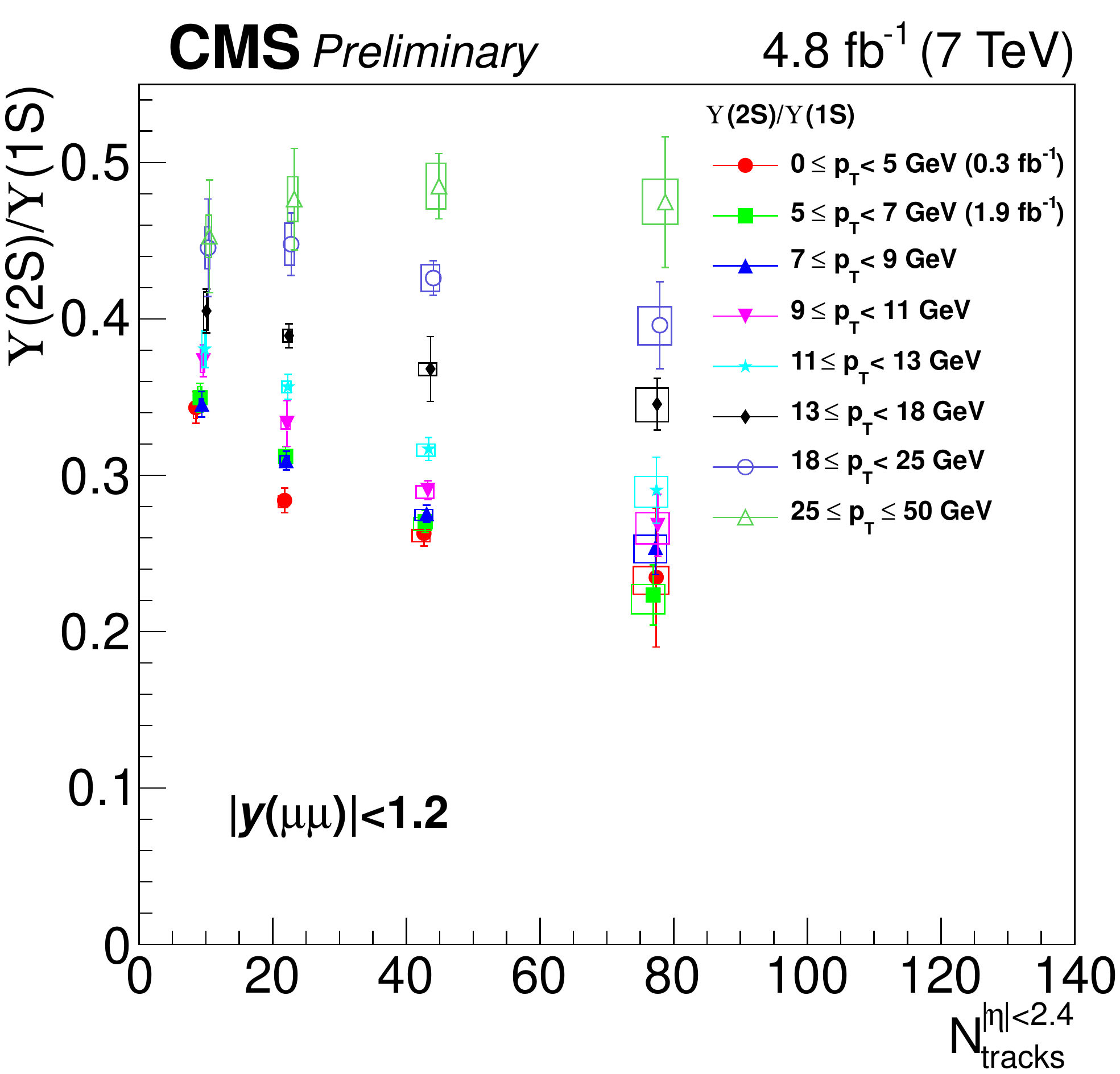}
\includegraphics[width=0.32\textwidth]{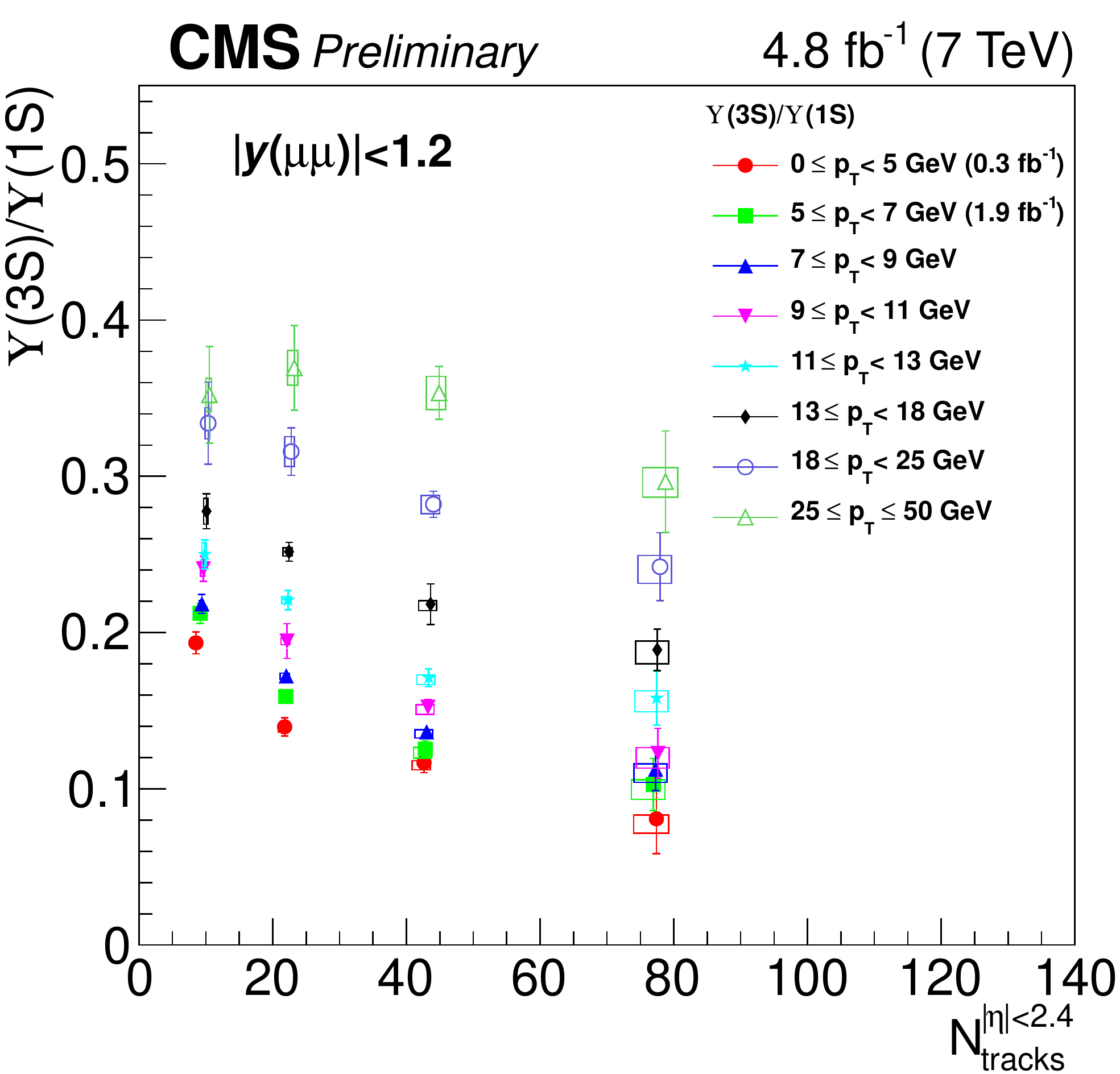}
\includegraphics[width=0.32\textwidth]{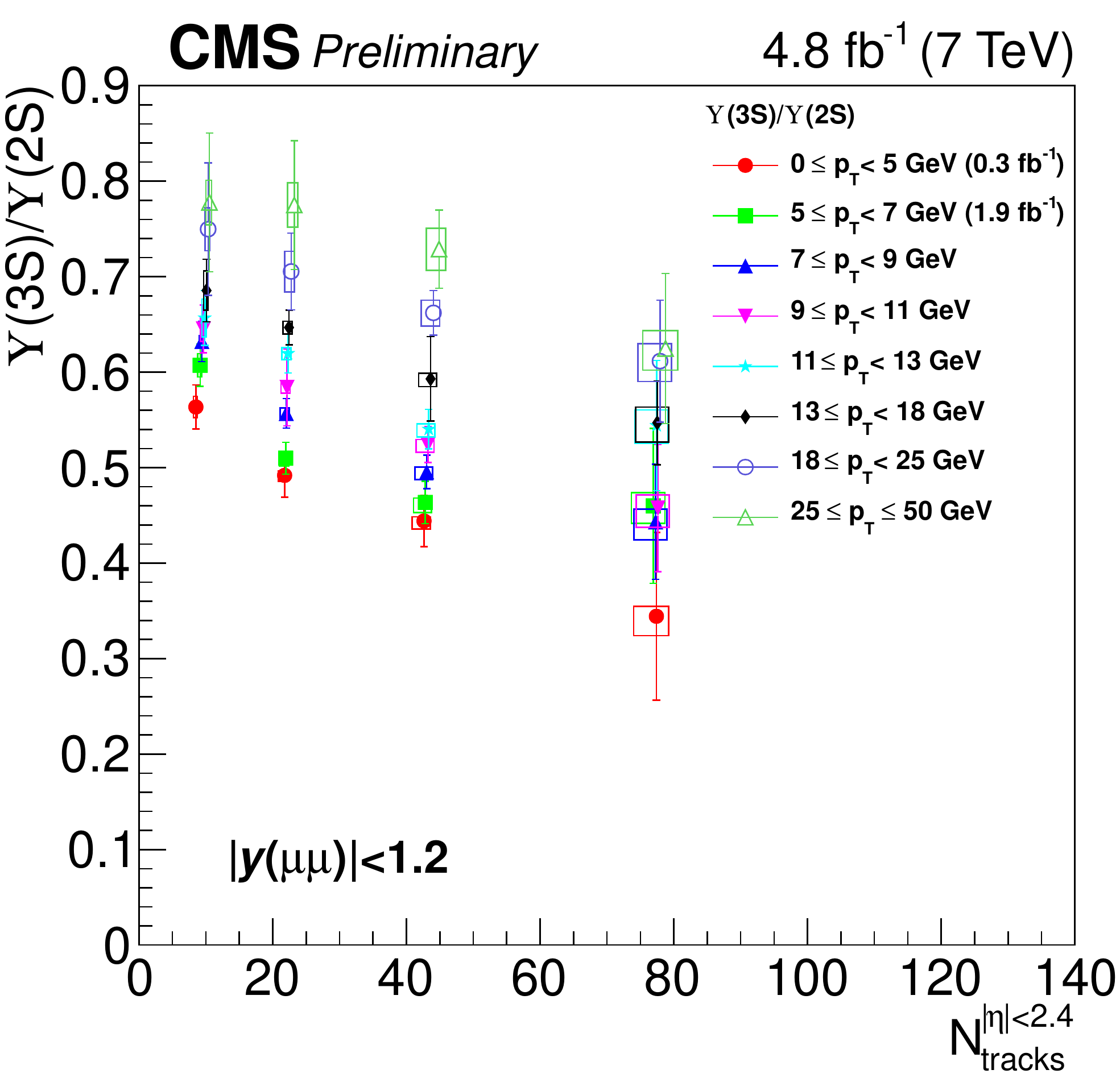}
\caption{Production ratios vs multiplicity in different regions of
$\Upsilon$ \pt, within $|y|< 1.2$. Error bars represent statistical
uncertainties, while empty squares show the systematic
uncertainties~\cite{BPH-14-009}. }    
\label{ratios-multiplicity-pt}
\end{figure}

The decrease of the $\Upsilon(2S)$/$\Upsilon(1S)$ ratio in
Fig.~\ref{evtActivity} may indicate the presence of some new
phenomena in quarkonium production in $pp$ collisions. Other
results recently published by the 
LHC experiments~\cite{collective1,collective2,collective3,collective4,collective5,
collective6} can also be interpreted as hints of collective effects in
the high multiplicity $pp$ environment at the attained 
energies~\cite{collective7,collective8}. However, it is still not clear
whether the medium produced in $pp$ collisions could undergo a phase
transition, as observed in PbPb collisions. 

A search was performed for possible new features in the bottomonium
yields as the particle multiplicities increase in $pp$
collisions~\cite{collective9,collective10}. Using the $pp$ collision
data collected at $\sqrt{s}=7$\,\TeV by the CMS experiment, the ratios of
the cross sections of the $\Upsilon(nS)$ mesons were investigated, as
shown in Fig.~\ref{ratios-multiplicity}~\cite{BPH-14-009}. The
production ratios for $\Upsilon(2S)/\Upsilon(1S)$,
$\Upsilon(3S)/\Upsilon(1S)$, and $\Upsilon(3S)/\Upsilon(2S)$ are
displayed as a function of the number of charged particles with $\pt >
0.4$\,\GeV and $|\eta| < 2.4$. All these ratios clearly decrease with
increasing multiplicity. Figure~\ref{ratios-multiplicity-pt} shows the
production ratios in different \pt regions~\cite{BPH-14-009}. The
decrease with increasing multiplicity is again present and is stronger
in the lower \pt region. The behavior at higher \pt is flatter, especially 
so for the $\Upsilon(2S)/\Upsilon(1S)$ ratio 
as indicated in Fig.~\ref{ratios-multiplicity-pt} (left). Overall, the
observed decrease in the production ratios at $\sqrt{s}=7$\,\TeV shows a
similar trend as that at $\sqrt{s}=2.76$\,\TeV (Fig.~\ref{evtActivity}). 


\section{$P$-wave states}

Measurements of $P$-wave quarkonium production can be used to help further
understand the hadron formation mechanism and test NRQCD predictions.
The production ratios $\chi_{c2}/\chi_{c1}$ have been measured by
CMS~\cite{chic-CMS}, LHCb~\cite{chic-LHCb}, and ATLAS~\cite{chic-atlas},
providing valuable insight into the quarkonium production mechanism. The
$\chi_{b1,b2}$ production cross section measurement is also important
since it is particularly sensitive to color-octet
contributions~\cite{chib-theory, chib-CMS}. The measurement is
challenging because of the small separation (only 19.4 MeV) between the
$\chi_{b1}(1P)$ and the $\chi_{b2}(1P)$ peaks, as well as their small
cross sections. The $P$-wave bottomonium production cross section ratio,
$\sigma(\chi_{b2}(1P))/\sigma(\chi_{b1}(1P))$, was measured with the
20.7 fb$^{-1}$ $pp$ collision data taken at $\sqrt{s}=8$
TeV~\cite{chib-CMS}. The $\chi_{b1}(1P)$ and $\chi_{b2}(1P)$ states are
detected through their radiative decays
$\chi_{b1,2}(1P)\rightarrow\Upsilon(1S)\,\gamma$, where the
$\Upsilon(1S)$ decays to two muons, and the photon is reconstructed
through its conversion to a $e^+e^-$ pair in the inner layers of the
tracker. Although the yield of reconstructed conversion photons is small, the
four-momentum of the photon can be precisely determined through a fit to
the electron and positron tracks in the tracker. With this strategy, the
resulting mass resolution (of the order of 5 MeV) of the $\chi_b$
candidates is sufficient to separate the $\chi_{b1}(1P)$ and
$\chi_{b2}(1P)$ peaks. 

\begin{figure}[t]
\centering
\includegraphics[width=0.45\textwidth]{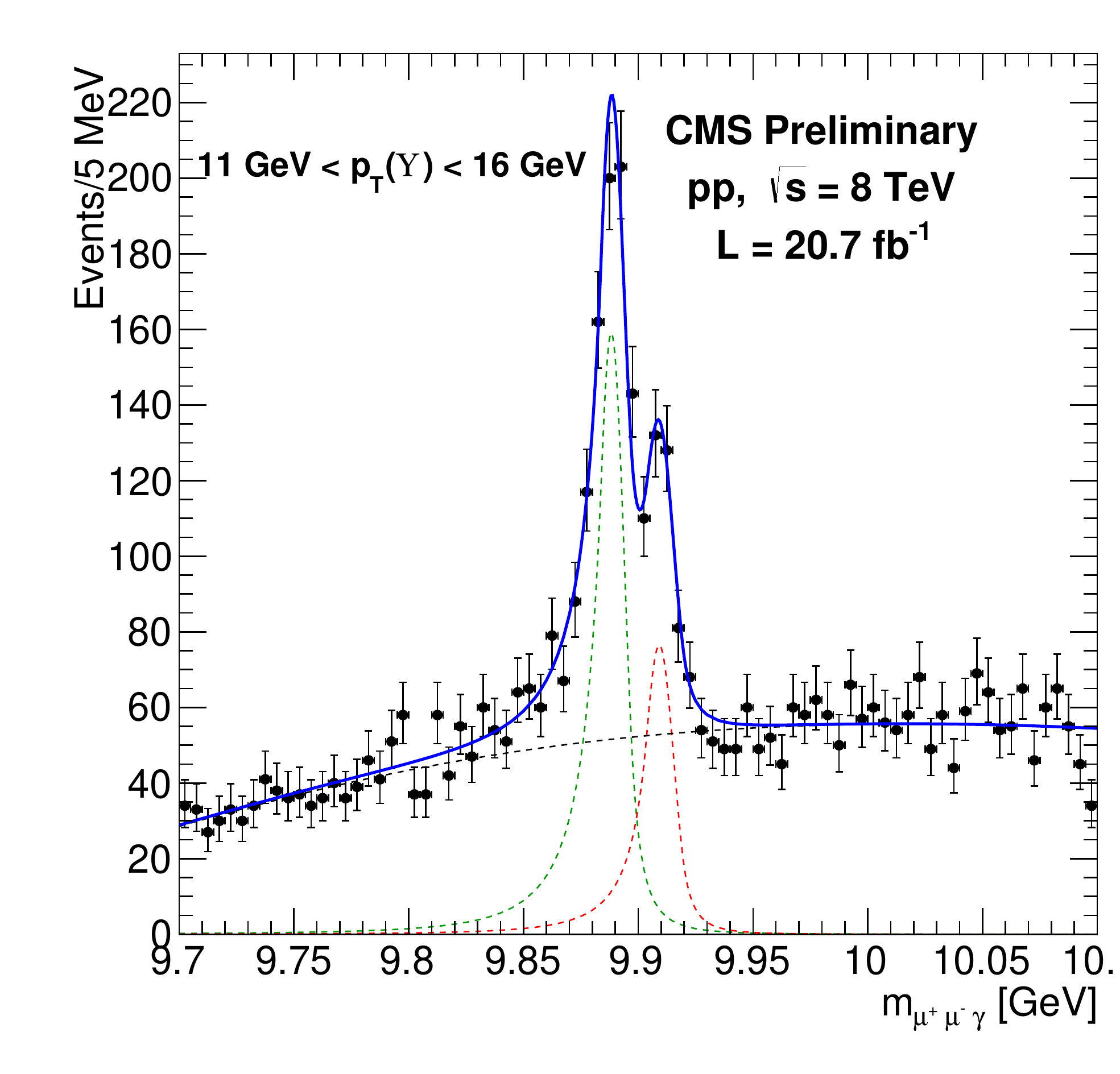}
\includegraphics[width=0.45\textwidth]{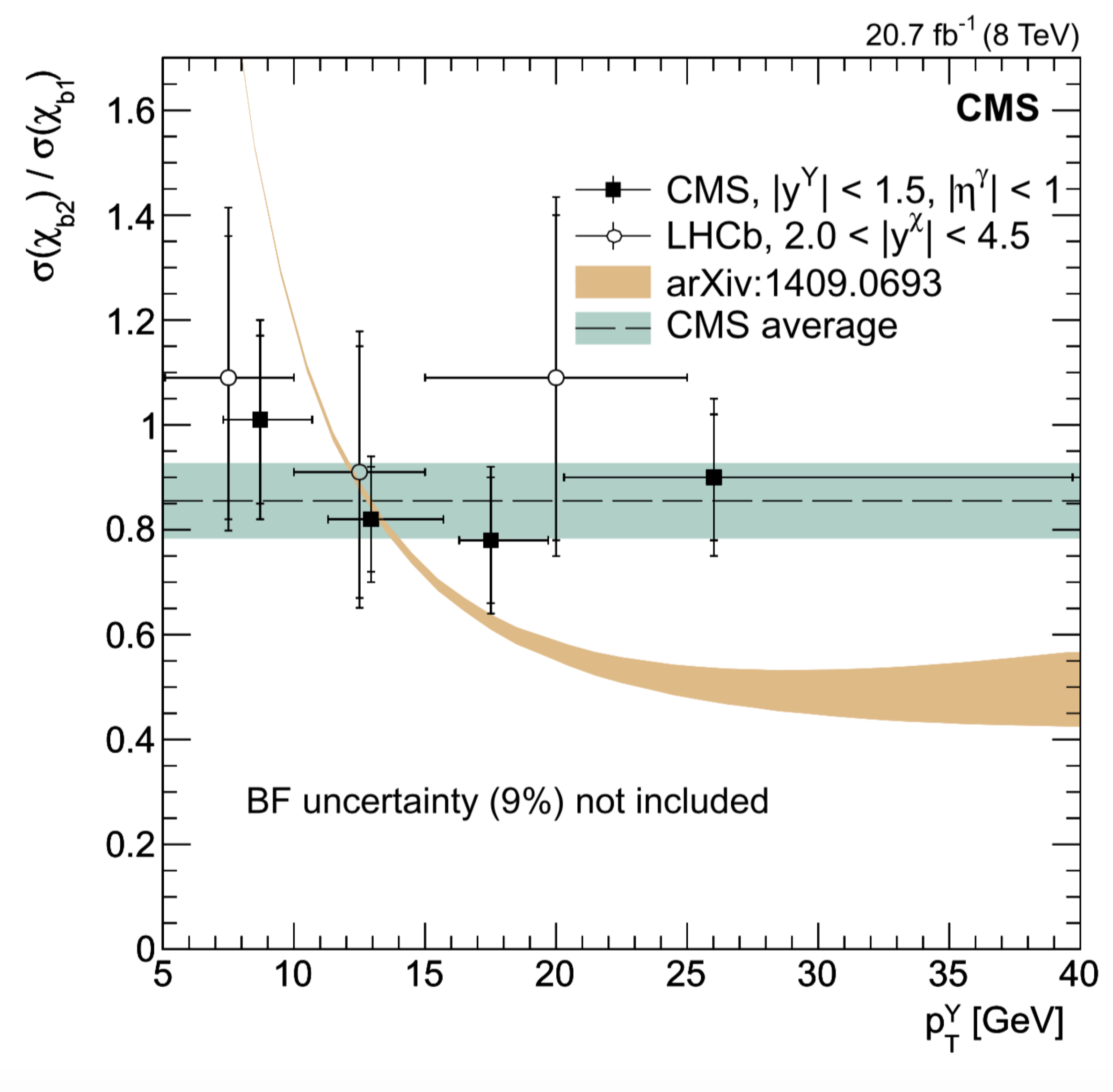} 
\caption{Left: the $\mu\mu\gamma$ invariant mass distribution. 
The dashed green and dashed-dotted red lines indicate the $\chi_{b1}(1P)$ and $\chi_{b2}(1P)$ contributions. 
Right: the $\sigma(\chi_{b2}(1P))/\sigma(\chi_{b1}(1P))$ ratio as a function of $\Upsilon(1S)$ \pt, 
measured by LHCb~\cite{chib-LHCb} and CMS\cite{chib-CMS}. The vertical error bars represent 
the statistical (inner bars) and total (outer bars) uncertainties. The dashed horizontal line is a fit of 
a constant to the CMS result and the horizontal band is the total uncertainty of the fit, 
which does not include the 9\% uncertainty in the ratio of branching fractions. 
The band represents the ratio predicted by a theoretical calculation~\cite{chib-NRQCD}. \label{chib}}
\end{figure}

The $\sigma(\chi_{b2}(1P))/\sigma(\chi_{b1}(1P))$ ratio was measured as
a function of $\Upsilon(1S)$ \pt.  The invariant mass distribution of
the $\mu\mu\gamma$ candidates is shown in
Fig.~\ref{chib}~(left)~\cite{chib-CMS}, where the $\chi_{b1}(1P)$ and
$\chi_{b2}(1P)$ states are visible. Figure~\ref{chib} (right) gives the
production cross section ratio as a function of $\Upsilon(1S)$ \pt
measured by CMS, and a comparison with the LHCb
result~\cite{chib-LHCb} and an NRQCD theoretical
calculation~\cite{chib-NRQCD}. Because of the lack of $\chi_{b}$
measurements, the calculation is based on the previous charmonium
measurements of $\sigma(\chi_{c2})/\sigma(\chi_{c1})$~\cite{chic-CMS,
chic-LHCb, chic-atlas}, but scaled for the case of bottomonium. Neither
CMS nor LHCb results agree with the band predicted by the theory. More
studies are needed in order to thoroughly test NRQCD predictions in the
$P$-wave bottomonium sector.

\section{Pair production}
\label{sec-Ypair}

The measurement of quarkonium pairs originating from a common vertex in
$pp$ collisions provides additional insight into the underlying
mechanisms of particle production at the LHC. The study of quarkonium
pair production is essential to understand contributions of SPS
(single-parton scattering) and DPS (double-parton scattering) processes.
It also forms the baseline for the search of quarkonium pair resonances,
which are predicted by several
studies~\cite{Berezhnoy:2011xy,Berezhnoy:2011xn}. In this section, the
first observation and cross section measurement of $\Upsilon(1S)$ pair
production, reported by CMS based on 20.7 fb$^{-1}$ of $pp$ collisions
at $\sqrt{s}=8$ TeV, is discussed~\cite{Khachatryan:2016ydm}.

\begin{figure}[h!]
\begin{center}
\includegraphics[width=0.48\textwidth]{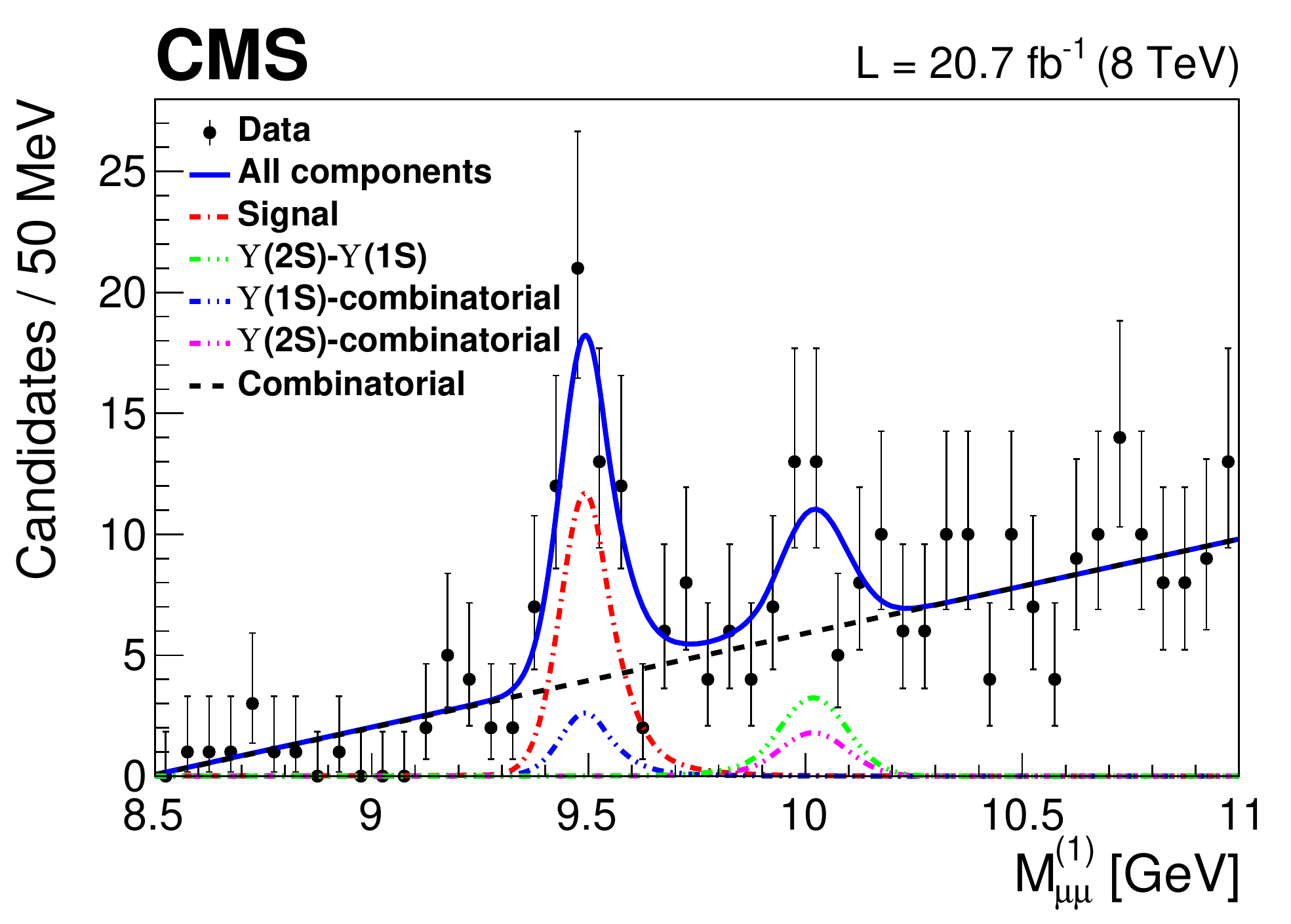}
\includegraphics[width=0.48\textwidth]{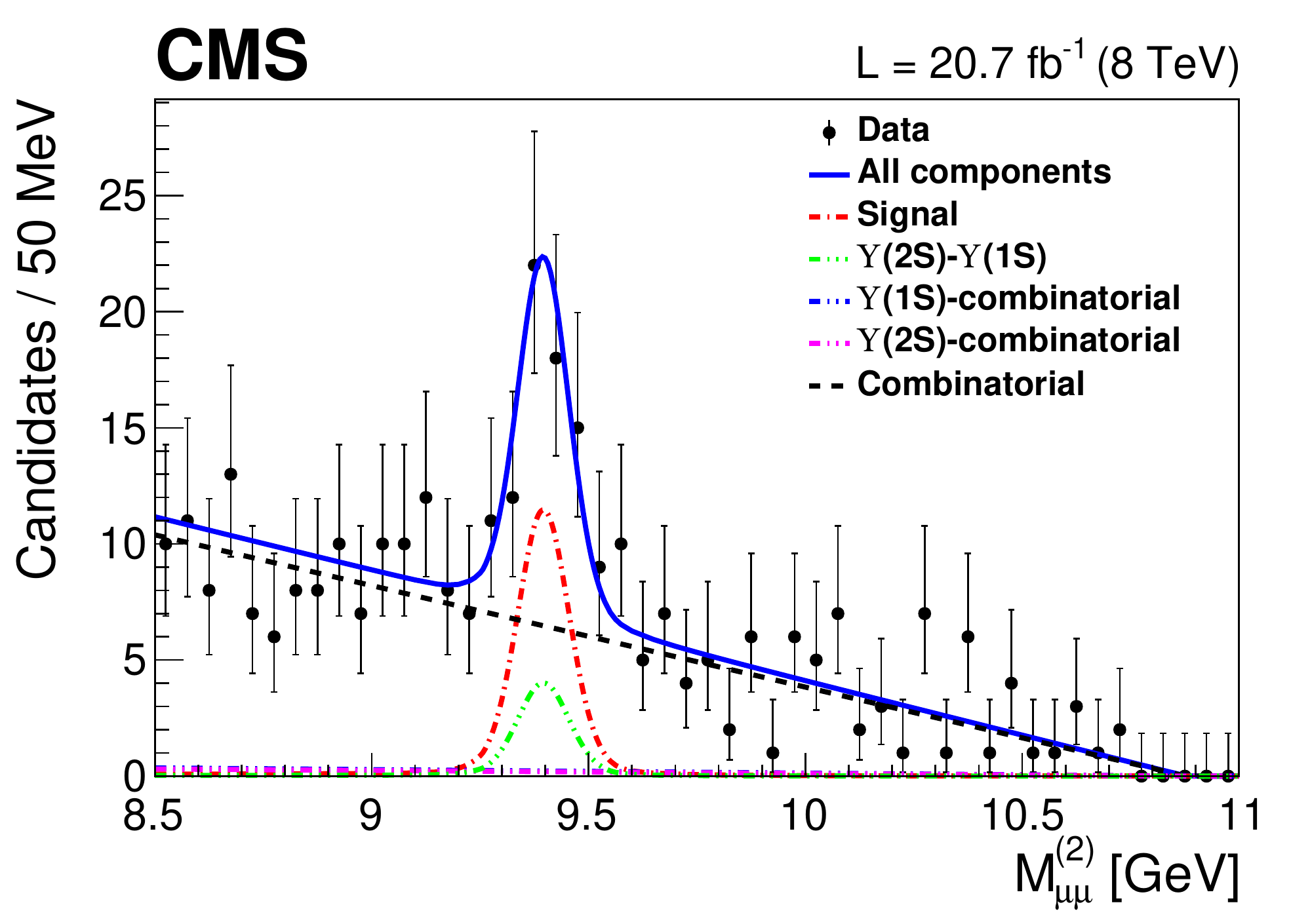}
\caption{Invariant mass distributions of the higher-mass muon pair
(left) and the lower-mass muon pair (right).  The different curves show
the contributions of the various event categories from the fit.}
\label{2d_fit_result}
\end{center}
\end{figure}

Proton collisions at hadron colliders are described by parton models.
Each colliding hadron is characterized as a collection of free
elementary constituents. In a single hadron-hadron collision, two
partons often undergo a single interaction (SPS), although the composite
nature of hadrons permits multiple distinct interactions
(multiple-parton interactions or MPIs) to occur, the simplest case being
DPS. The SPS mechanism for heavy quarkonium pair production can be
described to a good extent by the NRQCD COM approach~\cite{COM-1}. 
However, contributions from the DPS mechanism are not easily addressed within this
framework~\cite{Ko:2010xy}, and DPS or MPIs are sometimes invoked to
account for observations that cannot be explained
otherwise~\cite{Berger2010}. Heavy quarkonium states are expected
to probe the distribution of gluons in the proton since their production
is dominated by gluon-gluon
interactions~\cite{Baranov:2012re,Berezhnoy:2012tu}. The large event
samples at the LHC enable a search for exotic states decaying into heavy
quarkonium, such as tetra-bottom or tetra-charm quark states, and a
measurement of the double $\Upsilon$ cross section provides a benchmark
measurement for these searches~\cite{Berezhnoy:2011xn, KYI}.

In the CMS pair production study~\cite{Khachatryan:2016ydm}, each $pp$ collision is scanned
for a signature of four-muon candidates with the sum of muon charges equal to zero. 
To ensure a uniform muon acceptance and a well-defined
kinematic region, selected muons are required to be within
$|\eta^{\mu}|<2.4$ and have $\pt > 3.5$\,\GeV. An $\Upsilon$ candidate is
reconstructed by combining two oppositely charged muons that originate
from a common vertex. In constructing the $\Upsilon\Upsilon$ candidates, the two
pairs of muons are assigned as (i) the $\mu^{+}\mu^{-}$ invariant mass of
the higher-mass $\Upsilon$ candidate, $M^{(1)}_{\mu\mu}$, (ii)
the $\mu^{+}\mu^{-}$ invariant mass of the lower-mass $\Upsilon$
candidate, $M^{(2)}_{\mu\mu}$.

The signal yield of $\Upsilon(1S)$ pair events is extracted by
constructing a two-dimensional (2D) unbinned maximum likelihood fit to
the invariant mass of the reconstructed $\mu^{+}\mu^{-}$ combinations, 
$M^{(1)}_{\mu\mu}$ and $M^{(2)}_{\mu\mu}$.
Figure~\ref{2d_fit_result} shows~\cite{Khachatryan:2016ydm} the
projection of the invariant mass distributions of the higher and lower
mass muon pairs in the data with the superimposed 2D fit. The efficiency
and acceptance of the detector are computed with a data-embedding method
that repeatedly substitutes the measured muon four-momenta into
different simulated events, which are then subjected to the complete CMS
detector simulation and reconstruction chain.

The cross section of $\Upsilon(1S)$ pair production, 
measured for events in which both $\Upsilon(1S)$ mesons have $|y^{\Upsilon}|<2.0$, 
is determined to be 
\begin{eqnarray}
\begin{array}{rcl}
\displaystyle \sigma(pp \rightarrow \Upsilon(1S)\Upsilon(1S)X)
 =\frac{N_{\mathrm{signal}} \, \bar{w}}{\mathcal{L} \, [B(\Upsilon(1S) \rightarrow \mu ^{+} \mu ^{-})]^{2}}\\[8pt]
 = 68.8 \pm 12.7 \text{(stat.)} \pm 7.4 \text{(syst.)} \pm 2.8 \, (B) \, \text{pb}\,, 
\end{array}
\end{eqnarray}
where $N_{\mathrm{signal}}=38\pm7$ is the measured $\Upsilon(1S)$ pair
yield, $\bar{w}=23.06$ is the average correction factor that accounts
for all the inefficiencies in the measurement, $\mathcal{L}=20.7$
fb$^{-1}$ is the integrated luminosity. The world-average branching
fraction of the $\Upsilon(1S)\rightarrow \mu^{+}\mu^{-}$ decay, $B =
(2.48\pm0.05)\%$, is used, and its uncertainty is quoted separately in
the result. The $\Upsilon(1S)$ mesons are assumed to decay
isotropically. Compared to an isotropic $\Upsilon(1S)$ meson decay, the
cross section is expected to vary up to 40\%, for fully longitudinal or
transverse polarizations of the $\Upsilon(1S)$ meson.

An effective area parameter, $\sigma_\mathrm{eff}$, may be defined~\cite{Baranov:2012re,D0-JpsiY} 
that accounts for the geometric size and the transverse spatial distribution of 
the partonic matter inside the proton. While $\sigma_\mathrm{eff}$ is assumed to 
be independent of the scattering  process, it may depend on the parton flavor. 
The cross section for  $\Upsilon(1S)$ pair production is related to $\sigma_\mathrm{eff}$ as
\begin{equation}
\label{xsection_form}
\sigma_{\text{eff}}  =\frac{[\sigma(pp \rightarrow \Upsilon(1S)X)]^2}{2f_{{\text{DPS}}}\,\sigma(pp \rightarrow \Upsilon(1S)\Upsilon(1S)X)}\,, 
\end{equation}
where $f_{\mathrm{DPS}}$ is the fraction of DPS contribution to the observed cross section. 
An experimental measurement of the fraction $f_{\mathrm{DPS}}$, 
which may be extracted from an analysis of the rapidity or azimuthal angle difference 
between the two produced quarkonia, has not been yet produced, 
given the reduced number of $\Upsilon$ pairs in the current data set.  
Using the single $\Upsilon$ cross section~\cite{xsecPLB} in Eq.~(\ref{xsec40ipb}), 
restricted to the fiducial region ($|y^{\Upsilon}| < 2.0$) adopted in the measurement, 
and assuming $f_{\mathrm{DPS}}$ to be in the range 10--30\%~\cite{Berezhnoy:2012tu}, 
a result of $\sigma_\mathrm{eff}$ in the rage 2--7 mb would be obtained. 
Such an estimate of $\sigma_\mathrm{eff}$ is consistent with the range of values 
obtained from heavy quarkonium studies~\cite{D0-doubleJpsi,D0-JpsiY}, 
but smaller than that from multi-jet studies~\cite{jet1,jet2,jet3,jet4}. 
This variation may be interpreted as resulting from a dependence of $\sigma_\mathrm{eff}$ 
on the dominant partonic interactions in each type of process; 
namely, gluon-gluon interactions for quarkonium production, and quark-antiquark 
and quark-gluon parton interactions for the jet-related channels. 
More data at the LHC will improve the determination of SPS and DPS contributions 
in associated particle production.

\section{Search for new and exotic states}
\label{sec-exo}

\begin{figure}[h]
\centering
\includegraphics[width=0.38\textwidth]{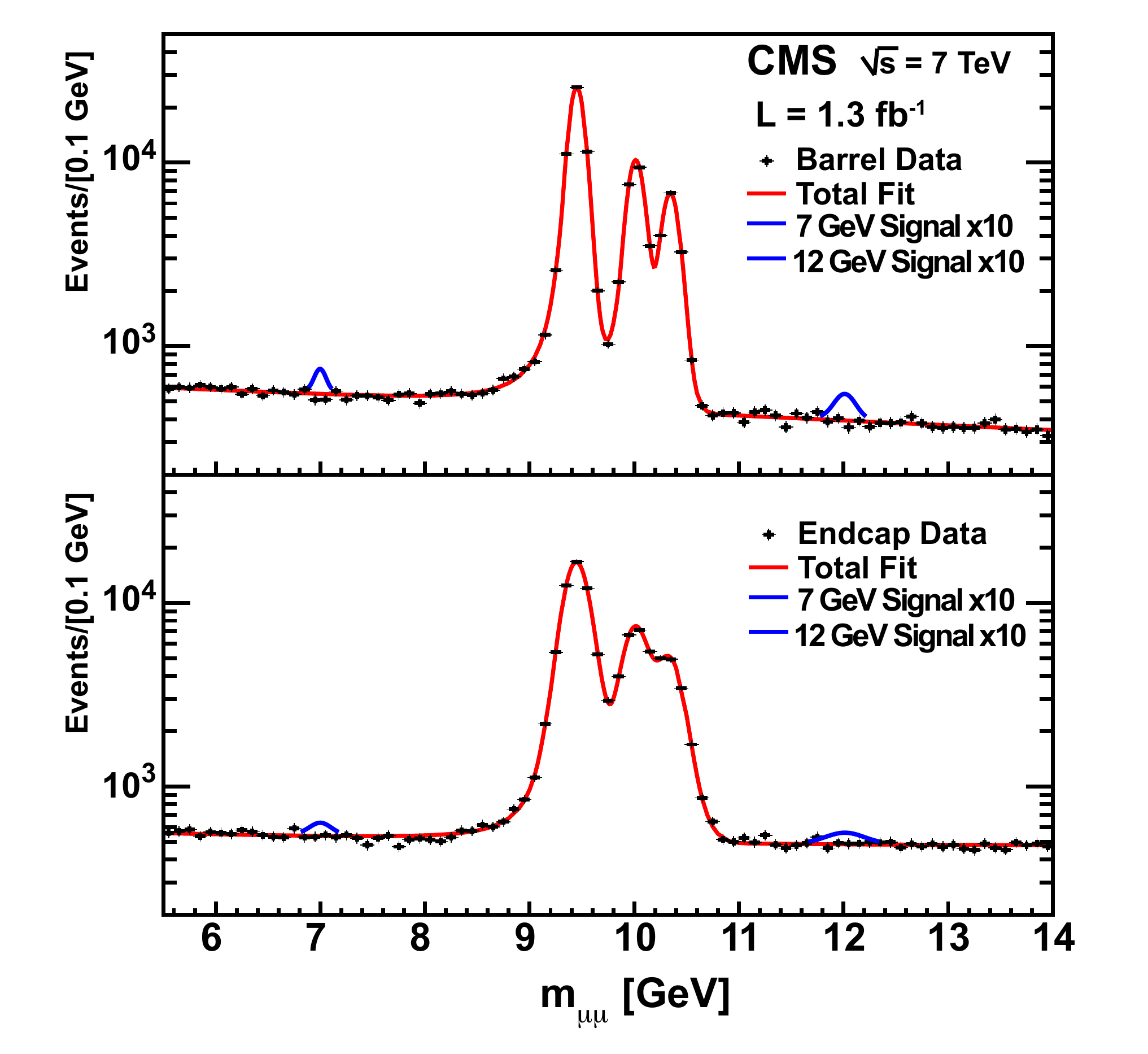}
\includegraphics[width=0.52\textwidth]{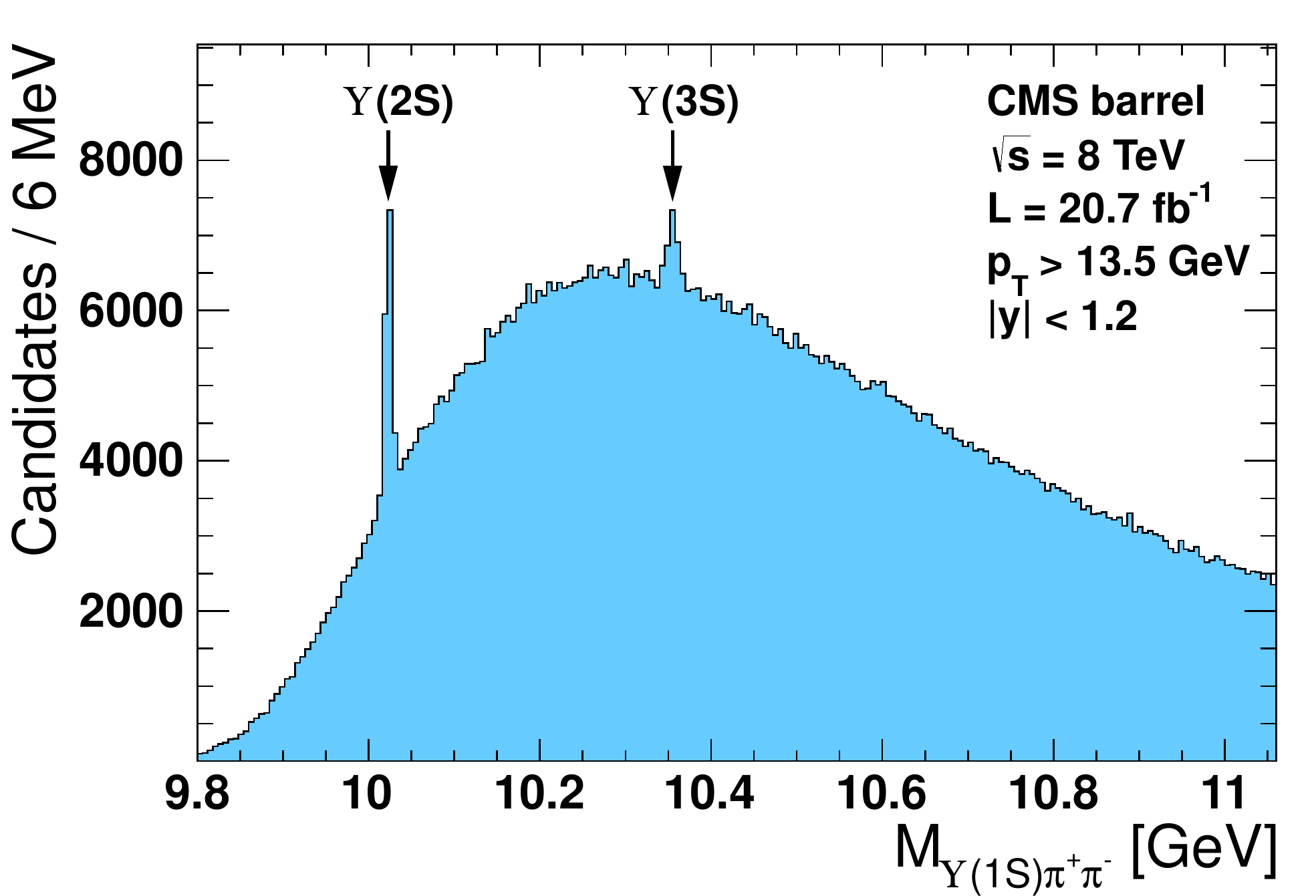}
\caption{Left: search for light states near the $\Upsilon$ mass, showing
hypothetical signals from pseudoscalar Higgs bosons of masses 7 and 12 GeV. Right: Search for an exotic bottomonium state in the $\Upsilon(1S) \pi^+\pi^-$ invariant mass spectrum.
The two pronounced peaks correspond to the $\Upsilon$ excited states decays $\Upsilon(nS) \to \Upsilon(1S) \pi^+\pi^-$.   \label{amumu}}
\end{figure}

Quarkonium states are ``standard candles'' that are used extensively for detector and trigger calibration,
and for extending measurements and searches down to the limit of the detector kinematic acceptance.
The low-mass region is directly probed for new particle states.  
For example, searches are performed for light pseudoscalar states around the $\Upsilon$ mass~\cite{amumu-cms},
as predicted in scenarios such as next-to-minimal supersymmetric models (Fig.~\ref{amumu}, left).
In addition, quarkonium states can be explored as final states in the search for new particles and rare processes. 

Studies of di-quarkonium production, such as reported in Sec.~\ref{sec-Ypair}, open new opportunities 
to search for possible exotic resonance states decaying to quarkonium pairs. 
In the bottomonium case reported above, one may probe for four $b$-quark bound states.~\cite{Berezhnoy:2011xn,KYI}

Various unexpected quarkonium-like states have been identified
since the discovery of the X(3872) state by BELLE more than a decade ago~\cite{x-belle}.
This exotic charmonium was discovered in the final state $J/\psi \pi^+\pi^-$,
tagged through $B$ meson decays, and its prompt production has been confirmed  
by various experiments, including CMS~\cite{x-cms}. 
A search is conducted for an exotic bottomonium counterpart~\cite{y2pi-cms}. 
This is done probing the corresponding final state $\Upsilon(1S) \pi^+\pi^-$,
as shown in Fig.~\ref{amumu} (right). 
Besides the excited $\Upsilon$ states that are expected (Fig.~\ref{Bottomonium}) and appear prominently in the spectrum,
no excess is detected over background, despite the analysis having the
sensitivity for detecting such a state if its relative strength were   
comparable to the corresponding value for the X(3872). 
The search yielded the first exclusion limits on the production of the exotic bottomonium state at a hadron collider.

\section{Conclusions and outlook}

The CMS experiment has already contributed a set of very significant
results towards an improved understanding of the heavy quarkonium sector
at the LHC. The CMS detector acceptance, trigger, and reconstruction
capabilities have allowed the prompt collection of large data sets
containing dimuon signals, down to low transverse momenta, resulting in
the ability to probe a wide mass spectrum as shown in
Fig.~\ref{dimuonMass}. While the muon detectors contribute particle
identification and the seeds for online selection, the silicon tracker
allows for precise kinematic and topological reconstruction of particle
decays. In particular, the precision achieved allows for separately
identifying lightest three $\Upsilon(nS)$ states -- a capability that
is fully exploited in the measurements reported here.  

CMS has contributed the first bottomonium measurements at the LHC. Cross
sections are among the first results extracted with initial data, both
in LHC's Run 1 and Run 2 data-taking periods. Measurements in $pp$
collisions at the unprecedented center-of-mass energies of 2.76, 5.02,
7, 8, and 13\,\TeV have been undertaken, within the rapidity window of
$|y|<2.4$, and dimuon momenta up to 100\,\GeV.
Cross sections and cross section ratios have been measured for $S$-wave
and $P$-wave states. In addition, the angular distribution of the
final-state muons has been analyzed in complementary reference frames,
resulting in detailed measurements of the polarization parameters. This
measurement has been reported for all three $S$-wave states. The cross
sections and the polarizations, combined in global fits, have shed
considerable light on the QCD production mechanisms,
contributing towards the resolution of the ``quarkonium
puzzle''.  
Analysis of new LHC data will extend the kinematic reach of the
measurements and, in addition, extend the cross section and
polarization measurements to the $P$-wave states. These will be important
pieces in the puzzle, and will allow the disentanglement of the $P$- to
$S$-wave feed-down contributions, thus granting access to directly
produced bottomonia.  

Ground breaking results have also been achieved in collisions involving
heavy ions. The considerable jump in collision energy and detector
capability, as compared to previous heavy ion experiments, has placed
CMS in a privileged position to explore the bottomonium sector in
nuclear collisions. In particular, the experiment has delivered the
first complete measurements of the individual states of the $\Upsilon$
family in collisions involving heavy ions. Foremost, the ability to
separately identify the individual $\Upsilon(nS)$ states has been
explored to probe their relative suppression. Such a novel and robust
analysis has experimentally established the pattern of sequential
suppression, wherein the excited states are more suppressed than the
smaller, more strongly bound states. The absolute suppression, in PbPb
with respect to $pp$, of the individual states has also been assessed.
Detailed measurements of bottomonia, in PbPb but also $p$Pb and $pp$,
have probed (hot and cold) nuclear and environment effects. 
The analysis of larger LHC data sets will explore $P$-wave states,
quantifying the suppression of higher-mass states and their feed-down
contribution to the inclusive measurements of the lighter ones. It will
allow also to explore new observables, such as azimuthal anisotropies
and polarizations through angular analyses, and to perform further studies
of kinematic, angular, and environment dependencies. The continued
exploration of the bottomonium sector, through more precise and new
measurements, across different collisions systems and energies, will
provide a more complete understanding of the underlying processes
contributing to quarkonium production and suppression, and a more
complete characterization of the properties of the hot medium attained
with LHC collisions. 
Bottomonia can be explored in addition via photoproduction in collisions of protons or ions at the LHC. 
In such exclusive collisions, the incoming  hadrons remain intact after the interaction and 
no other particles are produced, e.g. $pA \to pA\Upsilon$~\cite{CMS-PAS-FSQ-13-009}. 
Such studies will offer a clean probe of the structure of the target hadron (A).

Production of bottomonium pairs has been observed for the first time and
the cross section of $\Upsilon(1S)$ pair production was reported.
Extended and more precise studies of quarkonium pair production will
more precisely quantify the effect of multi-parton interactions at the
LHC. Additional studies of bottomonia associated production, involving
other quarkonium states and, more generally, other hadrons, jets, and
vector bosons, will contribute new perspectives towards a more complete
understanding of the mechanisms of hadron production. At the same time,
such studies of associated production open an interesting window into
the spectroscopy realm, forming the basis for the search for new and
exotic states. Rare decays of heavier particle states, e.g. $Z$ and Higgs,
to heavy quarkonium provide interesting channels that should be explored with
increased precision in future higher luminosity LHC runs.  

Precision heavy flavor measurements with a general purpose detector at
the LHC are challenging, required an understanding of fine-grain effects
and calibration of the detector in a low-\pt regime, with a thorough
appreciation of the corresponding increasing online rate limitations. 
Building upon the solid achievements already attained, and benefitting
from detector and data acquisition systems upgrades, a dedicated and
promising program of heavy-quarkonium physics will continue to be
pursued in future LHC runs, aiming at contributing to a more complete
understanding of the nature of strong interactions and the search for
new phenomena.

\section*{Acknowledgments}

We congratulate our colleagues in the CERN accelerator departments for
the excellent performance of the LHC and thank the CMS institutes for their
contributions to the success of the CMS effort.
Z. Hu acknowledges partial support by the United States Department of Energy. 
N. T. Leonardo acknowledges partial support by Portugal's Foundation for Science and Technology  grant IF/01454/2013/CP1172/CT0003. 
In addition, we are grateful to Leonard Spiegel, Don Lincoln, Ao Liu, Kai Yi, Sergo Jindariani, Tingjun Yang, Zijun Xu, Wanwei Wu, 
as well as the CMS B-Physics group and the CMS Heavy-Ion group, 
for the useful discussions and valuable comments.

\end{document}